\documentclass{aa}
\usepackage{txfonts}
\usepackage{footnote}
\usepackage{epsfig}
\usepackage{natbib}

\begin{document}

\title{The AGB population of NGC 6822: distribution and the C/M ratio from JHK photometry}

\author{L. F. Sibbons\inst{1}
        \and 
        S. G. Ryan\inst{1}
        \and
        M.-R. L. Cioni\inst{1,2,}\thanks{Research Fellow of the Alexander von Humboldt Foundation}
        \and
        M. Irwin\inst{3}
        \and 
        R. Napiwotzki\inst{1}
        }

\offprints{L.Sibbons1@herts.ac.uk}

\institute{University of Hertfordshire, Physics Astronomy and Mathematics, Hatfield AL10 9AB, United Kingdom
           \and
           University Observatory Munich, Scheinerstrasse 1, 81679 M\"{u}nchen, Germany
           \and
           University of Cambridge, Institute of Astronomy, Madingley Rd., Cambridge CB3 OHA, United Kingdom
           }

 \date{Received 31 Oct 2011/ Accepted 3 Feb 2012}

 \titlerunning{AGB Population in NGC 6822}
 
 \authorrunning{Sibbons et al}

 \abstract{NGC 6822 is an irregular dwarf galaxy and part of the Local
   Group. Its close proximity and apparent isolation provide a unique
   opportunity to study galactic evolution without any obvious strong external
   influences.}{This paper aims to study the spatial distribution of the
   asymptotic giant branch (AGB) population and metallicity in NGC
   6822.}{Using deep, high quality $JHK$ photometry, taken with WFCAM
   on UKIRT, carbon- and oxygen-rich AGB stars
   have been isolated. The ratio between their number, the C/M ratio,
   has then been used to derive the [Fe/H] abundance across the
   galaxy.} {The tip of the red giant branch is located at $K_0 = 17.41 \pm 0.11$ mag and
   the colour separation between carbon- and oxygen-rich AGB 
   stars is at $(J-K)_0 = 1.20 \pm 0.03$ mag (i.e. $(J-K)_{2MASS} \sim 1.28$ mag). A C/M ratio of 
   $0.62 \pm 0.03$ has been derived in the inner $4$ kpc of the galaxy, 
   which translates into an iron abundance of [Fe/H] $= -1.29 \pm 0.07$
   dex. Variations of these parameters were investigated as a function of distance from the
   galaxy centre and azimuthal angle.}{The AGB population of NGC 6822 has been detected 
   out to a radius of $4$ kpc giving a diameter of $56$ arcmin. It is metal-poor, but there is no obvious gradient in 
   metallicity with either radial distance from the centre or azimuthal angle. 
   The detected spread in the TRGB magnitude is consistent with that of a galaxy surrounded 
   by a halo of old stars. The C/M ratio has the potential to be a very useful 
   tool for the determination of metallicity in resolved galaxies but a better 
   calibration of the C/M vs. [Fe/H] relation and a better understanding of the 
   sensitivities of the C/M ratio to stellar selection criteria is
   first required.\thanks{Tables. 2, 3 and 4 are only
   available in electronic form at the CDS via anonymous ftp to
   cdsarc.u-strasbg.fr (130.79.128.5) or via
   http://cdsweb.u-strasbg.fr/cgi-bin/qcat?J/A+A/}} 

\keywords{techniques: photometric - stars: AGB and post-AGB - stars:
  carbon - galaxies: irregular - galaxies: dwarf}
 
\maketitle

\section{Introduction}
\label{intro}
\begin{figure}
\resizebox{\hsize}{!}{\includegraphics[scale=0.3]{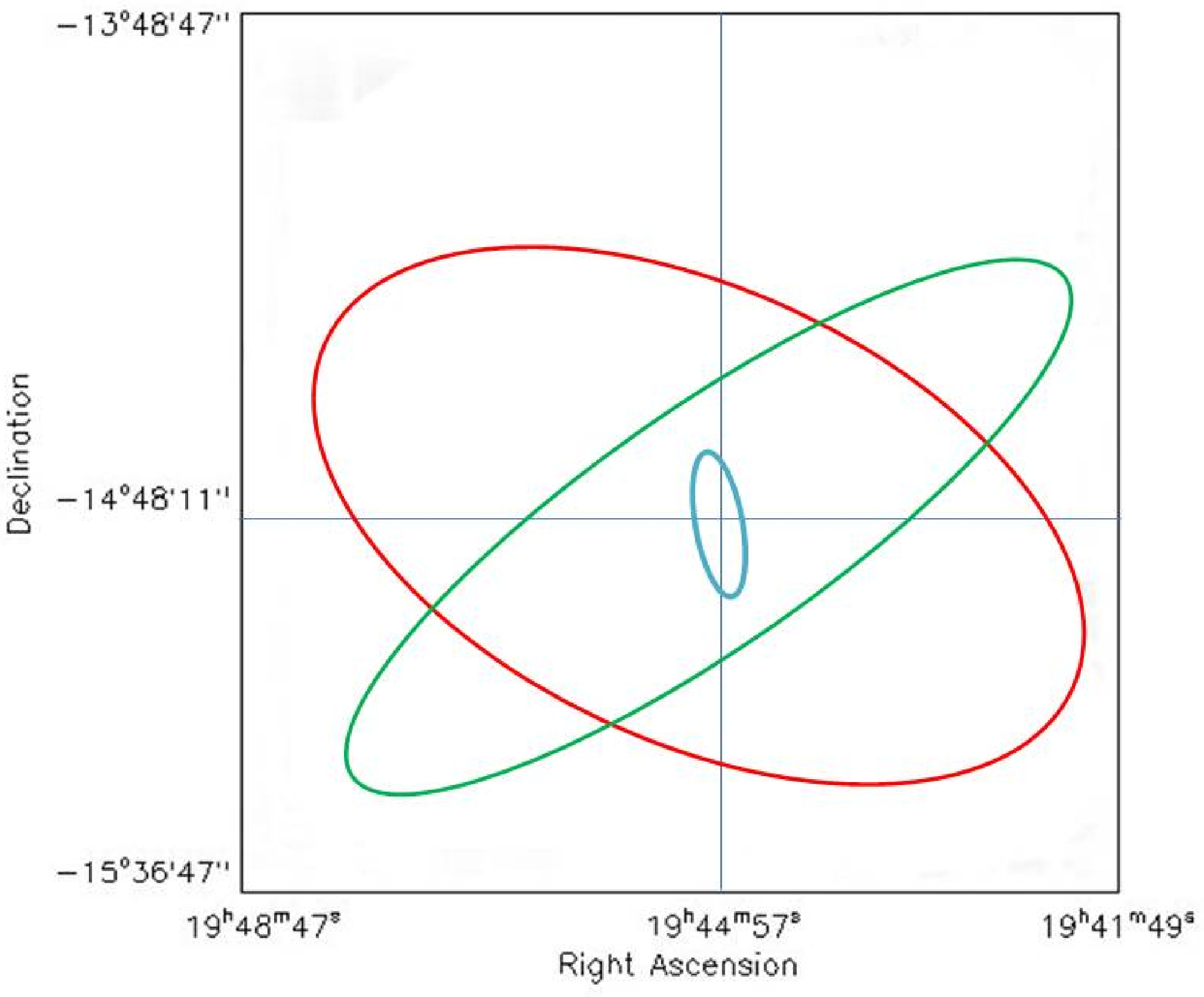}}
\caption{Schematic diagram showing the relative scale and
orientation of the three main structural components of
NGC 6822. The bar of the galaxy (blue)
\citep{1991ApJ...379..621H}, the HI envelope (green) 
\citep{2000ApJ...537L..95D,2006A&A...451...99B} and the spheroid
(red) \citep{2006A&A...451...99B}}
\label{cartoon}
\end{figure}

Forming part of the Local Group, NGC 6822 is an irregular dwarf galaxy
(dIrr) similar to the Small 
Magellanic Cloud (SMC). At a distance of $\sim 490$ kpc,$(m-M)_{0}=
23.45 \pm 0.15$ mag, \citep{1998ARA&A..36..435M,1993ApJ...417..553L} it is
the closest `independent' dIrr galaxy beyond the Magellanic
Clouds. Its close
proximity and apparent isolation have made NGC 6822 a popular candidate
for studies of galactic evolution, without the strong gravitational
influences of other systems \citep{2006A&A...451...99B}.
The morphology of the galaxy can be broadly divided into three
structural components (Fig.\ref{cartoon}); firstly, a central bar
which contains much of the young stellar population is clearly visible
at optical wavelengths and is orientated almost
in a north-south direction
\citep{1991ApJ...379..621H,2009A&A...497..703K}. This bar is embedded
in a large envelope 
of neutral hydrogen oriented in a roughly SE-NW direction. Although
this kind of HI structure is not unique in the Local Group -- IC 1613
and IC 10 have similar structures
\citep{2003MNRAS.340...12W,1972IAUS...44...12R} -- NGC 6822 is unusual
in that the HI envelope is so much more extended than the main optical
component. A third, halo-like structure made up of old- and
intermediate age stars has been detected by
\citet{2002AJ....123..832L} and is approximately $\sim 1$ degree along the
major axis. This elongated spheroidal structure is positioned
orthogonally to and is dynamically decoupled from the HI 
envelope \citep{2006A&A...456..905D,2006AJ....131..343D,2006A&A...451...99B}.

The detection of RR Lyrae stars in NGC 6822
\citep{2004ASPC..310...91B,2003ApJ...588L..85C} indicates the presence
of an old stellar population $\sim 11$ Gyr old whereas the many HII and OB
associations confirm that star formation is still ongoing. NGC 6822
has been the focus of numerous investigations of its stellar content
\citep[e.g.][]{2006AJ....131..343D,2003ApJ...590L..17K,1996AJ....112.1928G}
and several estimates have been made of its iron 
abundance. Using optical photometry, \citet{1996AJ....112.1928G} obtained
[Fe/H] $= -1.50 \pm 0.3$ dex from the slope of the red giant branch (RGB). Further analysis of
the RGB population by \citet{2001MNRAS.327..918T} yielded a value of [Fe/H] $=
-1.0 \pm 0.3$ dex from the strength of CaII absorption lines, in agreement 
with the result of \citet{2003PASP..115..635D} who derived the same
[Fe/H] value from the slope of the RGB in the
near-infrared (NIR). Studies of the AGB population by
\citet{2005A&A...429..837C} and \citet{2006A&A...454..717K} using the 
ratio of C- to M-type stars,  the
C/M ratio, have detected a spread in the metallicity of the population
of $\Delta$[Fe/H]$ = 1.56$ dex and $\Delta$[Fe/H]$ = 0.07 - 0.09$ dex 
(between $0.93 \pm 0.03$ and $1.02 \pm 0.03$ dex),
respectively. The difference between the two values is attributed to
differences in the size and location of the observed area. \citet{2003ApJ...588L..85C}
obtained a value of [Fe/H] $= -1.92 \pm 0.35$ dex from the average period
of old RR Lyrae variables. Looking at the younger stellar
population, \citet{2001ApJ...547..765V} derived an 
average value of [Fe/H] $= -0.49 \pm 0.22$ dex from the optical
spectroscopy of A-type supergiants. These results suggest, as expected,
that the chemical enrichment of the interstellar medium in NGC 6822 has been
a continual process due to multiple stellar generations, since
star formation began. 

Cool AGB stars trace the old- and intermediate-age population in
galaxies, and as they are among the brightest objects they are detected
well in the NIR, providing a sample that is relatively
unobscured by dust along the line of sight. The purpose of this paper is to study the
distribution of AGB stars and the metallicity (iron abundance) across the
galaxy. During the AGB phase, mixing mechanisms dredge up triple-$\alpha$ processed
material from the He-burning shell and can cause the dominant metal abundance
in the stellar atmosphere to change from oxygen to carbon. Oxygen-rich
stars have an excess of oxygen atoms in their atmosphere relative
to carbon, which leads to the formation of O-rich molecular
species (i.e. TiO, VO, H$_2$O). Carbon-rich stars have a higher
abundance of carbon atoms relative to oxygen, leading to the formation
of carbonaceous molecules (i.e. C$_2$, CN, SiC). These two types of
stars are known as M- (C/O $<$ 1) and C-type (C/O $>$ 1). Stars in
which the number of carbonaceous molecules equals the number of oxygen
rich molecules (i.e. C/M $\sim 1$) are S-type stars. At lower
metallicities the transformation from an initially O-rich atmosphere to a C-rich
one is easier as fewer dredge-up events are required
\citep{1983ARA&A..21..271I,1978Natur.271..638B}, therefore the ratio
between stars of spectral type C and M should provide an indirect
measure of the local metallicity at the time those stars 
formed. 

The paper is organised as follows: Sect. \ref{obs} presents the
observations and the data reduction process, Sect. \ref{Analysis}
analyses the data and defines the sample of C- and M-type AGB stars,
results are presented in Sect. \ref{results}, followed by a discussion
and conclusions in Sect. \ref{diss} and Sect. \ref{concl},
respectively.

\section{Observations and data reduction}
\label{obs}

Observations were obtained using the Wide Field Camera (WFCAM) on the
$3.8$m United Kingdom Infrared Telescope (UKIRT) in Hawaii during two
runs, in April $2005$ and November $2006$, as part of a large project
to survey the AGB content of Local Group galaxies in the Northern
Hemisphere (PI Irwin). WFCAM comprises four non-contiguous Rockwell-Hawaii-II
infrared detector arrays (HgCdTe $2048 \times 2048$) that can be
utilised to observe an area of $0.75$ deg$^2$ (a tile) on the sky with
a scale of $0.4^{\prime\prime}$ per pixel. A mosaic of four tiles was
obtained in three broad-band filters ($J$, $H$ and $K$)
covering a contiguous area of $3$ deg$^2$ centred on the optical coordinates
of NGC 6822 ($\alpha = 19^h 44^m 56^s, \delta = -14^{\circ} 48^{\prime}
06^{\prime\prime}$). We refer the reader to
\citet{2001ASPC..232..357C} for a more 
detailed description of the WFCAM instrument. The exposure time of
each tile in the $J$
band was $150$ sec, from the co-addition of $3$ exposures of $10$ sec
each taken in a dithered pattern of $5$ positions. In the $H$ and $K$
bands the 
exposure time was $270$ sec from the co-addition of single $10$ sec
exposures, in a $3 \times 3$ micro-stepping following a dithered
pattern of $3$ positions.  
The total exposure time per pixel over the two runs was then $300$ sec
in $J$ and $540$ sec in $H$ and $K$.  

Reduction of the data, including all the standard steps for
instrumental signature removal --flat fielding,
crosstalk, sky-correction and systematic noise-- was completed using
the WFCAM pipeline at the Institute of Astronomy in Cambridge. Sources
extracted using the pipeline were given a morphological classification
from which assorted quality control measures are computed. Astrometric
and photometric calibrations were performed based on the 2MASS
point source catalogue 
\citep{2009MNRAS.394..675H,2004SPIE.5493..411I}. The photometric 
measures are based on aperture photometry, with \textit{zeropoints}
calibrated  
against 2MASS although they are \textit{not} transformed into the
2MASS system  
\citep{2009MNRAS.394..675H}. In other words the magnitudes and colours 
we quote are on the WFCAM instrumental system; transformation equations 
are given in \citet[eq. $4-8$]{2009MNRAS.394..675H}. 

\begin{figure} 
\resizebox{\hsize}{!}{\includegraphics[scale = 0.5]{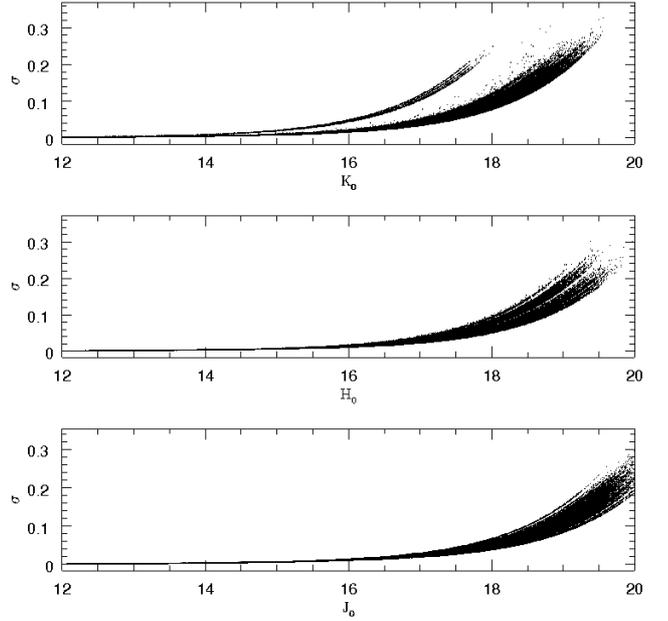}}
\caption{Magnitude vs. error for each photometric
    band. The branch that splits away from the main body of sources at
    $\sim 15$ mag in the $K$-band shows the increased error associated with
    the subset of data from the NE region that was collected during a technical fault.}
\label{comp2}
\end{figure}

Duplicated sources were removed 
using the photometric error and the morphological classification to
select a `best' unique entry per object, to produce a final catalogue
containing $\sim 375,000$ sources. Most of which are, as we show in
Sect. \ref{foreground}, Milky Way (MW) foreground sources. With the
exception of one pointing in the NE that 
suffered from technical difficulties in the form of jittering causing
oblong images in one set of the K-band observations,
the typical seeing across the two observing runs was between $\sim
0.9-1.1^{\prime\prime}$. Figure \ref{comp2} shows error vs magnitude
for each source in all three bands. The effect  of the technical fault on
the $K$-band observations can be seen in the top panel, sources from
that tile have a higher error for a given magnitude.\\Total  
reddening values across NGC 6822 have been found to vary widely from
E(B-V) = 0.24 in the outer regions to E(B-V) = 0.45-0.54 in the centre
\citep{2009A&A...505.1027H,1995AJ....110.2715M}. Here, no corrections were made for
internal reddening. Corrections for the foreground component were
made using the extinction map of \citet{1998ApJ...500..525S}. All
magnitudes and colours are  
presented in their extinction-corrected (dereddened) form, denoted by
the subscript `$0$'.

\begin{figure} 
\resizebox{\hsize}{!}{\includegraphics[scale = 0.3]{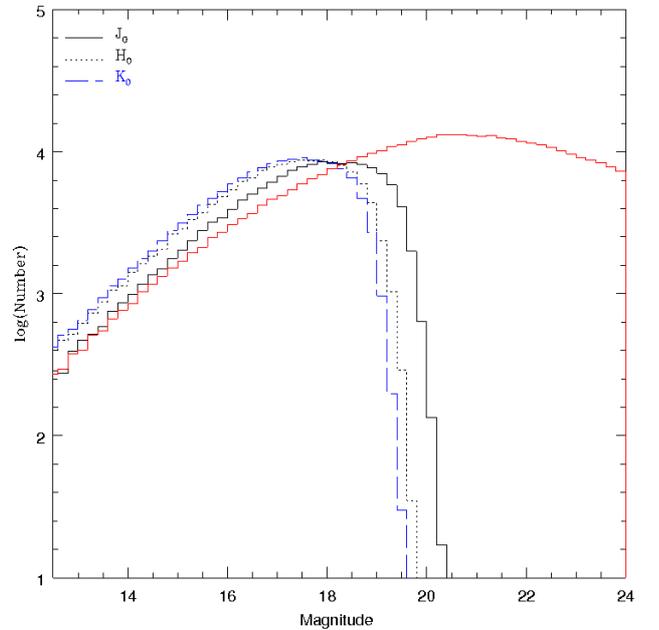}}
\caption{Log number vs. linear magnitude distribution 
of sources in each photometric band in 0.2 mag bins. The completeness
level of each band has been inferred and is discussed in the text - Sect. \ref{obs}.}
\label{comp}
\end{figure}

The maximum depth reached in each photometric band was 
$20.61, 20.00$ and $19.61$ mag in $J, H$ and $K$ respectively. The 
completeness limit for each band has been inferred from
Fig. \ref{comp} which shows the logarithmic distribution of 
the magnitudes in each band for our sample and the magnitude 
distribution of a synthetic MW foreground generated using the 
population synthesis code TRILEGAL \citet{2005A&A...436..895G}. 
The shape of the sample distribution will be slightly 
different to that of the MW distribution due to the presence of 
NGC 6822 sources but the distributions do show some similar features. 
Both continue to increase along a similar line until they reach 
a peak and then start to decline. Whilst in the synthetically 
generated population this is due to some change in the population, 
in the sample distribution we believe the rollover, which occurs at 
brighter magnitudes, is the effect of decreasing completeness in the sample 
after the peak. Therefore, the data are assumed to be complete up to 
the peak of the distribution and we estimate the completeness at fainter 
magnitudes by normalising to the observed star counts at the peak of each band. 
In the $J$-band we are $100\%$ complete to a depth of $17.9$ mag, falling  
to a completeness of $50\%$ between $19.3-19.5$ mag. In the $H$-band we are  
$100\%$ complete to a depth of $17.9$ mag, falling to the $50\%$ level 
between $18.7-18.9$ mag and in the $K$-band (including the poorer quality 
data) we are $100\%$ complete to a depth of $17.5$ mag, declining to $50\%$ 
complete between $18.5-18.7$ mag. For comparison, the completeness 
levels for the $K$-band in the NE only are $100\%$ down to $17.3$ mag 
falling to the $50\%$ level between $\sim 18.5-18.7$ mag. AGB sources at 
the distance of NGC 6822 are expected to have an apparent magnitude
brighter than $K_0 = 17.5$ mag, therefore we are confident that 
our sample is sufficiently complete for the purposes of this study.

\begin{figure*}
\resizebox{\hsize}{!}{\includegraphics[scale=0.3]{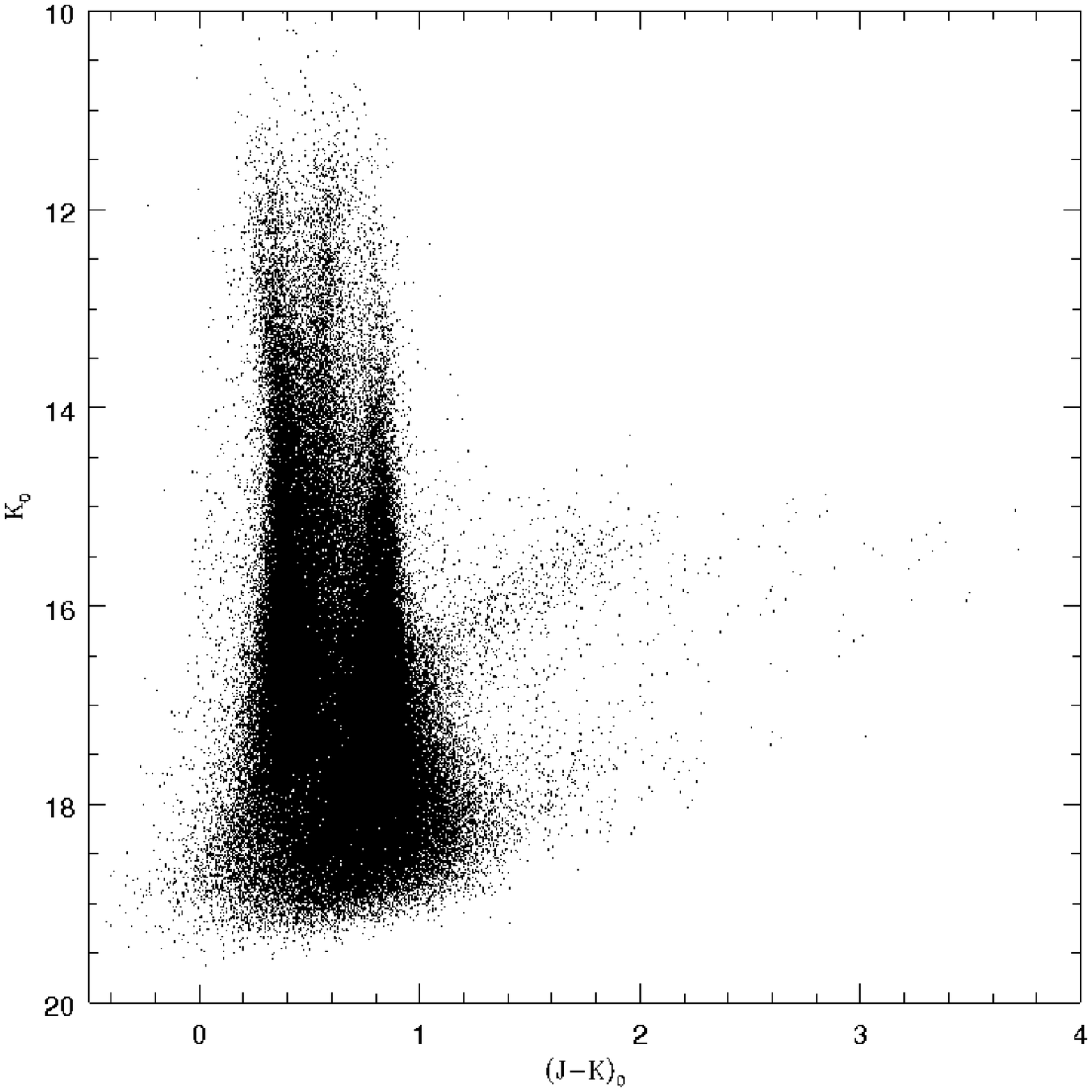}
\includegraphics[scale=0.3]{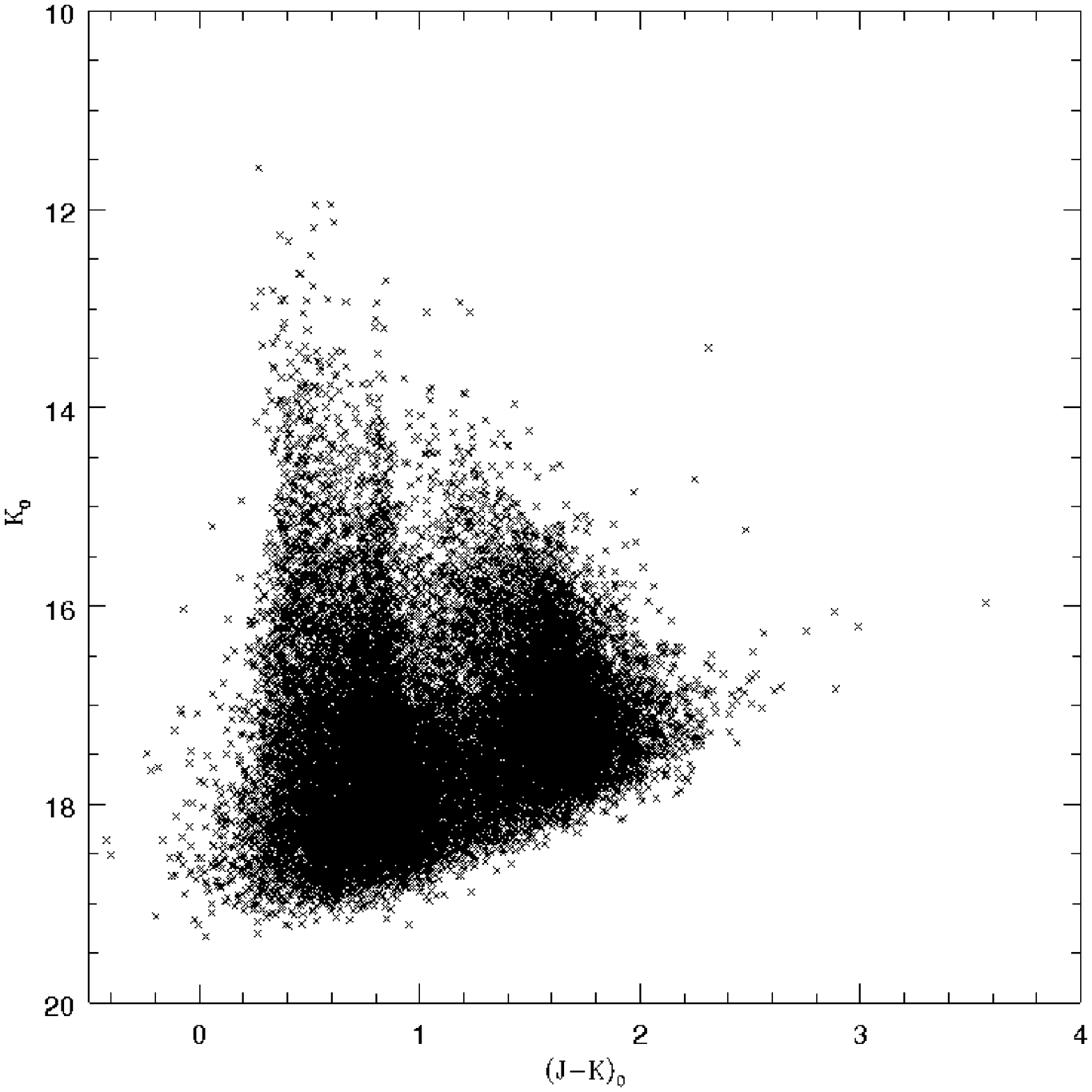}
\includegraphics[scale=0.3]{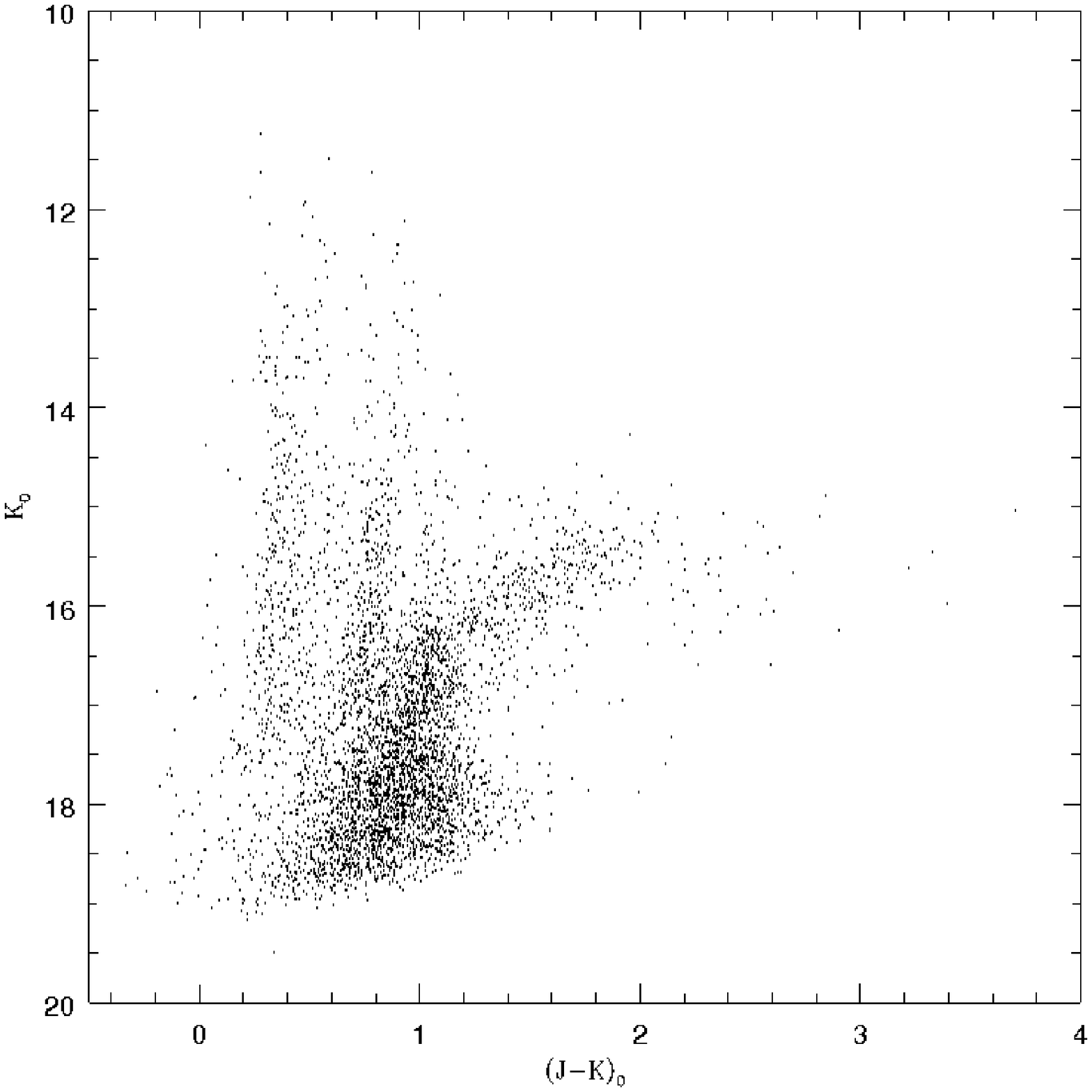}}
\caption{Left: CMD of all the stellar and probably-stellar 
sources, including Milky Way foreground and genuine NGC 6822 stars. 
Middle: CMD of the non-stellar sources from the original photometric 
catalogue in the direction of NGC 6822. Right: CMD of the stellar and
probably-stellar sources from the centre of the observed area - MW
foreground contamination is still present but the CMD is dominated by
sources belonging to NGC 6822. The peak belonging to the M-type stars
and the diagonal branch generated by the C-type stars are more easily
distinguished here.} 
\label{flags1}
\end{figure*}

Sources in the photometric catalogue are flagged as 
stellar, probably stellar, compact but non-stellar, noise like,
saturated, a poor-match with the astrometric data or non-stellar, in
each band. This source classification is based on the flux 
curve-of-growth for a series of apertures; a similar method has been 
used in the IPHAS survey and is discussed in \citet{2008MNRAS.388...89G}. 
Considering only those sources with the same flag in all
three bands, no compact non-stellar or poorly-matched sources
remained; $469$ saturated objects ($K<12.75$), $1703$ noise-like sources,
$21400$ non-stellar, $449$ probably-stellar and $139900$ stellar
sources were left. To ensure a reliable data set for the subsequent
analysis only sources consistently detected in all three photometric
bands and classified as stellar or probably-stellar in each band were used. 
A colour-magnitude diagram (CMD) of these sources across the full observed 
area is shown in the left-hand panel of Fig. \ref{flags1}. Some of 
the sources listed as non-stellar are probably in fact stellar; 
comparing the first two panels of Fig. \ref{flags1}, many of the 
non-stellar sources with $(J-K)_0 < 1.20$ mag occupy the same region of the CMD 
diagram as the stellar sources. However, due to the problems of crowding 
or their being close to the detection limit of the data, it was not 
possible to resolve them adequately into individual stars and so they were classified 
as non-stellar and removed from the sample.

\section{Analysis}
\label{Analysis}

\subsection{The foreground}
\label{foreground}
As NGC 6822 ($l = 25^{\circ}.34$, $b = -18^{\circ}.39$) is close to
the Galactic plane of the Milky Way  observations suffer from heavy  
foreground contamination. The extent of this contamination is clearly seen 
in the left-hand panel of Fig. \ref{flags1}. Three vertical fingers at colours
$(J-K)_0 \sim 0.35$, $0.60$ and $0.80$ mag, have been associated with the
following features following \citet{2000ApJ...542..804N}: the bluest finger 
is due to blue supergiants of NGC 6822 and Galactic F-K dwarfs; the second is due
to Galactic foreground stars including K-type dwarfs and giants as
well as young supergiants of NGC 6822; the third results from
Galactic M-type dwarfs and K-, M-type giants, as well as K- and M-type
giants of NGC 6822. This feature merges with a less distinct vertical sequence, relating to the M-type 
AGB population of NGC 6822, up to about $(J-K)_0 \sim1.20$ mag. At 
$(J-K)_0 > 1.20$ mag and brighter than $K_0 = 17$ mag, the C-type 
AGB stars of NGC 6822 occupy a diagonal sequence on the right of the 
CMD - these features are more clearly seen in the right-hand 
panel of Fig. \ref{flags1}. At  magnitudes fainter than $K_0 \sim 17$ mag where
the distinction between the vertical sequences becomes blurred, the 
sources are a mixture of Galactic G-, K- and M-type dwarfs as well
as RGB and early-AGB stars belonging to NGC 6822. 

\begin{figure}
\resizebox{\hsize}{!}{\includegraphics[scale=0.3]{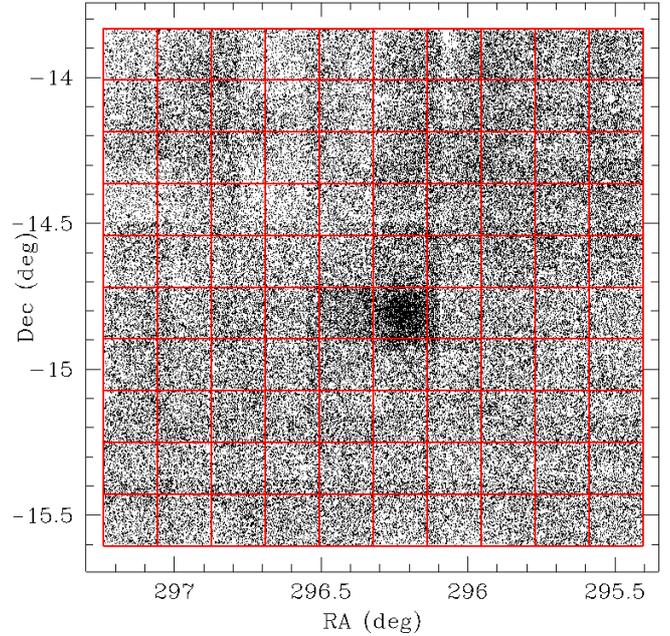}}
\caption{The division of the observed area into a grid of $100$ 
regions, each of dimensions $10' \times 10'$, prior to foreground 
removal. The sources shown have been defined as stellar or 
probably-stellar in all three photometric bands, no other selection 
criteria have been applied. Some of the patchiness seen here is 
the result of varying limiting magnitudes in the observations 
- particularly in the NE footprint.}
\label{grid}
\end{figure}

\begin{figure}
\resizebox{\hsize}{!}{\includegraphics[scale=0.3]{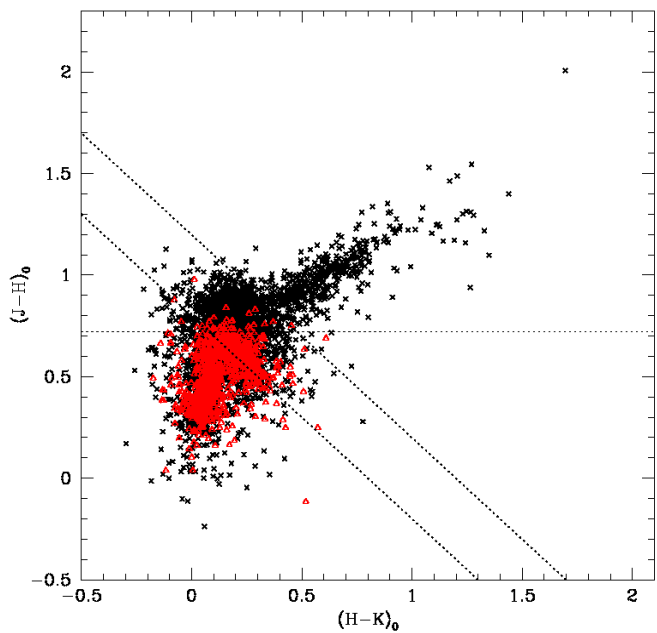}}
\caption{Colour-colour diagram of sources selected from two 
grid regions in Fig. \ref{grid}. Sources in black 
are assumed to be representative of genuine NGC 6822 and MW 
foreground, whilst sources in red are assumed to be representative 
of MW sources only, as described in the text. The 
horizontal line at $(J-H)_0 = 0.72$ mag has been used to separate 
the galaxy and the foreground stellar population. The upper diagonal 
line at $(J-K)_0 = 1.20$ mag represents the colour boundary between 
C- and M-type AGB stars applied to the sample (see Sect. \ref{jk}). 
The second diagonal line at $(J-K)_0 = 0.80$ mag relates to an alternative 
foreground removal method discussed in Sect. \ref{cioni}.}
\label{fore}
\end{figure}

\begin{figure}
\resizebox{\hsize}{!}{\includegraphics[scale=0.3]{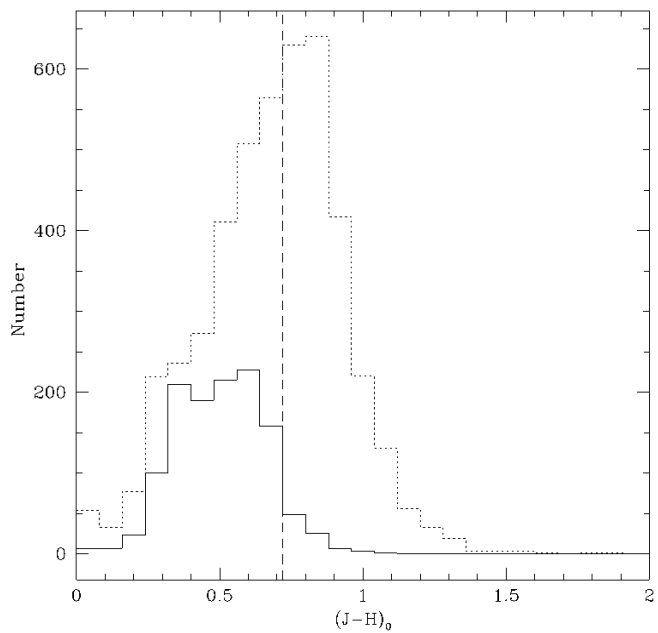}}
\caption{Colour histograms with bin size 0.1 mag, of the same sources as 
in Fig. \ref{fore}. The dotted-line histogram corresponds to 
the central region of the galaxy and the solid-line histogram 
corresponds to the outer region. The dashed vertical line is at 
$(J-H)_0 = 0.72$ mag.}   
\label{fore2}
\end{figure}

\begin{figure*} 
\resizebox{\hsize}{!}{\includegraphics[scale = 0.5]{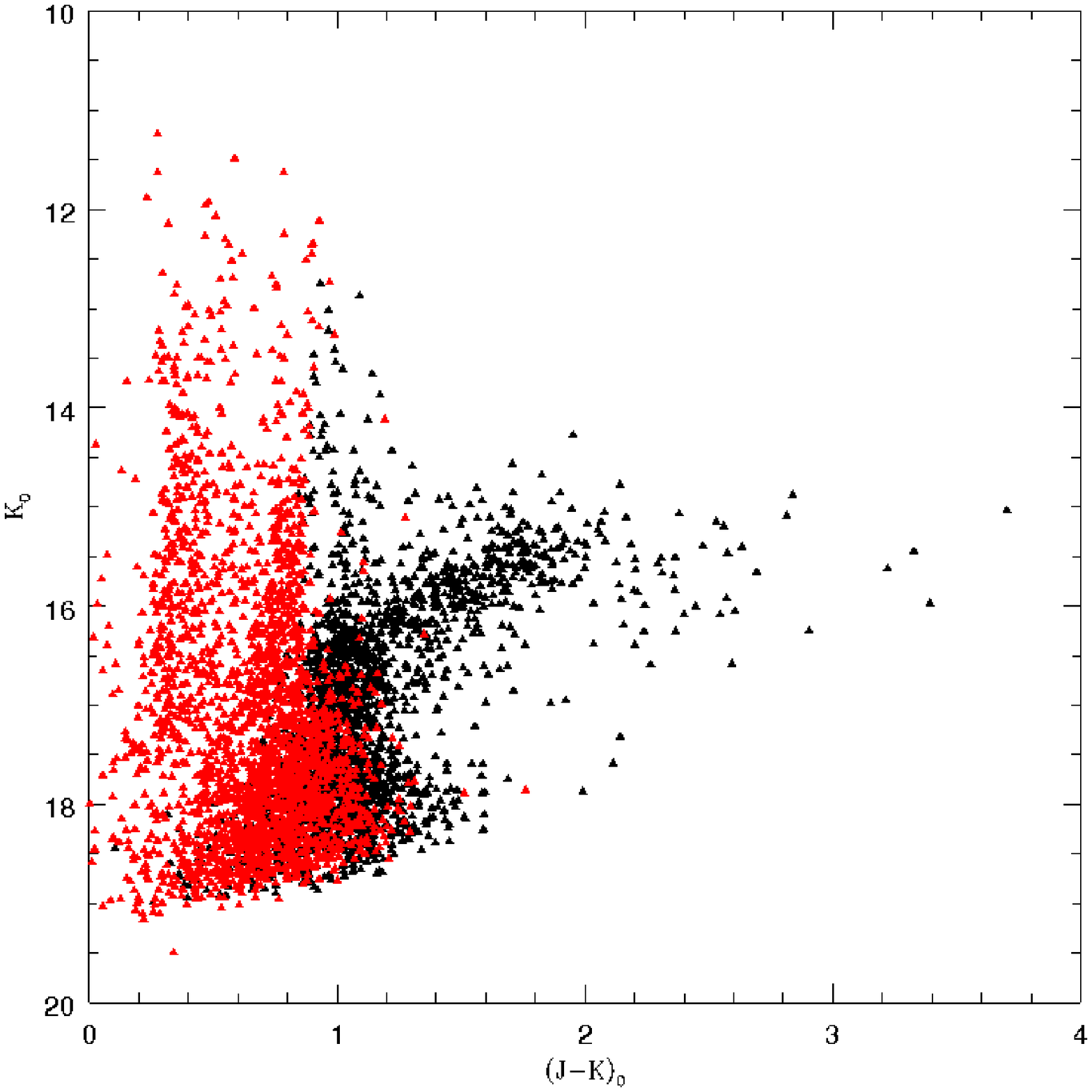}
\includegraphics[scale = 0.5]{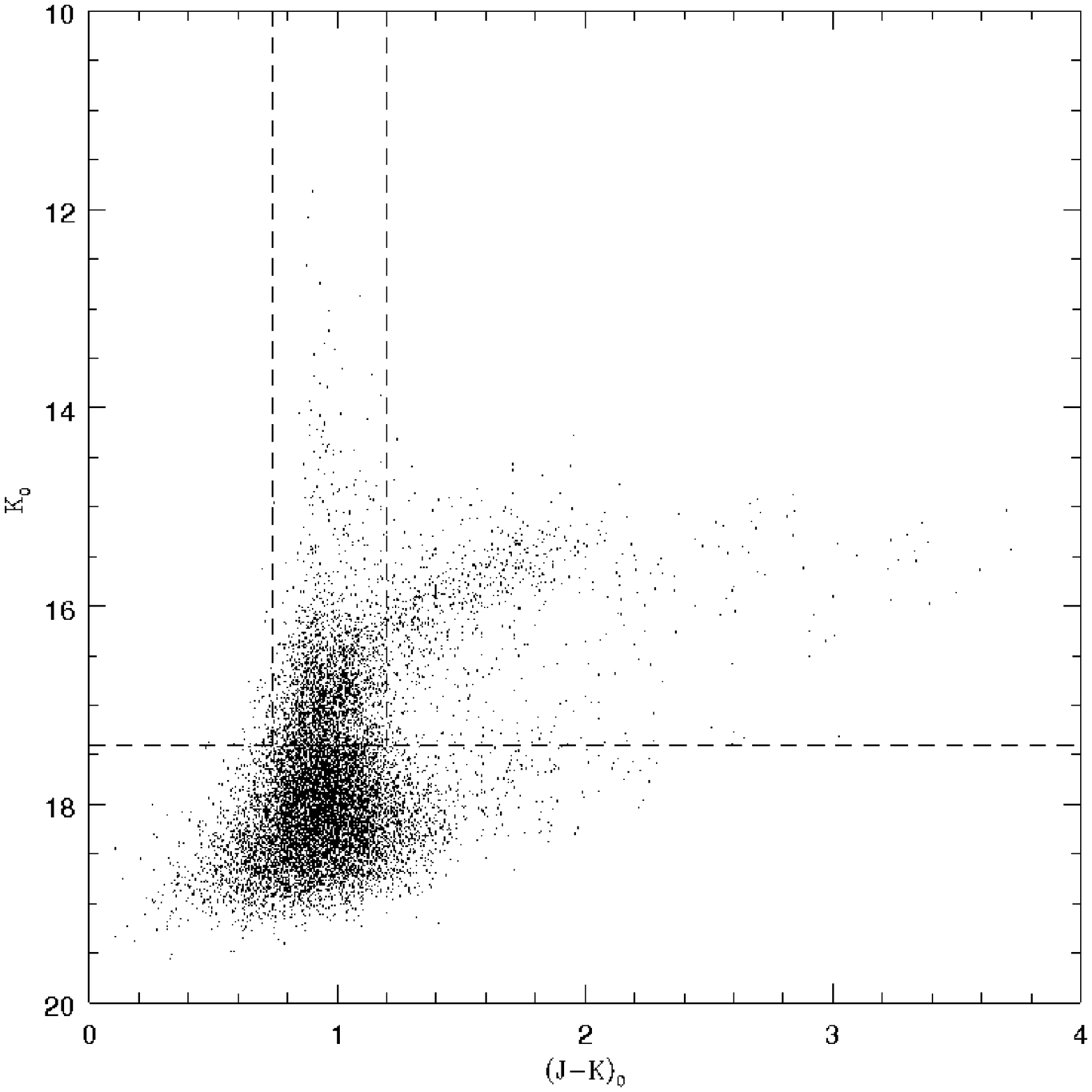}}
\caption{Left: CMD of the 
sources shown in black in Fig. \ref{fore}. Sources shown in black 
are believed to be genuine NGC 6822 sources whilst those sources shown 
in red have  $(J-H)_0 \leqslant 0.72$ mag and have therefore been removed 
from our sample as MW foreground contamination. Right: CMD of the
sources with $(J-H)_0 > 0.72$ mag i.e. mostly belonging to NGC 6822,
across the full observed area. The vertical and diagonal sequences
relating to the M- and C-type AGB stars, above the TRGB at $K_0 \sim
17-17.5$ mag, are clearly visible. The horizontal and vertical lines
mark the position of the determined TRGB and the colour selection
criteria (Sect. \ref{jk}).} 
\label{postforecmd}
\end{figure*}

The foreground contamination was substantially removed using 
colour-selection criterion based on the work of \citet{1988PASP..100.1134B}. 
The appropriate colour 
selection criterion was determined as follows; the full observed 
area was subdivided into a grid of $100$ regions, each with dimensions 
of $\sim 10'\times 10'$ (Fig. \ref{grid}). Sources from the grid 
region with the highest number density of sources - i.e. containing
the majority of NGC 6822 - were plotted on a 
colour-colour diagram ($(H-K)_0$, $(J-H)_0$). Sources from a region at the periphery
of the observed area (bottom left corner) that was assumed to be dominated 
by MW foreground stars were then plotted on the same colour-colour diagram in a different colour
(Fig. \ref{fore}). This process was repeated for each of 
the peripheral regions to ensure that a suitable average 
colour-selection criterion was adopted.\\In Fig. \ref{fore} a distinct 
separation in the colour distribution of the sources
from the centre and those from the outer region can be seen at $(J-H)_0
= 0.72$ mag. This is typical of all the peripheral regions. 
Assuming that foreground stars are
evenly distributed across field, the stars with
$(J-H)_0 > 0.72$ mag are likely members of NGC 6822. This technique relies 
on the separation of dwarfs and giants in $J-H$ to separate the foreground 
dwarfs from NGC 6822 giants \citep[Fig.~A3]{1988PASP..100.1134B}. The position of this
colour separation was confirmed using a colour histogram of sources
from the central and outer region (Fig. \ref{fore2}); the sharp decline
in the number of sources from the outer region at $(J-H)_0 > 0.72$ mag,
confirms the colour selection criterion.This method will also have 
removed some genuine NGC 6822 sources bluer than $(J-H)_0 = 0.72$ (mostly RGB stars) from our sample. 
However, as we are primarily interested in the detection and 
identification of AGB, rather than RGB, stars this colour selection technique 
is very effective (left-hand panel of Fig. \ref{postforecmd}) 
and quite suitable for our purposes. Foreground giants are extremely 
bright and are expected to have been removed as saturated sources ($K_0 <12.75$ mag) 
during the data selection. \citet{2008MNRAS.388.1185G} also used this technique to remove 
foreground contamination and to select C- and M-type stars above the 
tip of the red giant branch (TRGB).

Some MW contamination of 
our sample is likely to remain. The presence of a few sources in the
outer quadrant with $(J-H)_0 > 0.72$ mag suggests either a small leakage
of foreground stars 
into the NGC 6822 sample or a small NGC 6822 component out to large 
radii. If they are contaminating sources many of them are later
removed from the AGB sample by the  
application of a $K$-band magnitude criterion (see Sect. \ref{TRGB}). At 
most $1.4\%$ of the AGB stars retained in the central field will belong to 
the foreground, though this fraction will increase in the outer fields. Both 
alternatives are discussed further in Sect. \ref{StellDen}. The right-hand 
panel of Fig. \ref{postforecmd} presents a CMD of all the sources that 
remain after the application of the $J-H$ criterion and which are 
therefore are believed to be predominantly genuine members of NGC 6822. Although we are 
unable to absolutely identify individual sources in the CMD as MW contaminants, we note 
that the brightest sources extending up to $K_0 \sim 12.8$ mag - significantly above 
the C-type star branch - may be residual foreground contamination. We
base this on 
the colour and magnitude distribution of a simulated foreground in the 
direction of NGC 6822 generated using TRILEGAL \citep{2005A&A...436..895G}. 
Using the maximum magnitude of the C-type star branch as a guide $61$ 
sources with a magnitude of $K_0 <14.75$ mag have been isolated 
and their $J-H$ colour distribution examined. Approximately half lie
near the $(J-H)_0 = 0.72$ mag boundary and seem likely to be foreground
contamination, whilst the remainder lie significantly above this. After
the application of the $K$-band magnitude criterion discussed below, these potential 
contaminants make up less than $0.8\%$ of our sample. As we do not have 
conclusive evidence on which to reject these sources and due to their small number, it 
was decided not to use a bright magnitude limit or a more severe $J-H$
selection to remove them from our sample. However, the effect on our
final results of using both of these methods to eliminate potential
contaminants is discussed in Sect. \ref{diss}.

\subsection{The tip of the RGB}
\label{TRGB}
The TRGB is one of the most prominent features in the magnitude distribution 
of old- and intermediate-age populations as it causes a large discontinuity
between the RGB and AGB populations and is commonly used to identify AGB
stars in galaxies outside the MW \citep[e.g.][]{2006A&A...454..717K,2005ApJ...633..810M}. 
In principle the removal of some genuine RGB sources as discussed in Sect. 
\ref{foreground} may have affected our determination of the TRGB, however 
the general position of the TRGB at  $K_0 \sim 17-17.5$ mag still detectable in the 
left-hand panel of Fig. \ref{postforecmd} and is more obvious in the right-hand 
panel of the same figure.
\\The exact position of the TRGB discontinuity was identified 
from the magnitude distribution of sources within 
a central region of $17' \times 17'$ using the Sobel edge 
detection algorithm \citep{1993ApJ...417..553L}. This area was chosen 
as it contains a large portion of the galaxy and therefore minimises
the effects of any residual MW contamination and is unaffected by the 
poorer quality data collected in the outer NE field. The Sobel algorithm is a first derivative operator 
that computes the rate of change (gradient) across
an edge, producing a peak where there is a significant change of
slope. Due to the large discontinuity (change in slope) in the magnitude distribution at the TRGB, the largest peak corresponds 
to the position of the TRGB. As initially used by
\citet{1993ApJ...417..553L}, the Sobel 
filter had the disadvantage that the position of the resulting peak
is affected by both the bin sizing and the position of the bins in
the magnitude distribution. However, the improved analysis of
\citet{1996ApJ...461..713S} applies the edge detection filter to a
smoothed magnitude distribution which is constructed by replacing each
discretely distributed stellar magnitude with a Gaussian curve of unit
integrated area and standard deviation $\sigma$ equaling the magnitude error.  
This avoids the problems of binning the data and is the technique
applied here.

\begin{figure} 
\resizebox{\hsize}{!}{\includegraphics[scale = 0.3]{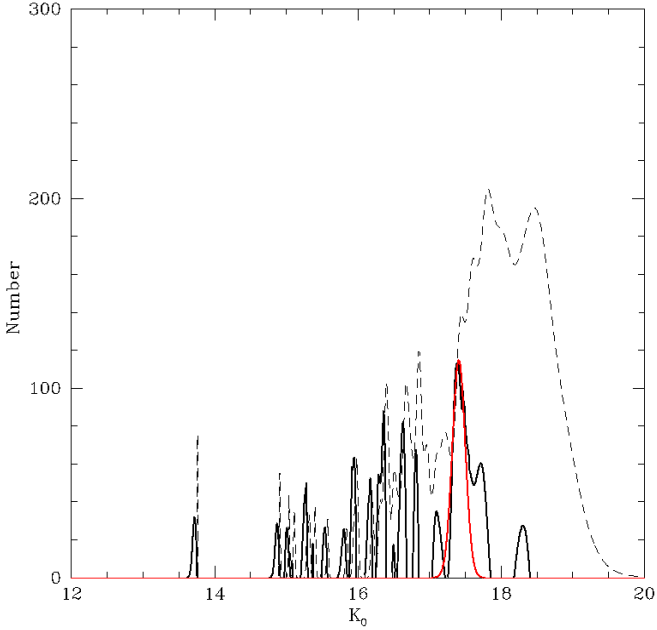}}
\caption{A smoothed $K$-band magnitude distribution(dashed line) 
for region of size $17' \times 17'$. The Sobel filter (solid black 
line) has been applied and a Gaussian curve (solid red line) has 
been fitted to the strongest peak to locate the TRGB magnitude and
error. The TRGB lies at $K_0 = 17.41 \pm 0.11$ mag. The distribution is
a generalised histogram and the vertical scale is arbitrary.}
\label{sobel}
\end{figure}

After the Sobel filter was applied, a Gaussian was fitted to the strongest peak. The mean 
and dispersion of the fitted Gaussian were taken as the TRGB magnitude and associated error
(Fig. \ref{sobel}). The following points should be noted: firstly, \citet{2000A&A...359..601C}
found the Sobel filter to be systematically biased towards fainter magnitudes, due to the
effects of smoothing the data. Magnitude
corrections were supplied by the same authors and have been applied here. Secondly, in order to ensure a
credible detection with the Sobel filter, there must be at least $100$
sources in the range extending one magnitude fainter than the TRGB, according to \citet{1995AJ....109.1645M} and
\citet{2002AJ....124.3222B}. As the TRGB is estimated by eye to lie 
between $K_0 \sim 17-17.5$ mag and the magnitude distribution 
contains in excess of $1300$ sources in the
range $17.5 < K_0 < 18.5$ mag, the criteria regarding 
the reliable use of the Sobel filter have been met.

A TRGB magnitude of $K_0 = 17.41 \pm 0.11$ mag was found 
(Fig. \ref{sobel}) and has been used for the purposes of isolating 
AGB stars in our photometric sample. A discussion of the variation of the TRGB 
magnitude across the surface of NGC 6822 follows in Sect. \ref{TRGBvar}.

\subsection{C- and M-type AGB stars}
\label{jk} 

\subsubsection{J-K colour selection}
\label{col}
\begin{figure} 
\resizebox{\hsize}{!}{\includegraphics[scale = 0.3]{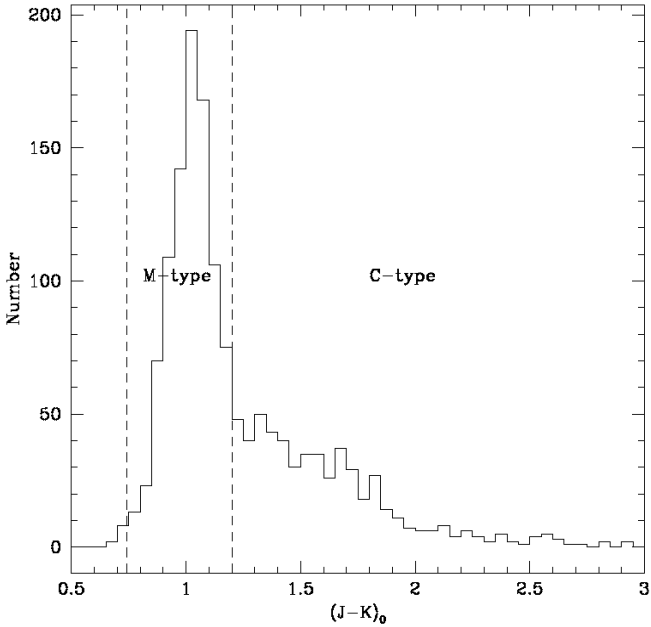}}
\caption{Colour histogram with a bin size of 0.05 mag, of 
    the AGB sources in the $17' \times 17'$ region used to determine the TRGB magnitude. 
    The vertical dashed lines mark the positions of the blue limit 
    at $(J-K)_0 > 0.74$ mag (Sect. \ref{blue}) and the colour
    separation between the C- and M-type stars at $(J-K)_0 = 1.20$ mag.} 
\label{finhist}
\end{figure}

AGB stars of spectral type C or M are easily identified on a CMD
(Fig. \ref{postforecmd}, right panel). M-type stars follow a vertical  
sequence above the TRGB with a large range of magnitudes at nearly
constant colour, whilst C-type stars display a smaller range of
magnitudes but a wider range of systematically redder colours,
resulting in a `red tail' extending diagonally upwards and away from
the M-type stars. The redder colours of C-type stars are due to the
increasing molecular opacity in the stellar atmosphere as more carbon
is brought to the surface, leading to a marked cooling and a larger
temperature gradient across the population as stars develop from
M-type to C-type \citep{2003A&A...403..225M}. An estimate from the CMD
places this separation at $(J-K)_0 \sim 1.10-1.20$ mag. However, as there is 
some overlap in the CMD between C- and M-type AGB stars, especially at 
fainter magnitudes, it is difficult to identify a precise colour separation 
between the two spectral types.\\The adopted position of the separation has 
been judged by eye from the discontinuity in the $J-K$ colour
  histogram (Fig. \ref{sobel}) of the AGB sources in the same $17'
  \times 17'$ region used to determine the position of the TRGB. The highest 
peak in Fig. \ref{finhist} relates to M-type stars, followed at redder colours by a 
significant drop and then a tail containing the C-type stars 
\citep{2005A&A...429..837C}. The colour separation was found to lie 
at $(J-K)_0 = 1.20 \pm 0.05$ mag. For $(J-H)_0 \sim 1.0$ mag and $(J-K)_0 \sim 1.0$ 
mag, $(J-K_s)_{2MASS} \sim (J-K)_0 + 0.08$ \citep[eq. 6 \&
8]{2009MNRAS.394..675H}, so the colour separation $(J-K)_0 = 1.20$
corresponds to $(J-K_s)_{2MASS} \sim 1.28$.

This value has been used for the purposes of identifying C- and M-type 
stars in the AGB sample. The application of such
a sharp colour selection criterion suggests a strict transition between
these two types of AGB stars; in reality this is unlikely
and the colour separation depends strongly on the metallicity of
the observed population \citep{2007A&A...474...35B}. The
identification and quantification of bias in our colour selection
criterion will be the subject of a future paper using spectroscopic
observations, but our selected value is strongly supported by 
the recent work of \citet{2012A&A...537A.108K}.

\subsubsection{A blue limit}
\label{blue}
The selected AGB sample spans a colour range of $0.48 < (J-K)_0 < 4.08$ mag. In
accordance with the findings of \citet{1988PASP..100.1134B} and 
\citet{2007A&A...474...35B}, it was decided to apply a `blue limit' to the 
selection of M-type stars in order to exclude late K-type stars. An empirical 
blue limit of $(J-K)_0 = 0.74 \pm 0.05$ mag was used based on the colour histograms. 
In Fig. \ref{finhist}, $(J-K)_0 \sim 0.80-0.90$ mag marks the beginning
of significant numbers of M-type AGB stars. To allow for some fluctuation in 
the position of this onset and the effects of binning the data, 
a slightly bluer limit has been selected to preserve genuine
sources. The use of a more severe (redder) limit may underestimate the
number of M-type stars; this is discussed further in Sect. \ref{blue2}. 

The final criteria applied for the selection of M-type AGB 
stars was $0.74 < (J-K)_0 < 1.20$ and $K_0 < 17.41$ mag. The upper colour limit 
is bluer than the limit applied in a study of the AGB population 
of NGC 6822 by \citet{2006A&A...454..717K} ($(J-K)_0 = 1.40$ mag) but 
it is in good agreement with the findings of \citet{2005A&A...429..837C} 
($(J-K_s)_{2MASS} = 1.24$ mag). This limit also agrees well with the 
analysis of \citet{2007A&A...474...35B}, who in reviewing the colour limits 
applied in various studies of the AGB population in several Local Group 
galaxies, concluded that the C- and M-type star boundary is ill defined 
but suggest that $(J-K)_0 = 1.20$ mag is an appropriate limit for NGC 6822. 
An upper colour limit was not applied to the selection of C-type stars. 
We would expect the intrinsic colours of C-rich AGB stars to reach $(J-K)_0 \sim 2.5$
mag; sources redder than this may still be AGB stars that are heavily dust 
enshrouded, which are more likely to be C-type than M-type
\citep{2006MNRAS.370.1961Z}.

\section{Results}
\label{results}

\subsection{The structure of NGC 6822}

\subsubsection{Spatial distributions}
\label{spadis}
\begin{figure*}
\centering 
\resizebox{0.8\hsize}{!}{\includegraphics[scale=0.4]{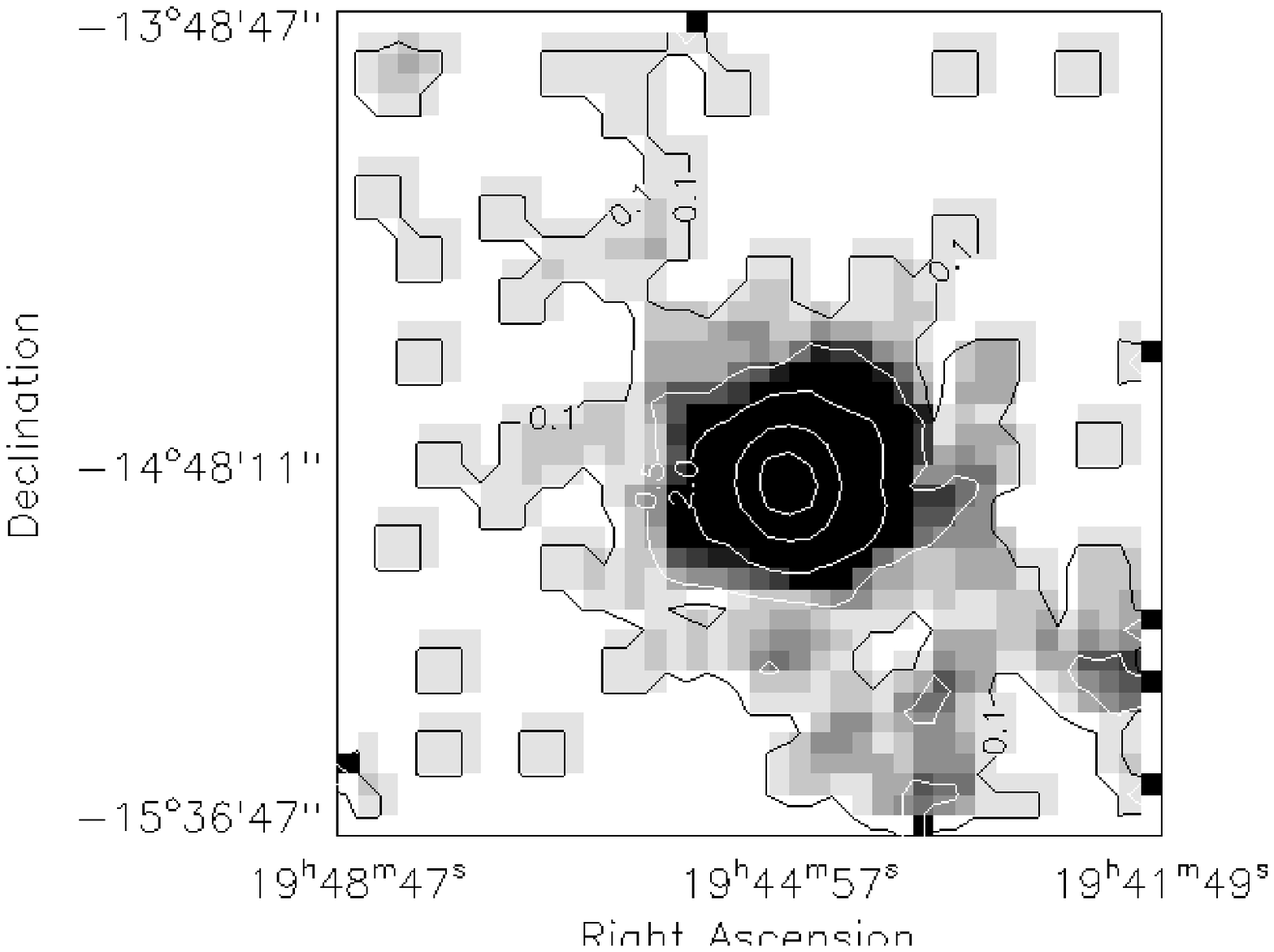}
\includegraphics[scale=0.4]{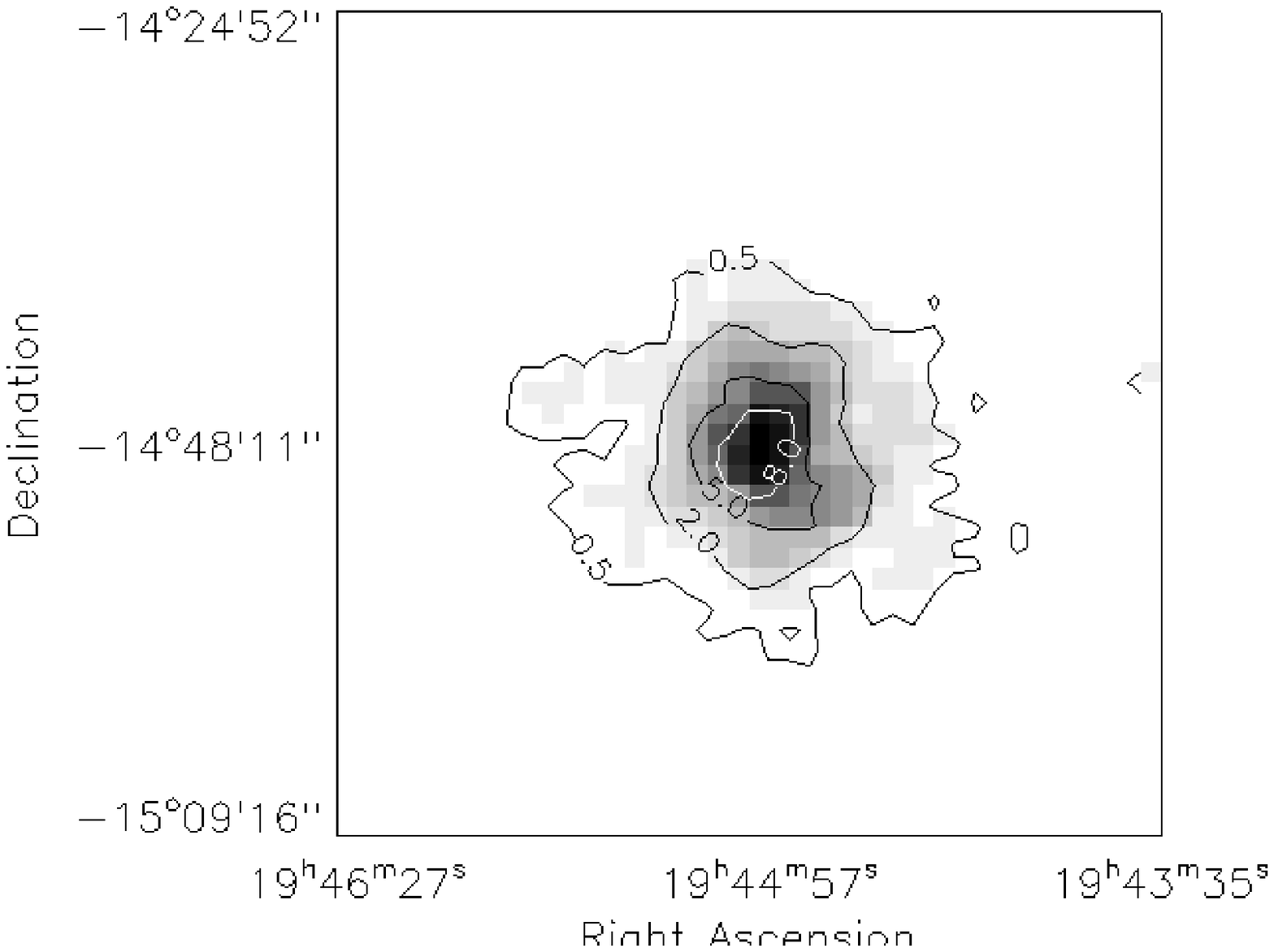}}
\resizebox{0.8\hsize}{!}{\includegraphics[scale=0.4]{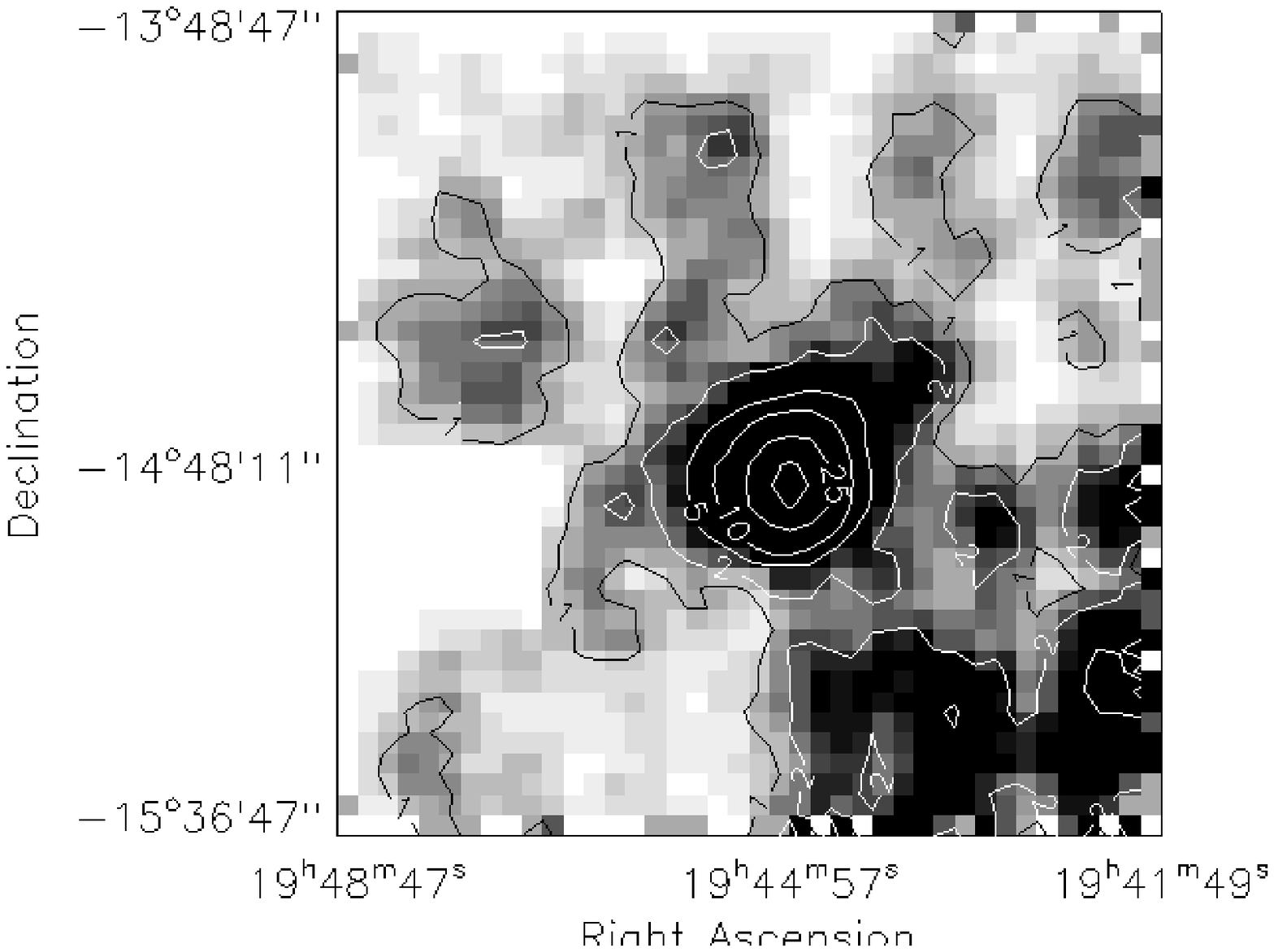}
\includegraphics[scale=0.4]{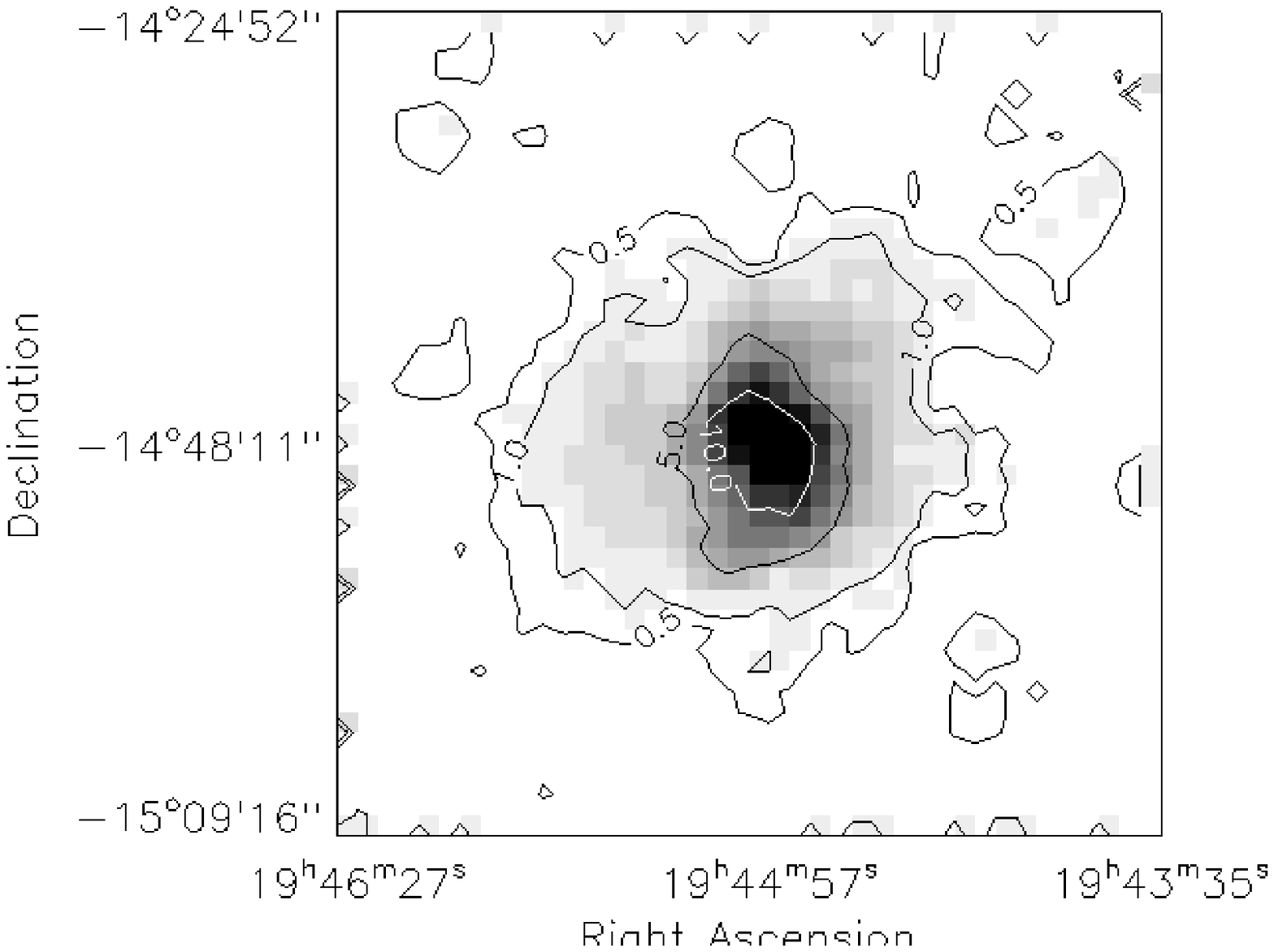}}
\resizebox{0.8\hsize}{!}{\includegraphics[scale=0.4]{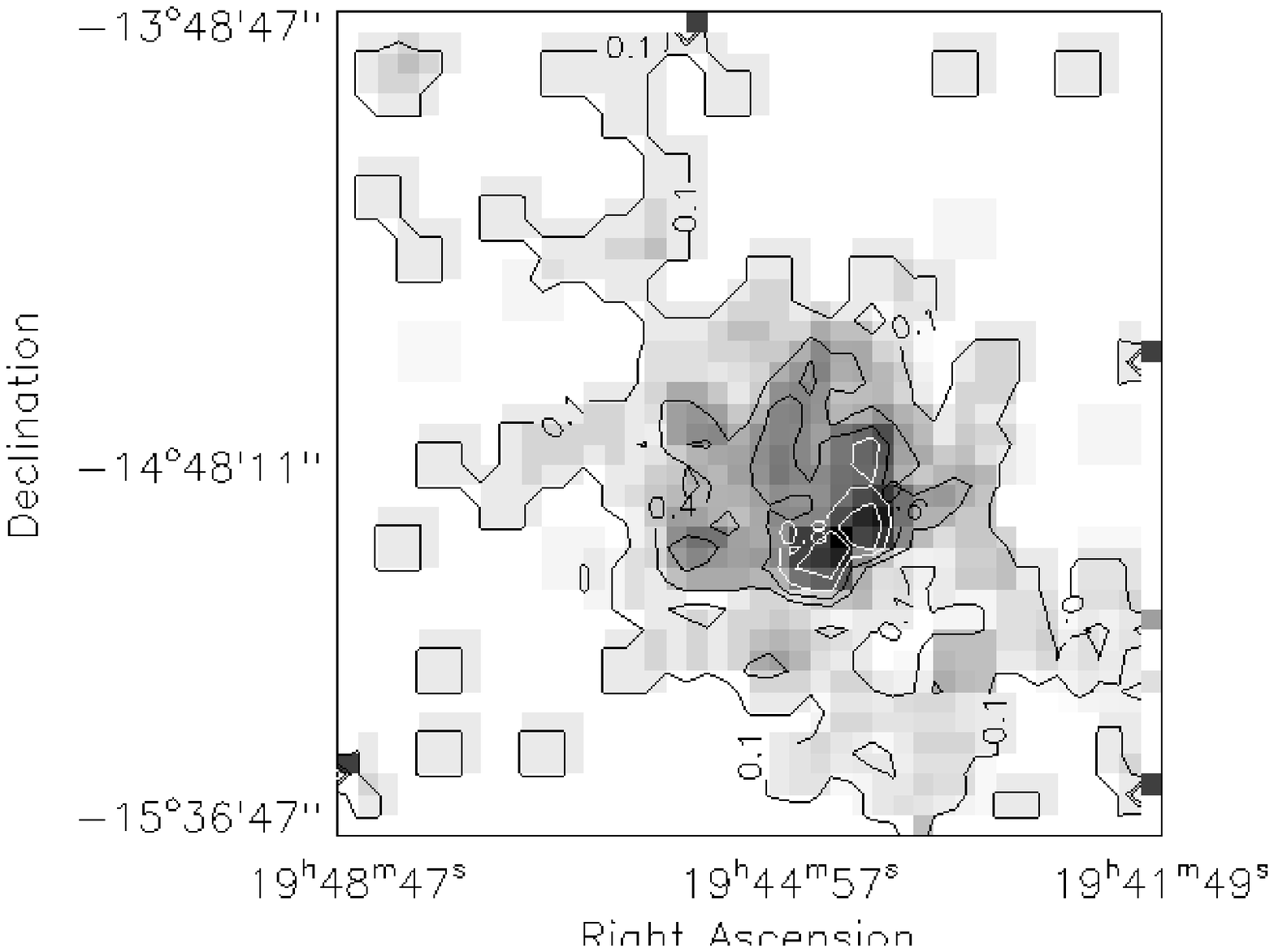}
\includegraphics[scale=0.4]{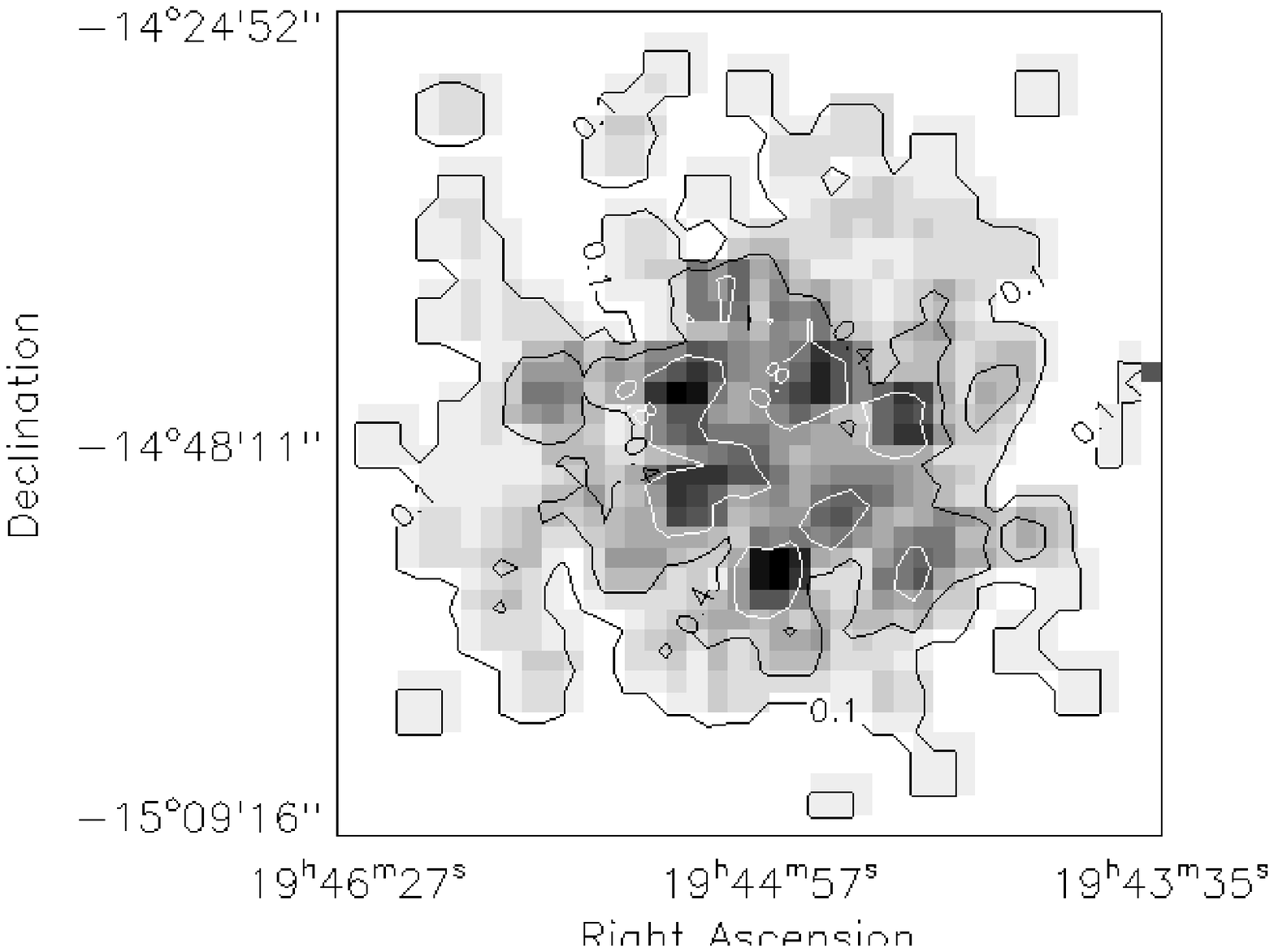}}
\resizebox{0.8\hsize}{!}{\includegraphics[scale=0.4]{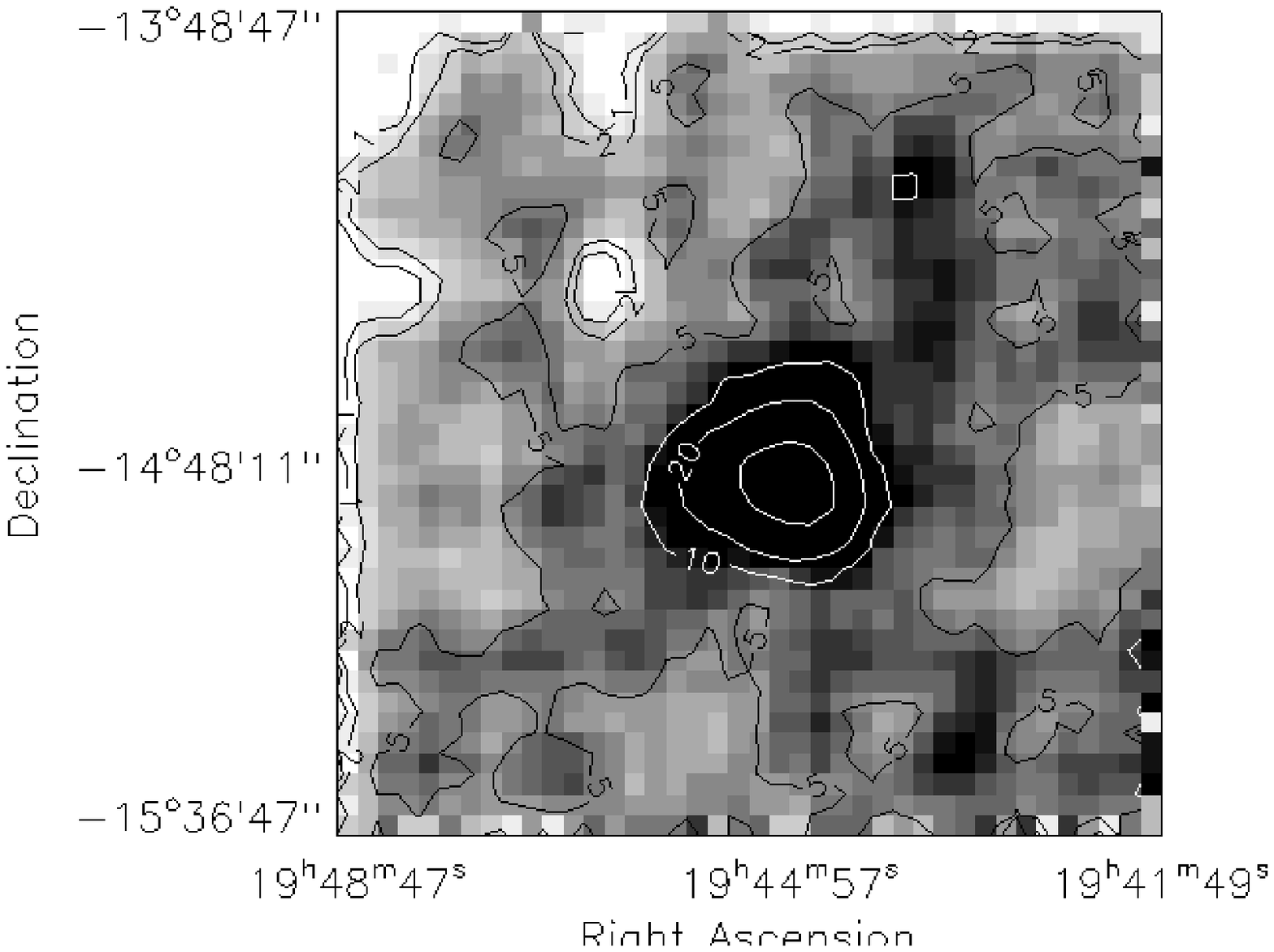}
\includegraphics[scale=0.4]{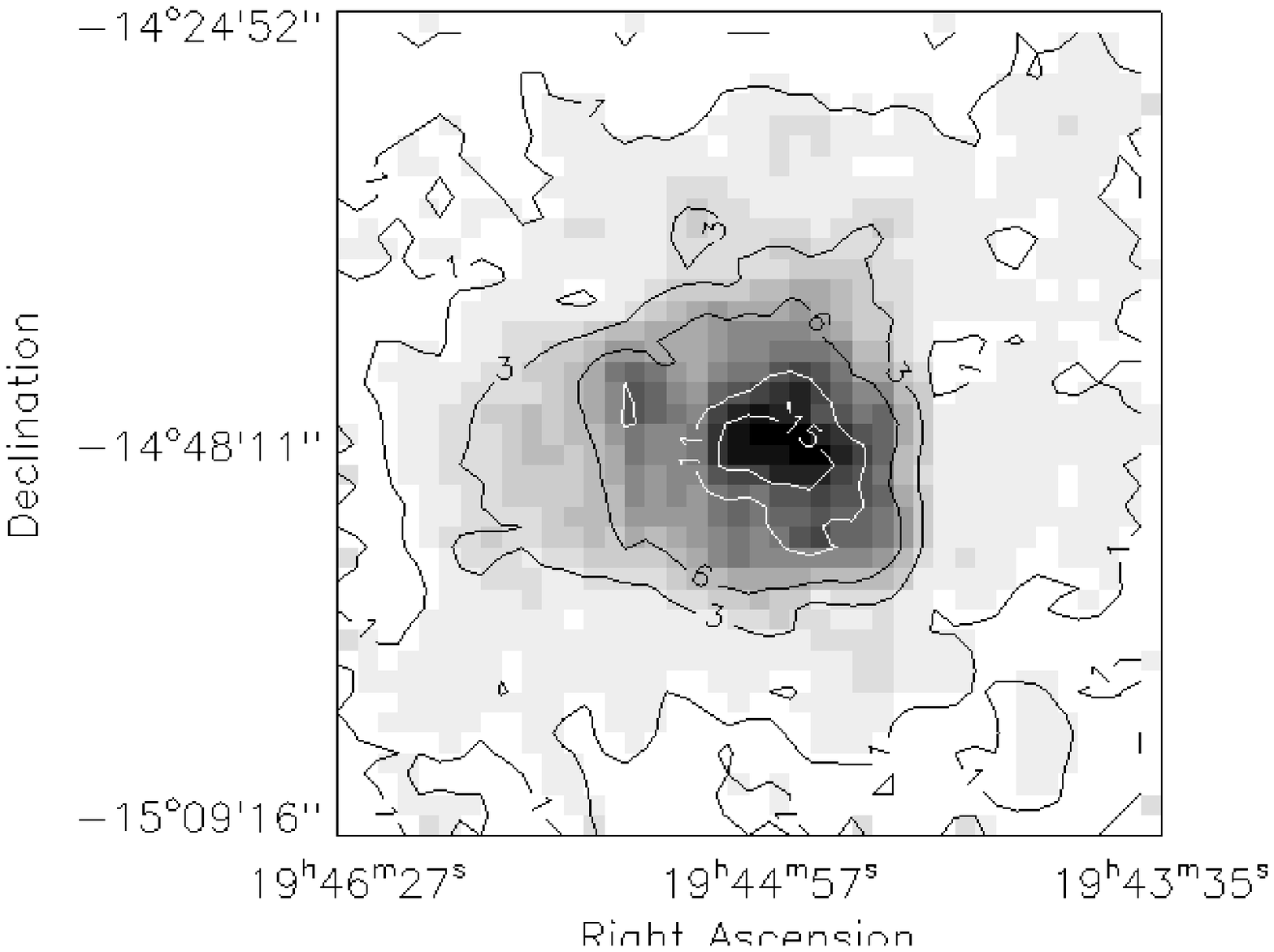}}
\caption{From top to bottom, density distribution of: carbon-rich
  stars, oxygen-rich stars, the C/M ratio and RGB stars. Each row
  shows, on the left a map of the full observed area using $1600$ bins
  and on the right a map of the central $34' \times 34'$ also using
  $1600$ bins. Contours are at: $0.1$, $0.5$, $2$, $10$
  and $15$ for carbon-rich stars in low resolution map and $0.5$, $2$,
  $5$ and $8$ in the high resolution map. For oxygen-rich stars
  contours are at: $1$, $2$, $5$, $10$, $25$ and $50$ in the low
  resolution map and $0.5$, $1$, $5$ and $10$ in the high
  resolution map. For the C/M ratio the contours are at: $0.1$, $0.4$,
  $0.6$, $0.8$ and $1$ in the low resolution and $0.1$, $0.4$ and $0.8$ in
  the high resolution map. For the RGB stars contours are at: $1$,
  $2$, $5$, $10$, $20$ and $50$ in the low resolution map and at $1$,
  $3$, $6$, $11$ and $15$ in the high resolution map. At a distance of
  $490$ kpc, the full field is $14.5 \times 14.5$ kpc$^2$, whilst the
  high resolution field is $5 \times 5$ kpc$^2$.} 
\label{maps}
\end{figure*}

\begin{figure*}
\resizebox{\hsize}{!}{\includegraphics[scale=0.5]{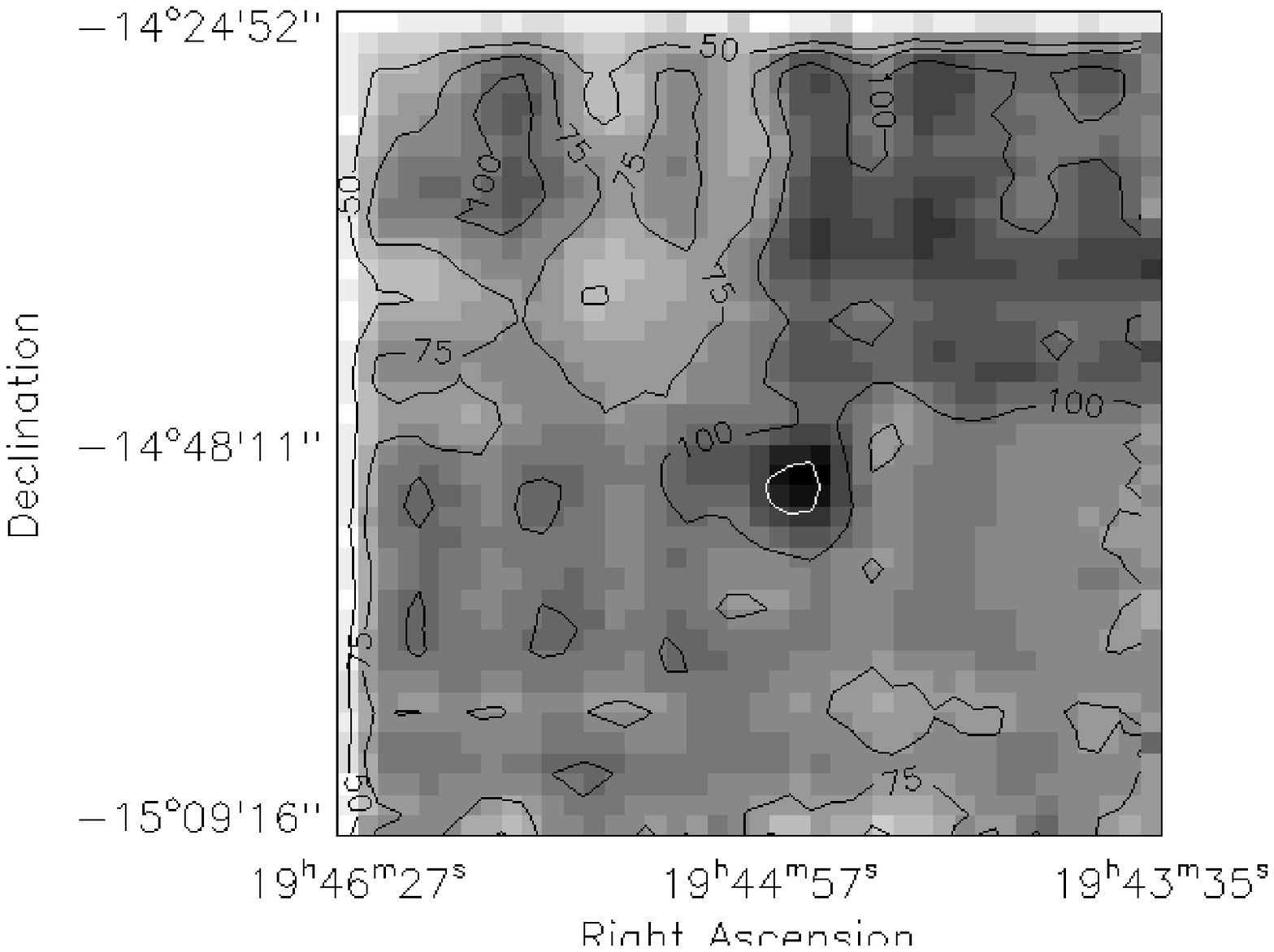}
\includegraphics[scale=0.5]{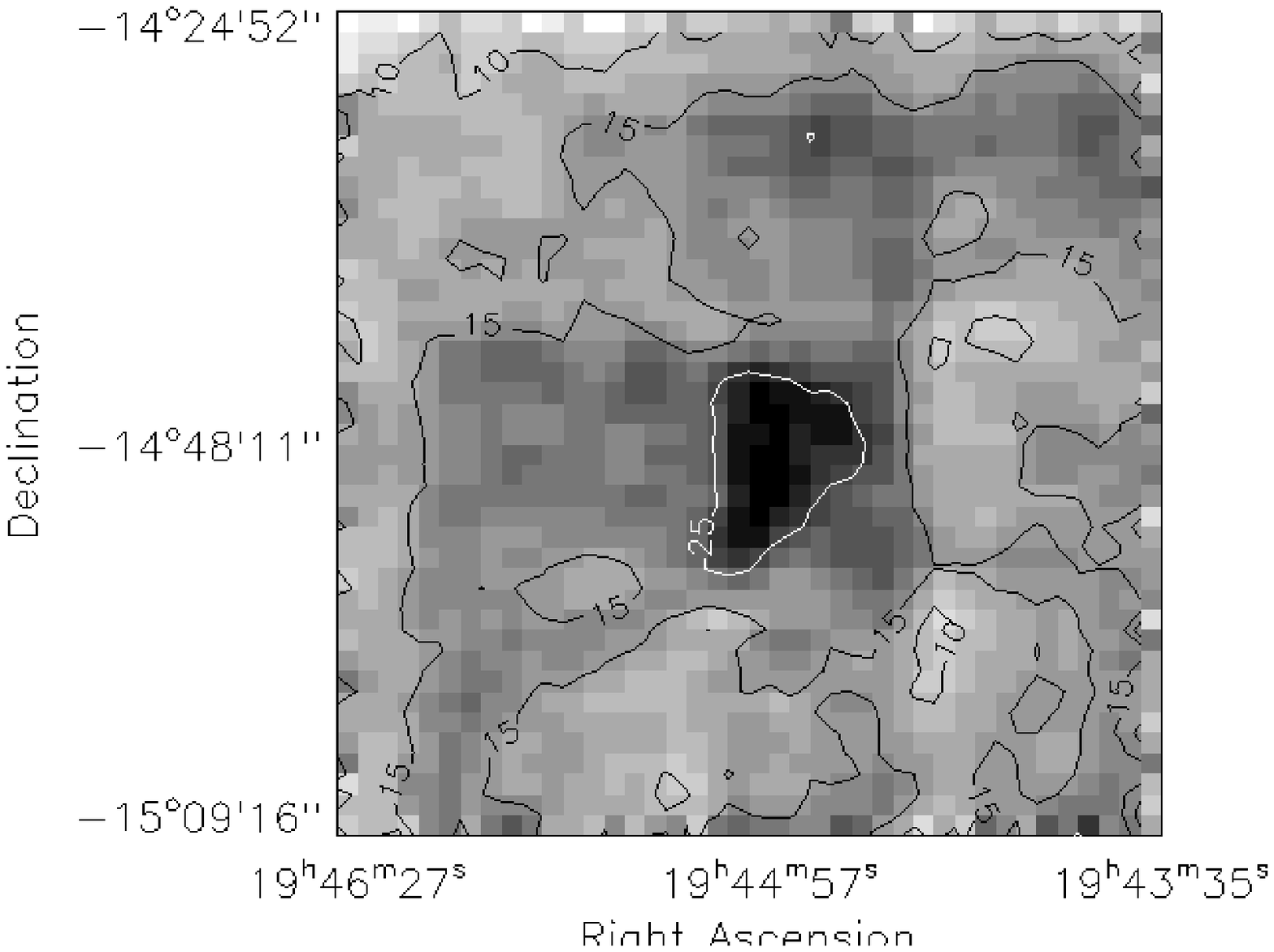}}
\caption{Density distribution of foreground sources removed from the
  data set as described in the text. On the left using $1600$
  bins across the full observed area contours are at $50$, $75$, $100$
  and $150$. On the right is the high resolution plot, contours are at
  $10$, $15$ and $25$. }
\label{mapsfore}
\end{figure*}

After the removal of much of the MW foreground and isolation of the C-
and M-type AGB stars, source density plots were constructed to examine
the distribution of these stars, the C/M ratio and the stellar population
below the TRGB (RGB sources) across the surface of NGC 6822 (Fig. \ref{maps}).
The distribution of the removed MW foreground is also shown in this
section (Fig. \ref{mapsfore}). The low resolution density maps have
been constructed by counting the number of sources of various types in
a 40 $\times$ 40 grid, where a single bin corresponds to $2'.55 \times
2'.55$, and then applying a box-car smoothing function of size
$2$. These maps were used to examine the large scale structure of the
galaxy. The same procedure was then repeated to produce a high
resolution map of the central $35' \times 35'$ of the galaxy in order
to examine any finer structure that was present.  

The distribution of C-type stars in both the low and high resolution
maps are shown in the top panels of Fig. \ref{maps}. The highest
concentration of C-type stars is in the area of the bar, although the
concentration is circular rather than following the elongated
bar structure. The C-type stars do not appear to trace the HI envelope
in any significant way.  

The M-type AGB population (Fig. \ref{maps}) is distributed similarly,
though the bar-like elongation is clearer and the density 
of M-type stars is higher than for C-type stars. There is a
 clear under-density of M-type stars in the SE that is most obvious
in the low resolution map and also in Table. \ref{tabA} which shows
the number counts of RGB, C-and M-type stars in the North-West,
South-West, South-East and North-East of the observed area. There is a
clear decline in the number of M-type AGB stars in the SE. This
under-density may be due to a structure similar to the super giant
shell, a large hole in the HI disk, described by
\citet{2000ASPC..218..357D}, which does not contain any AGB stars. The
high number of sources in the NW is attributed to the slightly better
completeness levels in this region. Remaining MW foreground
contamination would also affect M-type star and RGB counts in this
region, the C-type star counts would not be affected as C-type stars
are not seen in the MW foreground. This is discussed further at the in
the context of foreground removal at the end of this section. There is
also a clear overdensity in the SW  of the galaxy which we will return
to in Sect. \ref{strut}.  

\begin{table} 
\centering                  
\begin{tabular}{lcccc}
\hline
Type &$NW$ &$SW$ &$SE$ &$NE$\\
\hline
Carbon (C) &$215$ &$275$ &$189$ &$175$\\
Oxygen (M) &$652$ &$1127$ &$499$ &$623$\\ 
RGB &$2688$ &$2378$ &$2415$ &$2280$\\
\hline\\
\end{tabular}
\caption[]{RGB, C-and M-type star number counts.}  
\label{tabA} 
\end{table}

Figure \ref{maps} also shows the surface 
distribution of the C/M ratio across NGC 6822, where dark regions 
indicate a higher ratio. Variations in the C/M ratio are frequently assumed 
to reflect variations in the metallicity of the region, however, improvements 
in our understanding of stellar evolution and the effects of population age on 
the C/M ratio means that this traditional interpretation is no longer so 
straightforward. The C/M is not simply a function of 
metallicity, it is also dependent on the age of the population 
\citep{2010MNRAS.tmpL.129F,2003MNRAS.338..572M}. \citet{2010MNRAS.tmp..431H} 
suggest that the ratio may be much more sensitive to the age variations 
in the population then previously thought. During their study of 
LeoI dSph, \citet{2010MNRAS.tmp..431H} concluded that the number 
of C-type stars is much more dependent on age than the number of 
M-type stars. A conclusion  which is supported by the earlier work 
of \citet{2008MNRAS.388.1185G} on Leo II and by the work of \citet{2006A&A...448...77C}. 
\citet{2008MNRAS.388.1185G} and \citet{2010MNRAS.tmp..431H} both 
show plots of the production of C- and M-type stars as a function of 
age, with the C-type star count peaking in the first $\sim 2$ 
Gyrs and falling off to almost nothing 
at $\sim 7-8$ Gyrs, both of which would affect the interpretation of 
the C/M ratio as a metallicity indicator in the conventional sense. 
A similar plot is shown by \citet{2006A&A...448...77C} for the LMC 
with the useful addition of another plot showing how 
this effects the C/M ratio. It is clear from this plot that the C/M 
ratio is dependent on the age of the underlying population.   
This age dependence is consistent with our current 
understanding of how carbon-rich AGB stars evolve. C-type stars are 
only expected to form over a certain mass, and therefore age, range due 
to their dependence on the efficiency and effects of the third dredge-up 
(TDU), Hot Bottom Burning (HBB), mass loss on the AGB and molecular 
opacity \citep{2005MNRAS.356L...1S,2007A&A...467.1139M,2010MNRAS.408.2476V,2011apn5.confE.144K}. 
\citet{2003PhDT..01..286K} suggests that only AGB stars more massive than 
$1-1.5$M$_{\odot}$ undergo TDU and hence could become C-type stars. 
At present we do not have sufficient data to investigate the mass and 
age distribution of our candidate AGB stars. This work is been based 
on the classical interpretation of the C/M ratio as the calibration of 
\citet{2009A&A...506.1137C}, like previous calibrations, derives the 
iron abundance solely from the C/M ratio without detailed consideration 
of other population variables such as age (or AGB star mass). However, 
we draw the readers attention to the age dependence of the C/M ratio as 
an aid to any future interpretation of our results when such data has been obtained.

Returning to Fig. \ref{maps}, regions with the highest C/M ratio are located in and around 
the centre of the galaxy, although there is no clear enhancement defining the 
position of the bar. In fact the highest contour levels are slightly offset from 
the centre. The high resolution map, especially, demonstrates the clumpy and 
slightly elliptical distribution of the C/M ratio in the centre, set in two 
larger, more evenly distributed regions with lower C/M ratio's. Under traditional 
assumptions this clumpy distribution would suggest areas of lower metallicity 
in the galactic centre surrounded by regions with a higher metal content. Whilst 
the patchy distribution of the C/M ratio may be a real feature, the
metallicity distribution, associated with the traditional
interpretation of the C/M ratio, would be unusual as the central regions of a galaxy are typically 
expected to show a concentration of more metal-rich stars. It seems more 
probable that the irregular clumps in the C/M ratio are not the result of metallicity 
effects alone and other population parameters may also be important. \\The 
lower panels of Fig. \ref{maps} show the source density plots for all those
sources belonging to NGC 6822 that are below the TRGB ($K_0 > 17.41$ mag)
and therefore were not identified as AGB stars. These sources will
mostly be RGB stars. They are distributed more smoothly 
across the face of the galaxy and at a higher density than the AGB population. This makes the
under-densities due to the poor sensitivity in the NE more obvious,
however the under-density in the SE seen in the M-type AGB density
plots is less apparent in the low resolution RGB plot.                
In the low resolution RGB density plot there is also for the first time
a noticeable contour in the NW-SE direction that may be tracing the HI envelope.
This contour extends slightly to the SW as well, but a stellar overdensity 
in this region is not obvious.

\begin{figure} 
\resizebox{\hsize}{!}{\includegraphics[scale = 0.3]{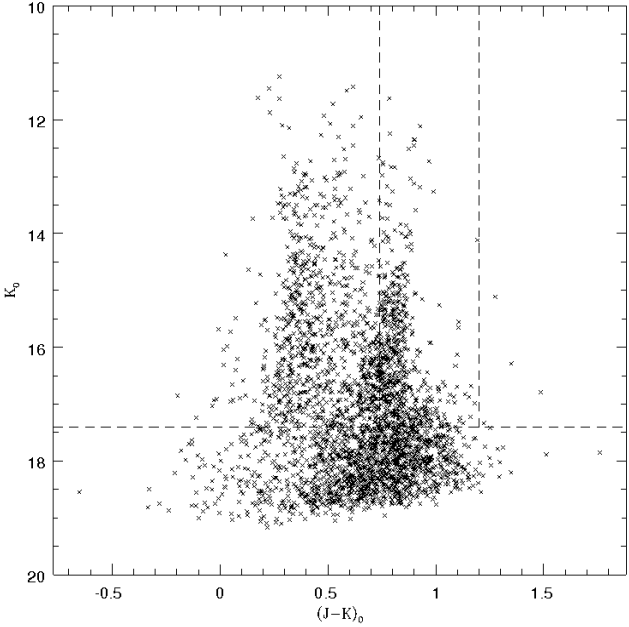}}
\caption{CMD of the MW sources located 
in the over-density in the centre of Fig. \ref{mapsfore}. The horizontal
and vertical lines represent the position of the TRGB and the colour
selection criteria for the selection of C- and M-type AGB stars,
respectively, as in Fig. \ref{postforecmd}.} 
\label{foreCMD}
\end{figure}

\begin{figure} 
\resizebox{\hsize}{!}{\includegraphics[scale = 0.3]{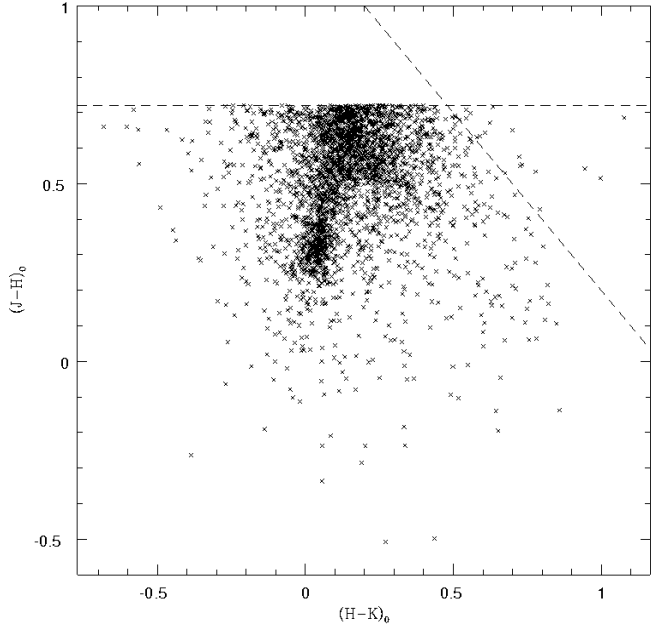}}
\caption{Colour-colour diagram of the same sources as 
Fig. \ref{foreCMD}. As these sources have been removed as foreground 
they all have a colour of $(J-H)_0 < 0.72$ mag - marked by the
horizontal. The turn over due mainly to the MW dwarf population is
clearly visible. The diagonal line shows the position of the $(J-K)_0
= 1.20$ mag colour cut.} 
\label{foreCCD}
\end{figure}

The source density plots in Fig. \ref{mapsfore} show the distribution
of the sources removed from our data set as MW foreground stars
(Sect. \ref{foreground}). Foreground stars are expected to be
distributed homogeneously across the observed area. However, there are 
overdensities in the centre and in the NW. This indicates that the 
foreground removal in the centre at least has been slightly too severe
and some genuine NGC 6822 stars have also been removed, however, a CMD
(Fig. \ref{foreCMD}) of the sources in the central overdensity
indicates that few genuine M- and C-type stars have been
removed. Although $\sim 640$ sources fall within the region occupied
by the M-type AGB stars and $\sim 15$ sources fall in the region
occupied by the C-type stars, these objects do not conform to the same
CMD as the NGC 6822 AGB sources in Fig. \ref{postforecmd} (right
panel) - there is no M-sequence at $(J-K)_0 = 0.9-1.0$ and no diagonal
C-branch. These sources are probably genuine MW stars correctly
subtracted. The greater number of sources in Fig. \ref{foreCMD} are
below the TRGB. Therefore of the genuine NGC 6822 sources that have
been wrongly removed the majority will be (K-type) RGB, rather than
AGB, stars. These genuine RGB sources will fall primarily in the
left-hand branch (i.e. bluer than  $(H-K)_0 \sim 0.15$) of the
inverted $U$, shown in the colour-colour plot (Fig. \ref{foreCCD}) of
the same region -  among genuine foreground sources. It is these RGB
sources that we believe are responsible for the overdensity seen in
the centre of the plots in Fig. \ref{mapsfore}. As RGB sources are not
our primary interest and as discussed in the Sect. \ref{TRGB} we do
not believe their loss will impact on our determination of important
variables, like the TRGB, it was decided to continue with the current
foreground removal technique. The overdensity seen in the NW of
Fig. \ref{mapsfore} corresponds roughly to a WFCAM tile and suggests
better observing conditions in this direction resulted in more faint
stars and hence more foreground objects being catalogued there. A
version of the low resolution plot  using only those sources with a
magnitude of $K_0 < 17.3$ shows a much smoother distribution of
sources with no obvious overdensity in the NW. Figures \ref{comp2} and
\ref{comp} were also produced separately for the NW and compared with
the average for the total observed area. Observations in the NW have
lower photometric errors and retain a $100\%$ completeness level to a
greater  depth ($K_0 \sim 17.7$ mag). These figures have not been
included here due to space restrictions, but seem to confirm the NW
overdensity in Fig. \ref{mapsfore} to be primarily an observational
effect. \\We acknowledge that the sharp $J-H$ colour separation
applied between the MW foreground and the M-type AGB stars of NGC 6822
(Fig. \ref{fore}) may result in some genuine AGB sources being
lost. The number of M-type sources counted and therefore the C/M ratio
is affected by the imperfect nature of the colour selection
criteria. This will be discussed further in Sect. \ref{forecol}.

\subsubsection{Stellar density profiles}
\label{StellDen}
\begin{figure}
\resizebox{\hsize}{!}{\includegraphics[scale=0.3]{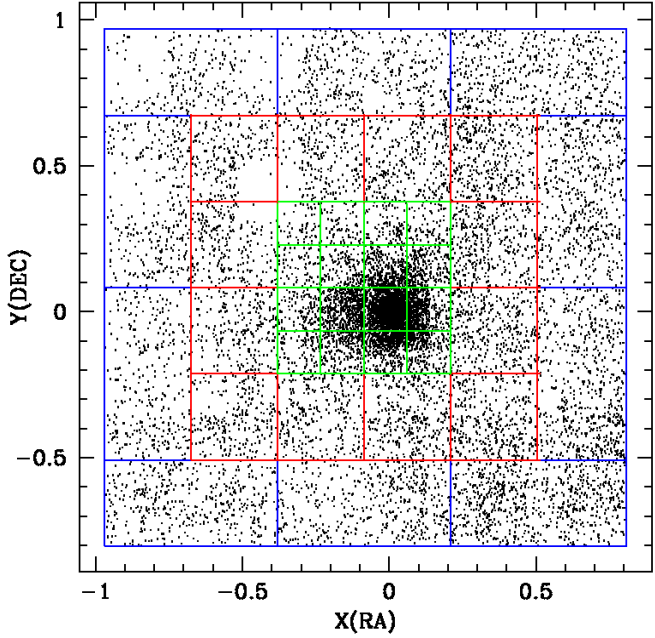}}
\caption{Multi-resolution grid overlaid on NGC 6822 sources,
    used for the construction of the stellar density profiles
    (Fig. \ref{stellden1}). Each region of the small (green), medium
    (red) and coarse (blue) grids have dimensions $8'.5 \times 8'.5$,
    $17' \times 17'$ and $17' \times 34'$ respectively. At a distance
  of $490$ kpc, $0.1^{\circ}$ subtends $0.9$ kpc.}    
\label{grid2}
\end{figure}

In order to further investigate the effectiveness of the foreground 
subtraction and the distribution of the NGC 6822 population, stellar 
density profiles of the AGB and the RGB (referring to all those sources 
below the TRGB) populations of NGC 6822 have been constructed. Using 
the position angle (PA) of the bar PA $= 10^{\circ}
\pm 3^{\circ}$ \citep{1977ApJS...33...69H} and its inclination $i =
0^{\circ}$ \citep{2005A&A...429..837C}, the distance in
kpc from the galactic centre and an angle $\phi$ in the plane of the
galaxy, measured anticlockwise from the major axis of the bar were
calculated for the central coordinates of each region in 
a multi-resolution grid (Fig. \ref{grid2}). The PA does not affect
the calculated distances in this case (as $i = 0^{\circ}$), it does
give the zero-point of the angle $\phi$. The PA and $i$ of the bar
were used as it was the most prominent feature in the surface density
plot of the AGB population.

\begin{figure} 
\resizebox{\hsize}{!}{\includegraphics[scale = 0.3]{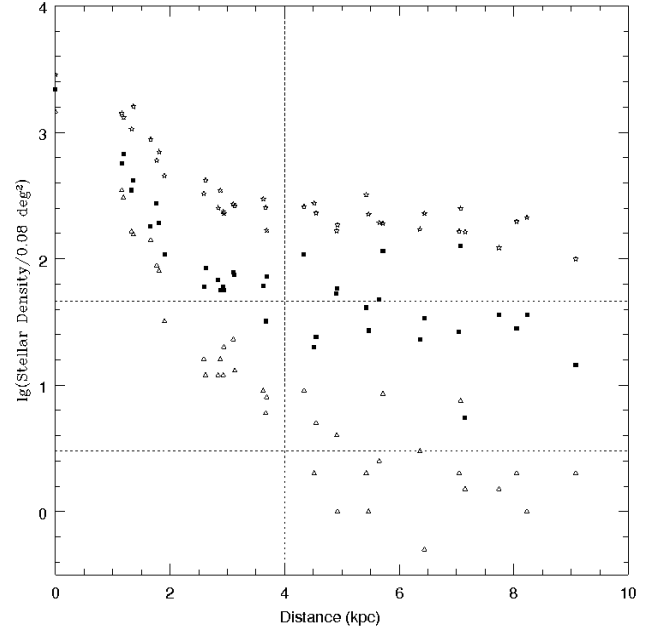}}
\caption{Stellar density profiles of the C-type (triangles)    
 and M-type (squares) AGB and RGB sources (stars) sources in each
 region in the multi-resolution grid (Fig. \ref{grid2}). The stellar
 density has been normalised to account 
 for the varying area of each grid region. The horizontal lines at
 $1.66$ and $0.48$ represent the level of the remaining foreground
 contamination in the M- and C-type samples respectively. The vertical
 line at $4$ kpc marks the limit of the detectable stellar component
of NGC 6822.} 
\label{stellden1}
\end{figure}

Fig. \ref{stellden1} shows the number density of C- and  M-type
AGB stars and RGB stars per unit area in each region of
Fig. \ref{grid2}, plotted against distance from the galactic
centre. The density of C-type stars show a fairly steep decline from 
the centre of the galaxy out to about $4.5$ kpc before leveling out. A 
similar pattern is repeated in the density profile of the M-type stars, 
leveling out closer to $3.5$ kpc. The density profile of the RGB stars
shows similar behaviour - a steep decline is seen out to $\sim
4.5$ kpc after which the source density is almost constant out to $\sim
10$ kpc. We note the slight downturn at $8-10$ kpc and attribute this to
the poorer quality data we received from the NE quadrant, this is discussed
further below.

We consider two possible interpretations for the observed decline and
leveling out of the source density profiles at $\sim 4$ kpc. The first is that the
leveling out at a more constant stellar density is the result of
remaining MW foreground sources becoming dominant in the sample at
this distance, as foreshadowed in Sect. \ref{foreground}. The second
alternative is that the stars beyond $4$ kpc constitute an extended halo
around NGC 6822 populated predominantly by RGB and some M-type AGB
stars, with few C-type stars. Under the former scenario the
decline in the C- and M-type star density profiles at similar radial
distances would then represent the decreasing stellar density of NGC
6822, whereas under the second scenario it would represent the
transition radius from the central region to the halo.

\begin{figure} 
\resizebox{\hsize}{!}{\includegraphics[scale = 0.3]{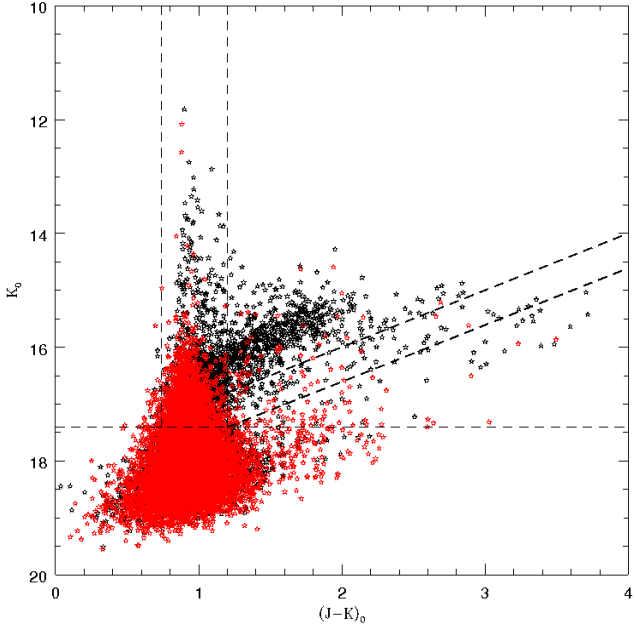}}
\caption{CMD of all the remaining sources after the 
removal of the foreground (black) and only those sources belonging 
to the medium and coarse grid regions (red) in Fig. \ref{grid2}. The horizontal
line represents the position of the TRGB at $K_0 = 17.41$ mag.  The
vertical lines represent the colour selection criteria at
$(J-K)_0 = 0.74$ and $1.20$ mag. The diagonal lines at $J_0 = 18.61$
and $18.0$ mag represent alternative selection criteria (see
Sect. \ref{jcut}).}  
\label{stellCMD}
\end{figure}

Figure \ref{stellCMD} also illustrates the declining number of AGB stars,
especially of C-type stars, with radial distance from the centre. It
shows a CMD of all the sources (in black) remaining after the 
removal of the foreground with those sources 
which are outside the small grid - Fig. \ref{grid2}, i.e.
$>2-3$ kpc from the galactic centre - shown in red. The sources 
in black show the peak belonging to the M-type AGBs and the diagonal
branch belonging to the C-type AGBs clearly. In comparison, sources
more than $2-3$ kpc from the centre (red) do not generate the strong
diagonal sequence. There are outer (red) sources
at $(J-K)_0 > 1.20$ mag but most seem to have merged upwards from below the
TRGB and do not look as though they belong to the C-star branch.
Sources from the outer parts of the galaxy between $0.74 < (J-K)_0 < 1.20$ mag, also
do not follow the strong peak belonging to the M-type AGB stars quite
as well. If the outermost stars are part of the MW foreground, then the
clearest indication of where the stellar component of NGC 6822 ends is
given by the stellar density profile of the C-type stars as they are
not present in the MW foreground. This interpretation is also consistent 
with the much lower density and more even distribution of sources outside the 
centre of the galaxy in Fig. \ref{maps}.

We have discounted the second scenario, an extended halo, as although extended stellar
halo's of this type have been detected in other galaxies, NGC 300
\citep{2005ApJ...629..239B} and M31 \citep{2005ApJ...628L.105I} for
example, in both cases the halo was discernible over several
scale lengths. In the case of NGC 6822,
if the slope in the stellar density profile of RGB population is
measured between $4-10$ kpc a scale length of $8.5$ kpc is
calculated. If the source density measurements from the NE
quadrant, which are responsible for the apparent further decline in
density at $8-10$ kpc, are removed the scale length of the proposed
extended halo increases to $20$ kpc. Thus although the idea of an
extended halo cannot be absolutely ruled out, we believe it is a
dangerous inference to make based on a measurement over $\sim 1$ scale
length or less. Therefore we adopt the first interpretation, that
beyond $4$ kpc our sample is dominated by the MW foreground and that
although genuine NGC 6822 AGB or RGB sources may still be present at
low densities beyond this limit, we are unable to reliably disentangle
them from the foreground with this data.

\subsection{AGB catalogue}
\label{cat}
As a result of the analysis in Sect. \ref{StellDen},
three catalogues are presented. The first, hereafter
Catalogue 1, presents \textit{candidate} C- and M-type AGB 
sources identified within $4$ kpc of the centre of NGC 6822. In 
order that our results may be verified we also present a second 
catalogue, Catalogue 2, containing all sources that met our 
reliability criterion (i.e. they have been classified as stellar or 
probably-stellar in all three photometric bands) across the full 
observed area - no other selection criteria have been applied to 
these sources. A third catalogue, 
Catalogue 3, is presented based on our findings during a comparative 
study of our work with the literature in Sect. \ref{cross} (we refer 
the reader to that section for further details) and contains sources
identified as stellar or probably-stellar in only two photometric  
bands. Catalogue 3 also covers the full observed area. Table. \ref{tab1} 
shows the first five lines of Catalogue 1, columns $1$
\& $2$ list the Right Ascension and Declination in degrees 
for the equinox J2000, columns $3$ \& $4$ list the dereddened $J$ magnitude and
associated photometric error, columns $5$ \& $6$ and $7$ \& $8$
contain the same information for the $H$ and $K$ bands
respectively and column $9$ classifies the star as either C- or M-type
based on its colour. Tables  \ref{tab3} \& \ref{tab4} which contain
similar data for Catalogues 2 and 3 but without the spectral type
classifications and with the addition of three columns 
giving the flag reference for each photometric band. The flags are as
follows; -1: stellar, -2: probably-stellar, -3: compact non-stellar,
-8: poor astrometry match, -9: saturated, 0: noise-like and 1: non-stellar.

\begin{table*} 
\centering                       
\begin{tabular}{ccccccccc}
\hline
$\mathrm{RA}$ &$\mathrm{Dec}$ &$J$ &$J \mathrm{-error}$ &$H$
&$H \mathrm{-error}$ &$K$ &$K \mathrm{-error}$
&$\mathrm{Type}$\\ 
$\mathrm{(deg)}$ &$\mathrm{(deg)}$ &$\mathrm{(mag)}$ &$\mathrm{(mag)}$
&$\mathrm{(mag)}$ &$\mathrm{(mag)}$ &$\mathrm{(mag)}$
&$\mathrm{(mag)}$ & \\     
\hline
296.284058 &-14.350628 &18.69 &0.09 &17.94 &0.06 &17.12 &0.04 &$\mathrm{C}$\\ 
296.081787 &-14.363892 &17.95 &0.05 &17.15 &0.03 &17.07 &0.04 &$\mathrm{M}$\\ 
296.149323 &-14.364606 &18.50 &0.07 &17.74 &0.05 &17.39 &0.05 &$\mathrm{M}$\\
296.248138 &-14.369500 &18.09 &0.05 &17.34 &0.04 &17.18 &0.04 &$\mathrm{M}$\\
296.397949 &-14.374426 &18.23 &0.07 &17.50 &0.05 &17.28 &0.05 &$\mathrm{M}$\\
... &... &... &... &... &... &... &... &...\\
\hline\\
\end{tabular}
\caption[]{The first five lines of Catalogue 1 - AGB sources belonging 
to NGC 6822 within $4$ kpc of the galactic centre.}  
\label{tab1} 
\end{table*}

\begin{table*} 
\centering                       
\begin{tabular}{cccccccccccc}
\hline
$\mathrm{RA}$ &$\mathrm{Dec}$ &$J$ &$J \mathrm{-error}$ &$J \mathrm{-flag}$ &$H$
&$H \mathrm{-error}$ &$H \mathrm{-flag}$ &$K$ &$K \mathrm{-error}$ &$K \mathrm{-flag}$\\ 
$\mathrm{(deg)}$ &$\mathrm{(deg)}$ &$\mathrm{(mag)}$ &$\mathrm{(mag)}$ &
&$\mathrm{(mag)}$ &$\mathrm{(mag)}$ & &$\mathrm{(mag)}$
&$\mathrm{(mag)}$ & \\     
\hline
296.377350 &-13.832600 &15.56 &0.009 &-1 &15.19 &0.008 &-1 &15.14 &0.010 &-1\\ 
296.366150 &-13.832656 &14.84 &0.006 &-1 &14.50 &0.005 &-1 &14.45 &0.006 &-1\\ 
296.564667 &-13.832878 &16.23 &0.010 &-1 &15.86 &0.010 &-1 &15.75 &0.040 &-2\\ 
296.771759 &-13.832906 &17.28 &0.030 &-1 &16.61 &0.030 &-1 &16.39 &0.070 &-1\\ 
296.739380 &-13.832977 &14.42 &0.004 &-1 &14.10 &0.004 &-1 &14.08 &0.010 &-1\\ 
... &... &... &... &... &... &... &... &... &... &...\\
\hline\\
\end{tabular}
\caption[]{The first five lines of Catalogue 2. See Sect. \ref{cat} for information
on the table contents.}   
\label{tab3} 
\end{table*}

\begin{table*} 
\centering                       
\begin{tabular}{cccccccccccc}
\hline
$\mathrm{RA}$ &$\mathrm{Dec}$ &$J$ &$J \mathrm{-error}$ &$J \mathrm{-flag}$ &$H$
&$H \mathrm{-error}$ &$H \mathrm{-flag}$ &$K$ &$K \mathrm{-error}$ &$K \mathrm{-flag}$\\ 
$\mathrm{(deg)}$ &$\mathrm{(deg)}$ &$\mathrm{(mag)}$ &$\mathrm{(mag)}$ &
&$\mathrm{(mag)}$ &$\mathrm{(mag)}$ & &$\mathrm{(mag)}$
&$\mathrm{(mag)}$ & \\     
\hline
296.335815 &-13.832008 &0 &0 &0 &16.34 &0.02 &-1 &16.23 &0.02 &-1\\ 
296.630035 &-13.832100 &0 &0 &0 &16.46 &0.02 &-1 &16.43 &0.07 &-1\\ 
296.654297 &-13.832395 &18.28 &0.07 &-1 &17.81 &0.08 &-1 &17.45 &0.17 &1\\ 
296.562927 &-13.832434 &19.05 &0.14 &-1 &18.28 &0.11 &-1 &0 &0 &0\\ 
296.705200 &-13.832597 &18.56 &0.09 &-1 &18.02 &0.09 &-1 &0 &0 &0\\ 
... &... &... &... &... &... &... &... &... &... &...\\ 
\hline\\
\end{tabular}
\caption[]{The first five lines of Catalogue 3. Columns are the same as for Table. \ref{tab3}.}  
\label{tab4} 
\end{table*}

Catalogue 1 contains $2368$ AGB stars, of which $769$ are C-type stars
and $1599$ are M-type stars. Applying our selection criteria to Catalogue 2 we find  
$3755$ candidate AGB sources of which $854$ are C-type stars and $2901$ are M-type
stars. However, it should be noted based on our findings in Sect. 
\ref{StellDen}, that we would expect both samples to contain foreground 
contamination.

\subsection{The C/M ratio}  
\label{c/m}

\subsubsection{Catalogue 1}
\label{cat1}
Conventionally the C/M ratio is used as an indirect indicator of the
metallicity of the environment in which AGB stars formed
\citep{2003MNRAS.338..572M}. A higher ratio is assumed to imply a
lower metallicity, because in low metallicity environments fewer
dredge up events  are required to create a carbon-rich atmosphere. The
C/M ratio is also affected  by the shift in the AGB evolutionary track
to higher temperatures at lower metallicities,  which reduces the
number of M-type AGB stars and increases the number of K-type stars
\citep{1999A&A...344..123M,1983ARA&A..21..271I}.  

The metallicity calibration of \citet{2005A&A...434..657B}, as refined by 
\citet{2009A&A...506.1137C} gives;

\begin{equation}
\mathrm{[Fe/H] = -1.39 \pm 0.06 - (0.47 \pm 0.10)log(\mathrm{C/M})}
\label{iron}
\end{equation}

Within $4$ kpc of the centre of NGC 6822 a C/M ratio of $0.48 \pm
0.02$ is derived. Using Eq. \ref{iron} this yields an overall
iron abundance of [Fe/H] $= -1.24 \pm 0.07$ dex. For the full 
observed area the C/M ratio is $0.29 \pm 0.01$, which
yields an iron abundance of [Fe/H] $= -1.14 \pm 0.08$ dex but the
foreground contamination of the M-type star population is not
negligible (Fig. \ref{stellCMD}). To obtain more reliable 
values for C/M and [Fe/H] further foreground removal is 
undertaken in Sect. \ref{addit}.

The error in the count of C- and M-type stars in each region has been 
calculated using Poisson statistics ($\pm \sqrt{N}$). This is
appropriate as the determination of the number of C- and M-type stars
is a counting exercise. Although we expect some variation in the number of stars of each
type we would expect the number per unit area to be around
some definite average rate ($\overline{N}$). The error associated with the 
C/M ratio and [Fe/H] has been calculated using the general formula of 
error propagation where the error associated with the count of C- and 
M-type stars have been treated as random and independent.

\subsubsection{Statistical foreground removal}
\label{addit}
Whilst the removal of foreground contamination by $J-H$ colour is 
a very useful technique, it is imperfect. 
\textit{Individual} sources with $(J-H)_0 < 0.72$ mag were removed as MW
foreground (Sect. \ref{foreground}) but based on our findings in Sect. 
\ref{StellDen} it was decided that further \textit{statistical} foreground removal was 
required. 
Using only those sources beyond the central $4$ kpc, i.e. 
those dominated by MW foreground, the average 
number density of C- and M-type stars that remained after the initial foreground 
removal was calculated. It was found that 
there were $\sim 46$ M-type stars and $\sim 3$ C-type stars remaining per 
$0.08$ deg$^2$ (or $\sim 8$ M-type stars and $\sim 0.5$ C-type 
stars per kpc$^2$). The count of AGB stars of each 
type inside a radius of $4$ kpc was then reduced accordingly. 
The effect of this statistical adjustment on the C/M ratio inside $4$ kpc is   
that it increases to $0.62 \pm 0.03$, giving an [Fe/H] of $-1.29 \pm 0.07$ dex.  
A comparison with previous estimates of the C/M 
ratio in NGC 6822 \citep{2006A&A...454..717K,2005A&A...429..837C} and
the sensitivity of the ratio to the initial foreground selection
criterion will follow in Sect. \ref{diss}.

\subsection{Gradients}
\label{spavar}
\begin{figure} 
\resizebox{\hsize}{!}{\includegraphics[scale = 0.3]{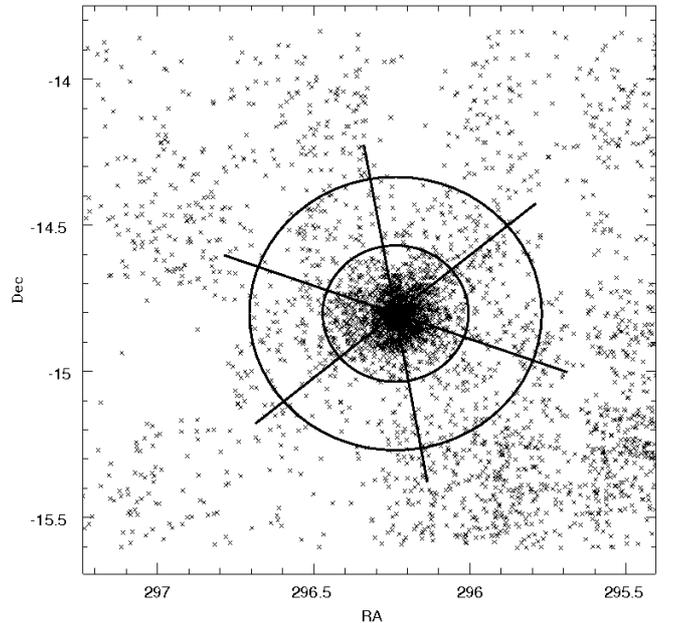}}
\caption{To study variations in the TRGB magnitude, C/M 
ratio and [Fe/H] as a function of azimuthal angle the region inside $4$ 
kpc has been divided into $12$ regions - $2$ annuli between $0-2$ kpc and $2-4$ kpc 
with each annuli further divided into $6$ regions at intervals of $60^{\circ}$.} 
\label{circ}
\end{figure}

Variations in the TRGB magnitude, C/M ratio and [Fe/H] across NGC 6822 
are examined in this section as a function of distance from the galactic 
centre and azimuthal angle. The region inside $4$ kpc has been divided 
into annuli extending between $0-2$ kpc and $2-4$ kpc. These annuli have 
then been further divided every $60^{\circ}$ (Fig. \ref{circ}) leaving $12$ 
regions that can be used to study any variation in the TRGB magnitude, C/M 
ratio and [Fe/H] with \textit{angle}. In order to study these parameters as a function 
of \textit{distance} from the galactic centre the area inside $4$ kpc has been separately 
divided into $4$ annuli at a spacing of $1$ kpc between $0-4$ kpc and the 
TRGB magnitude, C/M ratio and [Fe/H] measured/calculated for each annulus. Each region has been checked to ensure  
the criteria for the reliable application of the Sobel filter has been met.

\subsubsection{The metallicity gradient}
\label{metgrad}
\begin{figure} 
\resizebox{\hsize}{!}{\includegraphics[scale = 0.5]{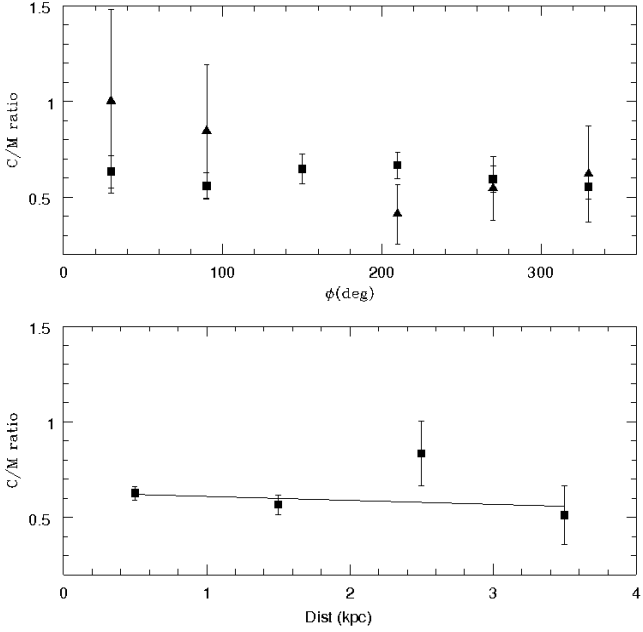}}
\caption{Top: Distribution of the C/M ratio in each region of the inner 
(squares) and outer (triangles) annuli 
(Fig. \ref{circ}) plotted against azimuthal angle. Bottom: The C/M 
ratio in each of $4$ consecutive annuli at a separation of $1$ 
kpc between $0-4$ kpc plotted against distance from the galaxy centre. 
An error-weight linear fit has been made to the data.}
\label{optBfig4}
\end{figure}

\begin{figure} 
\resizebox{\hsize}{!}{\includegraphics[scale = 0.5]{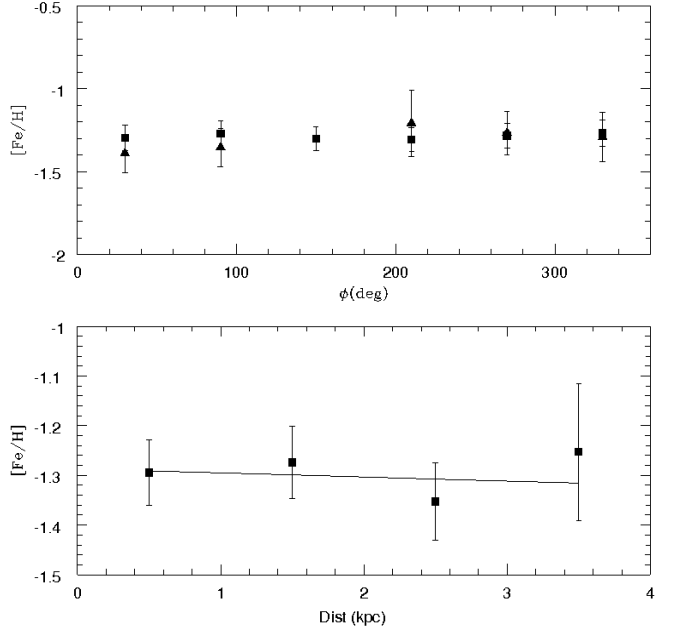}}
\caption{Top: [Fe/H] abundance in each region of the inner (squares) and outer (triangles) 
annuli (Fig. \ref{circ}) plotted against azimuthal angle. Bottom: [Fe/H] 
plotted against distance from the galaxy centre for the same annuli 
described in Fig. \ref{optBfig4}. An error-weight linear fit has been 
made to the data.}
\label{optBfig1}
\end{figure}

Prior to the calculation of the C/M ratio and [Fe/H] value the relevant 
statistically adjustments have been made to the counts of C- and M-type 
stars, as outlined in Sect. \ref{addit}.
The top panel of Fig. \ref{optBfig4} shows the distribution of the
C/M ratio as a function of angle for both the inner annuli ($0-2$ kpc) 
and the outer annuli ($2-4$ kpc) of Fig. \ref{circ}. In the bottom panel of the same 
figure the C/M ratio is plotted as a function of distance from the 
galaxy centre. Similar plots for the [Fe/H] abundance have 
also been made and are presented in Fig. \ref{optBfig1}.

A spread of $0.59$ has been detected in the C/M ratio out to a radius of 4 kpc, this
translates into a spread of 0.18 dex in the iron abundance between $−1.21$ dex and 
$−1.39$ dex. For the inner annuli there does not appear to be
any obvious dependence on angle in the distribution of the C/M
ratio (Fig. \ref{optBfig4}. For the outer annuli there is a significant scatter in
the ratio and a possible decline in the ratio with increasing angle, 
however the size of the
associated error bars, due to the small number of sources,
suggest that we can not draw any firm conclusions about the behaviour
of the C/M ratio with angle in the outer annuli. In the bottom panel
of Fig. \ref{optBfig4}, the C/M ratio has been plotted as a function
of distance for the $4$ annuli described above. A small negative
gradient appears to be present in the C/M ratio (C/M $= 0.63(\pm 0.06)
- 0.02(\pm 0.04)\times$dist/kpc), calculated using a weighted
least-squares fit but the size of the associated error suggests 
this is not significant, again due to the declining stellar density 
and hence small number statistics.

In the top panel of Fig.\ref{optBfig1} [Fe/H] is plotted as a function
of angle. Data points for both the inner and the outer annuli
are presented but there is no obvious variation of [Fe/H] with angle
in either case. In the bottom panel [Fe/H] is plotted as a function of
distance from the galactic centre. As with the C/M ratio a weighted
fit of the data gives a slightly negative slope of [Fe/H] $= -1.29(\pm
0.04) -0.008(\pm 0.023)\times$dist/kpc, which again implies no 
significant gradient. 

As we have selected the C- and M-type stars in each region of the galaxy on the basis of 
colour, reddening variations within NGC 6822 may also affect the distribution of the 
C/M ratio. \\The sensitivity of the C/M vs. [Fe/H] relation to changes in the 
selection criteria for C- and M-type stars, and the robustness of the C/M ratio 
as an indicator of metallicity, will be discussed further in Sect. \ref{C/Mvar}.

\subsubsection{TRGB variations}
\label{TRGBvar}
\begin{figure*}
\resizebox{\hsize}{!}{\includegraphics[scale=0.50]{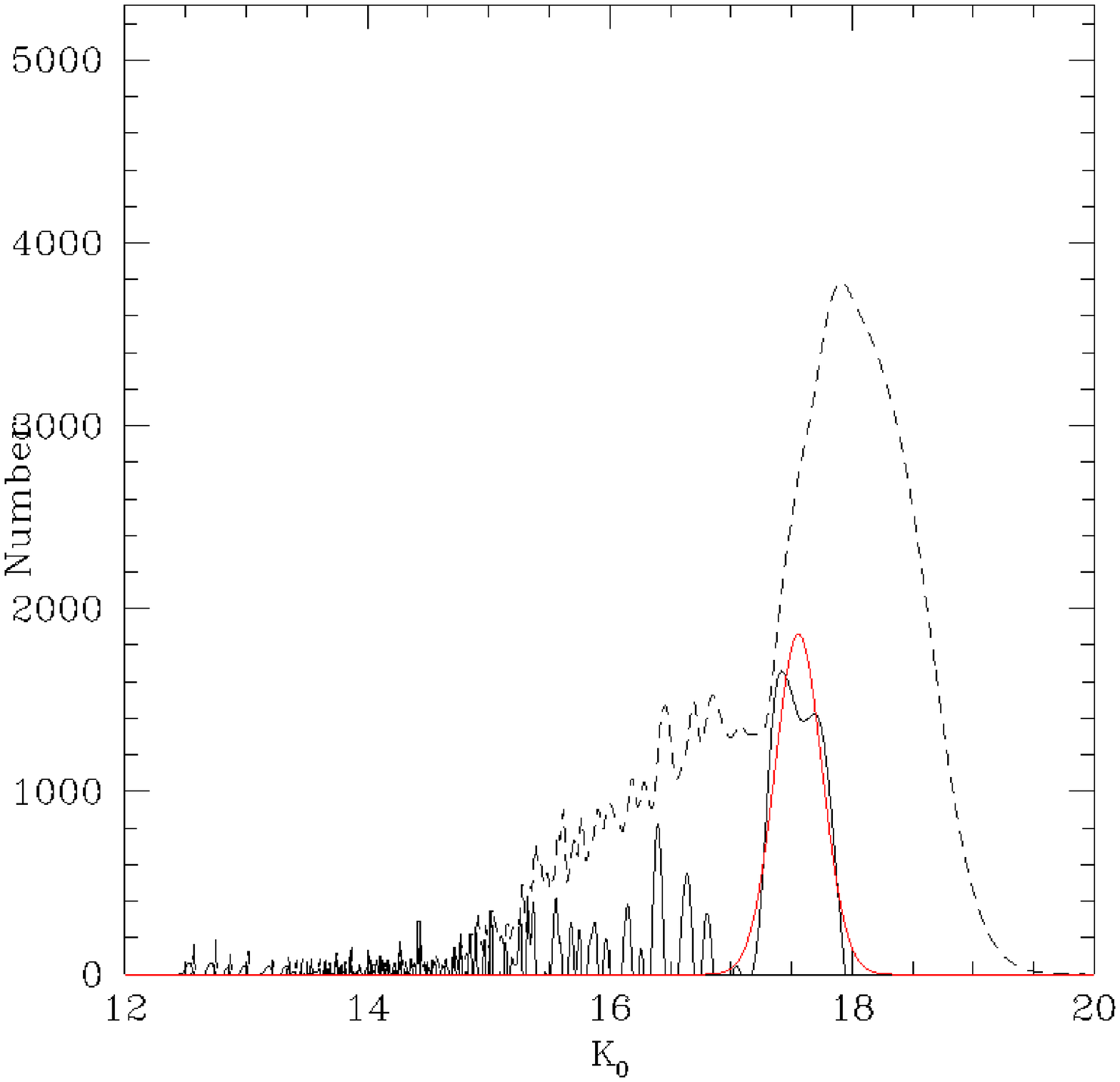}
\includegraphics[scale=0.50]{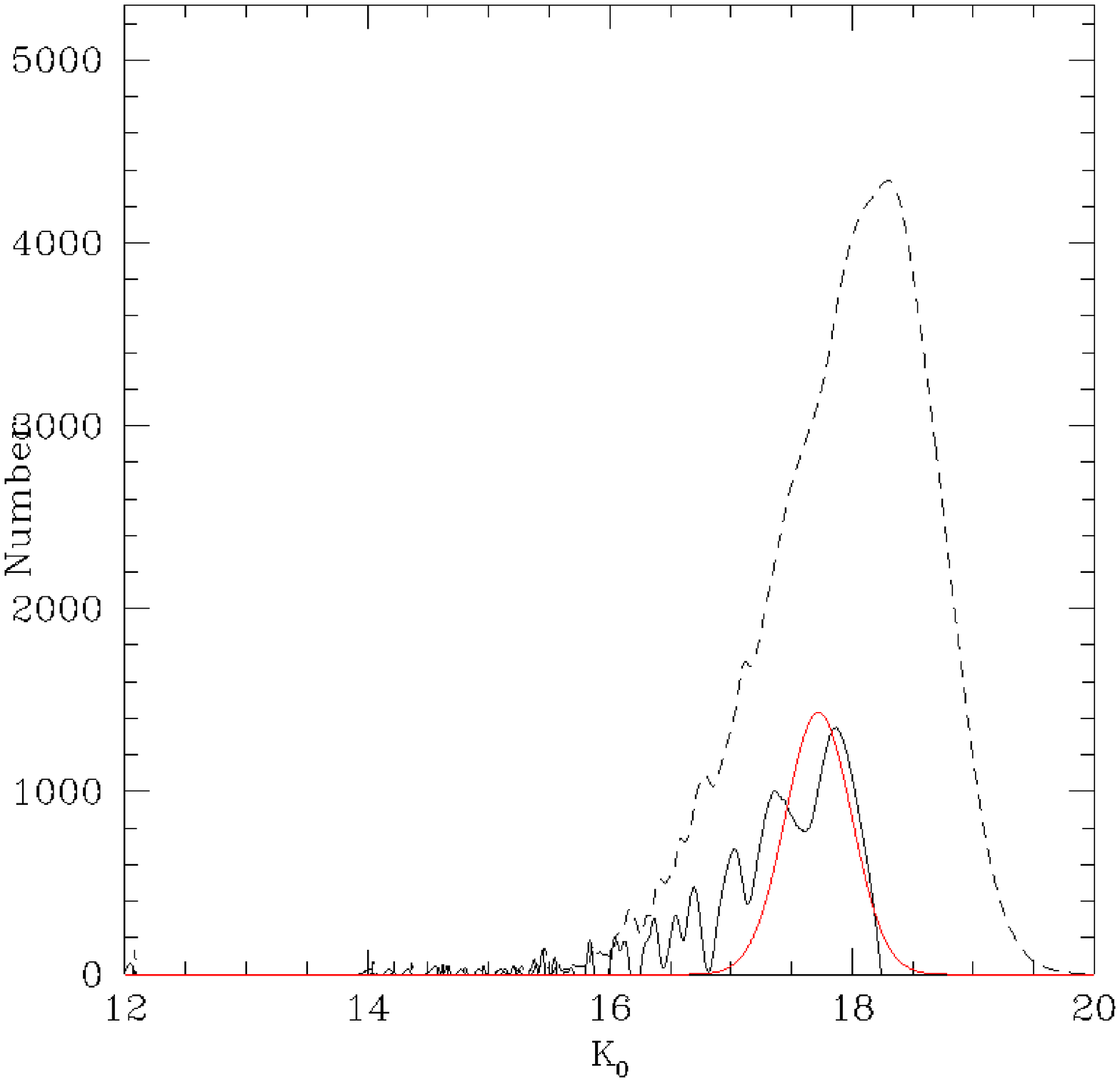}}
\caption{Left: Smoothed $K$-band magnitude distribution of sources that 
remain within the central $4$ kpc after the removal of foreground 
contamination $(J-H)_0$ colour. Right: Same as left panel but for all sources that remain 
outside the central $4$ kpc.}
\label{TRGBdist}
\end{figure*}

Within a radius of $4$ kpc (Fig. \ref{circ}) a variation of 
$\Delta K = 0.19$ mag was found in the magnitude of the TRGB,  
with an average and standard deviation of $K_0 = 17.46 \pm 0.05$ 
mag. When measurements made outside the $4$ kpc limit are included the spread of values  
in the position of the TRGB increases dramatically to $\Delta 
K = 0.99$ mag. The much greater spread in TRGB values detected outside the central
$4$ kpc is attributed to the decline in the number of genuine NGC 6822 sources,
hence we are not detecting a genuine TRGB here but simply a variation
in the magnitude distribution of the MW foreground. Figure
\ref{TRGBdist} shows the $K$-band magnitude distributions of stars
within (left) and beyond (right) $4$ kpc. The AGB population is
obvious in the inner sample at $15 < K_0 < 17.5$, but is inconspicuous
in the outer subsample, consistent with the galaxy being lost in the
MW foreground contamination beyond  $4$ kpc
(Sect. \ref{StellDen}). Hence TRGB measurements beyond $4$ kpc are
either poorly constrained or entirely spurious. The measurements arise
only because the Sobel filter reports the position of the greatest
change of slope in the magnitude distribution but if those stars are
dominated by the MW foreground and not the RGB population in NGC 6822,
then there may not be an RGB termination.

\begin{figure} 
\resizebox{\hsize}{!}{\includegraphics[scale = 0.5]{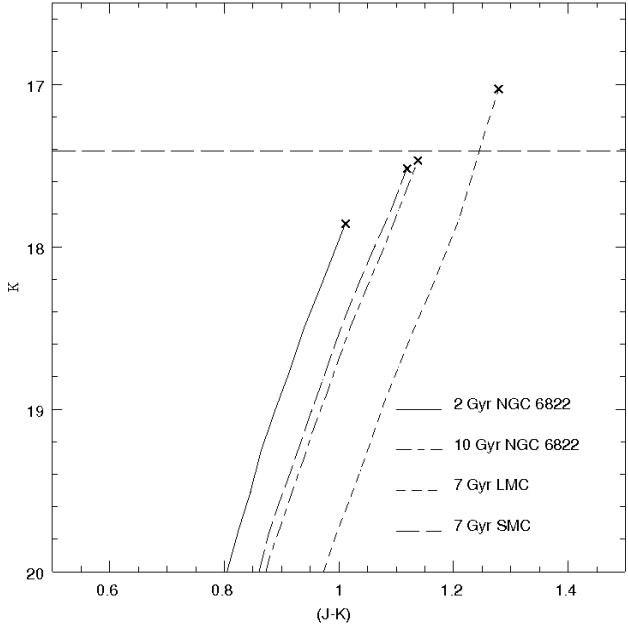}}
\caption{CMD for four isochrones plotted using data from the 
Padova evolutionary tracks. The stellar evolutionary tracks are as 
follows; 2Gyr (TRGB $K = 17.86$ mag) and 10Gyr (TRGB $K = 17.47$ mag) 
at Z = $0.002$ ($\sim$ NGC 6822; this work with an adjustment for 
the presumed abundance excess of $\alpha$-elements based on the work of \citealt{1991AJ....101.1865R}), 
and for Z = $0.002$ (SMC; \citealt{1999A&A...346..459M}) (TRGB $K 
= 17.52$ mag) and Z = $0.007$ (LMC; \citealt{1999A&A...346..459M}) 
(TRGB $K = 17.03$ mag) at an age of 7Gyr. The crosses mark the position 
of the TRGB for each isochrone and the horizontal line marks the 
position of the TRGB magnitude used in this study. The average of 
several [Fe/H] measurements of the LMC and the SMC have been used 
for convenience.} 
\label{iso2}
\end{figure}

$K$-band measurements of the TRGB magnitude are sensitive to both the age and metallicity 
of the population \citep{2005MNRAS.357..669S}. In a population of a single metallicity,
the TRGB in the $K$-band is fainter in the intermediate-age stars than in the older 
population. Whilst in a population of a single age the TRGB magnitude is brighter with 
increasing metallicity. Fig. \ref{iso2} clearly demonstrates the individual effects 
of these two variables. For a population of mixed age and metallicity the 
anti-correlation between these two affects makes it difficult to decipher the cause of
any observed magnitude spread \citep{2005MNRAS.357..669S,2008MNRAS.388.1185G}.

\begin{figure} 
\resizebox{\hsize}{!}{\includegraphics[scale = 0.5]{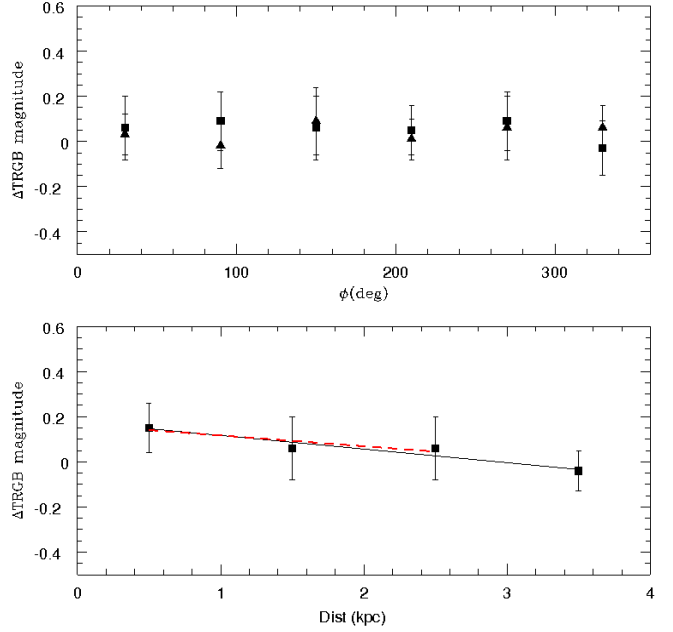}}
\caption{Top: Distribution of TRGB magnitudes for each 
region of the inner (squares) and outer (triangles) annuli (Fig. \ref{circ}) plotted 
against azimuthal angle. Bottom: TRGB magnitude measured in each of 
the $4$ equally spaced annuli described previously plotted as a function of 
distance from the galactic centre. A weighted linear fit has been made to 
all the data points (solid black line) and to only those points within
$3$ kpc of the centre (dashed red line).}
\label{optBfig2}
\end{figure}

In order to try and understand the metallicity and age distribution 
of the underlying AGB population and also the structure of the galaxy, 
the distribution of the TRGB magnitude has been investigated as a function 
of distance from the galactic centre and azimuthal angle. Negative 
values on the vertical axis represent a TRGB 
brighter than $K_0 = 17.41$ mag. The top panel of Fig. \ref{optBfig2} shows 
the variation in the TRGB magnitude with angle for both the inner and 
outer annuli (Fig. \ref{circ}). The data shows no obvious trend with angle 
for either annuli.

In the bottom panel of Fig. \ref{optBfig2}, a weighted fit to all 
the data shows a negative slope ($\Delta$TRGB $= 0.17(\pm 0.03) - 0.06(\pm 0.01)\times$dist/kpc) 
in the TRGB magnitude with radial distance. This is an interesting result but may be the result 
of increased distortion in the magnitude distribution as remaining foreground 
contamination becomes more severe in the outer annuli. Therefore a weighted fit was also 
made to the inner three data points only and a negative slope was again 
found ($\Delta$TRGB $= 0.16(\pm 0.04) - 0.05(\pm 0.02)\times$dist/kpc), although the slope is 
reduced from $6\sigma$ to a $2.5\sigma$ detection. Such a slope
could be consistent with the presence of a halo of older stars  
around NGC 6822. Assuming an environment with relatively little variation in 
metallicity (Fig. \ref{optBfig1}) and given that there seems to 
have been little recent star formation in the outer galaxy 
\citep{2006A&A...451...99B,2009MNRAS.395.1455S}. A shift in the TRGB to brighter 
magnitudes could be indicative of the increasing age of the population 
\citep{2005MNRAS.357..669S}. Therefore, although we advise caution due to the 
size of the associated errors in Fig. \ref{optBfig2}, the shift to a
brighter TRGB magnitude in the outer galaxy may be a genuine feature.    

An alternative possibility that the variation in the TRGB 
is due to the inclination of the galaxy to our line of sight, has been 
rejected as it would require NGC 6822 to have a depth of 
$\sim 45$ kpc. The effects of reddening have been shown to vary across 
NGC 6822, this may also account in part for the variation that has been detected.  
The \citet{1998ApJ...500..525S} maps show an extinction range of E($B-V$) $= 0.15$ 
to $0.39$ in the direction of NGC 6822 and a spread of $0.3$ is found in the literature
(Sect. \ref{obs}). However, even the larger range of values is insufficient 
on its own to account for the TRGB magnitude spread.

As all the 
TRGB values measured within $4$ kpc of the galactic centre are 
within $2\sigma$ of the mean value ($K_0 = 17.46 \pm 0.05$ mag) the spread we 
observe may also simply be the result of random statistical variations. The 
position of the TRGB is not a single value but a range due to the width of 
the RGB in a composite population \citep{2010MNRAS.tmp..431H}.
The effect of the detected spread in metallicity ($0.18$ dex)
on the TRGB magnitude for populations of a single age has been considered but it 
is insufficient to account for the TRGB magnitude variation when considered alone. 
A sufficiently large spread in the age of the population could account for the 
variation we detect, however, at this time we are unable to constrain the age of 
the AGB population.

The most likely scenario is that the TRGB magnitude spread is the cumulative 
result of a number of factors including reddening, age, metallicity, some 
distortion by remaining foreground contamination and expected variations 
in the TRGB. The spread detected in the $J$- ($\Delta J = 0.20$) and $H$-bands 
($\Delta H = 0.21$), in conjunction with $I$- and $V$-band data that we have
yet to analyse, could be used to constrain the metallicity, age and extinction 
variations in NGC 6822 due to the different sensitivities of each waveband 
to these variables.

\section{Discussion}
\label{diss}

\subsection{The structure of NGC 6822}
\label{strut}
The large area covered by the data gives a good overview of the 
structure of the galaxy. The primary result of 
Sect. \ref{results} was the placement of the $4$ kpc limit 
on the stellar component of the galaxy. This radial limit is supported
by the source density plots (Fig. \ref{maps}),  
the density profiles (Fig. \ref{stellden1}) and the 
magnitude distribution plots (Fig. \ref{TRGBdist}).

\begin{figure} 
\resizebox{\hsize}{!}{\includegraphics[scale = 0.5]{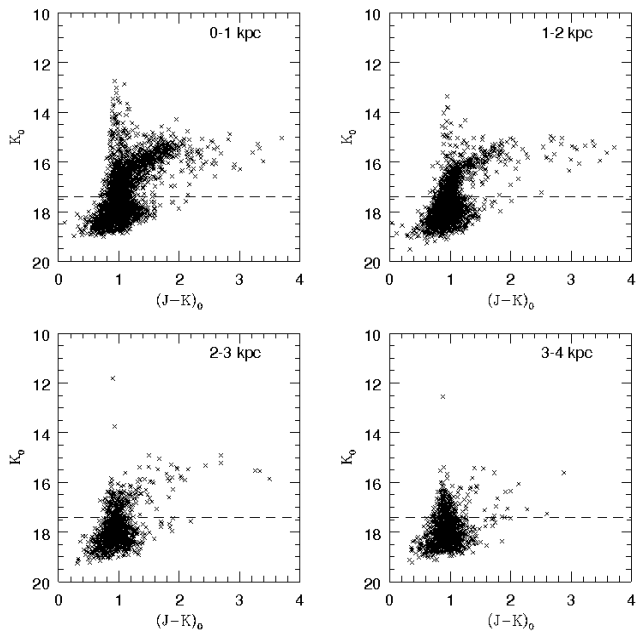}}
\caption{CMD of the sources contained in 
each of the $4$ annuli described in Sect. \ref{spavar}. 
The horizontal line in each frame marks the position of 
the TRGB at $K_0 = 17.41$ mag. No statistical adjustments 
have been made so foreground contamination that remained 
after the application of the $J-H$ colour selection is also 
present in the CMDs and is expected to become increasing significant with 
increasing radial distance. The total number of sources 
and the total number of AGB sources is $2683$ and $1325$ 
respectively in the $0-1$ kpc annuli, $1823$ and $580$ in 
the $1-2$ kpc annuli, $1052$ and $230$ in the $2-3$ kpc 
annuli and $1115$ and $233$ in the $3-4$ kpc annuli.} 
\label{rings}
\end{figure}

The source density plots in Sect. \ref{spadis} show that the majority 
of the AGB population is concentrated in and around the region of the central 
bar but extends beyond this, with decreasing density out to about $4$ kpc. 
This is supported by Fig. \ref{rings} which shows a CMD of the sources 
contained within each of the $4$ annuli used to examine the behaviour of various 
parameters with distance (Sect. \ref{spavar}). The vertical 
sequence belonging to the M-type stars and the diagonal branch generated by the 
C-type stars are clearly visible in the top two panels especially but deteriorate 
with increasing distance from the galactic centre due to the declining number of sources. 
There does appear to be some structure beyond the $4$ kpc limit in the SW, this is particularly 
apparent in the low resolution source density plot of the M-type AGB
population (Fig. \ref{maps}). The cause of this overdensity in the SW will be 
discussed below. Aside from the SW overdensity, we detect no significant 
structure beyond the central $4$ kpc.

A radial limit of $4$ kpc corresponds to an angular distance of $\sim 28'$ from 
the centre at a distance of $490$ kpc (a diameter of $\sim 56'$) and is 
comparable to previous estimates of the extent 
of the stellar component of the galaxy by \citet{2002AJ....123..832L}, \citet{2006A&A...456..905D} 
and \citet{2006A&A...451...99B}. Using R, I, CN and TiO filters, \citet{2002AJ....123..832L} 
surveyed C-type stars in an area of $42' \times 28'$ in NGC 6822 and were 
the first to propose the existence of a ``halo'' of old - and intermediate age stars around NGC 6822. 
\citet{2002AJ....123..832L} suggested that the spheroid had a major-axis length of $\sim 23'$
 (i.e. a radius of $1.65$ kpc) at a distance of $(m-M)_o = 23.49 \pm
 0.08$ mag. This is smaller than what we see in Fig. \ref{stellden1}; given the smaller observing
area of \citet{2002AJ....123..832L},
though by having data significantly further out (to a distance of $11$ kpc) we are able to see 
the extent of the structure more clearly and trace the AGB halo out to a radius of $4$ kpc. 
In a survey of area $2^{\circ} \times 2^{\circ}$ using $g', r'$ and $i'$ filters, 
\citet{2006A&A...451...99B} traced the density enhancement of RGB sources, 
selected from their CMDs, from the centre of NGC 6822 out to a semi-major axis distance of $36'$ 
($\sim 4.9$ kpc, assuming a distance of $470$ kpc). \citet{2006A&A...451...99B} also provide a 
surface density profile of the C-type stars identified from the SDSS colours to supplement the 
findings of \citet{2002AJ....123..832L}, although they admit that C-type stars cannot be 
unambiguously selected without appropriate corrections due to contamination from background 
galaxies in the data. They conclude that a non-negligible number of C-type stars are detected 
up to $\sim 40'$ ($\sim 5.5$ kpc). The extent of the RGB population detected by \citet{2006A&A...451...99B} 
is comparable with what is seen in Fig. \ref{stellden1}, however, we do not claim to reliably detect 
C-type stars out to such large distances as \citet{2006A&A...451...99B}. 
In a follow up to the work of \citet{2006A&A...451...99B},
\citet{2006A&A...456..905D} observed two regions of $34'.8 \times 34'.8$ in the 
J and K$_s$ bands along the major-axis of the spheroidal halo to further assess the extent 
of the C-type population. They present a surface density profile that is consistent with the 
work of \citet{2006A&A...451...99B} clearly showing the C-type population extends at least out 
to an angular distance of $\sim 30'$ ($4.3$ kpc at $490$ kpc),
and possibly beyond. \citet{2012A&A...537A.108K} 
also provides evidence for a radial limit of $\sim 4-5$
kpc on the stellar component of NGC 6822, they found that
all their candidate M- and C-type AGB stars outside the elliptical spheroid
of \citet{2006A&A...451...99B} for which they were able to collect
low-resolution spectra were in fact MW dwarf stars. 

We have detected the AGB population out to a distance of
$\sim 4$ kpc. However, \citet{2005nfcd.conf..181L} 
presented preliminary results reporting the discovery of a star cluster 
belonging to NGC 6822 at a distance of $12$ kpc from the galactic centre. 
\citet{2011ApJ...738...58H} expand on these findings with new star clusters 
associated with NGC 6822 spanning an area of $120' \times 80'$,
this is much larger than the area examined here or 
by \citet{2006A&A...451...99B}. This suggests
that the structure of NGC 6822 is complex and cannot be traced by a single 
stellar population. Such complex structure has also been detected 
in other dwarf irregular galaxies like Leo A \citep{2004ApJ...611L..93V}, the LMC 
\citep{2003Sci...301.1508M,2000A&A...358L...9C} and IC 10 \citep{2004A&A...424..125D}. 
\\The reliable detection of the AGB population of NGC 6822, the M-type star
population in particular and of the extent of the stellar halo has been hampered 
by heavy foreground contamination. Our selection of the $J-H$ criterion for 
the initial removal of the contaminating foreground will be discussed further 
in Sections \ref{over} and \ref{forecol}. However, we are confident that we have 
isolated well the C- and M-type AGB population of NGC 6822 within $4$ kpc of the 
galactic centre and for the first time provided NIR observations across the whole of
the AGB stellar component.

\subsection{The SW overdensity}
\label{over}
\begin{figure} 
\resizebox{\hsize}{!}{\includegraphics[scale = 0.5]{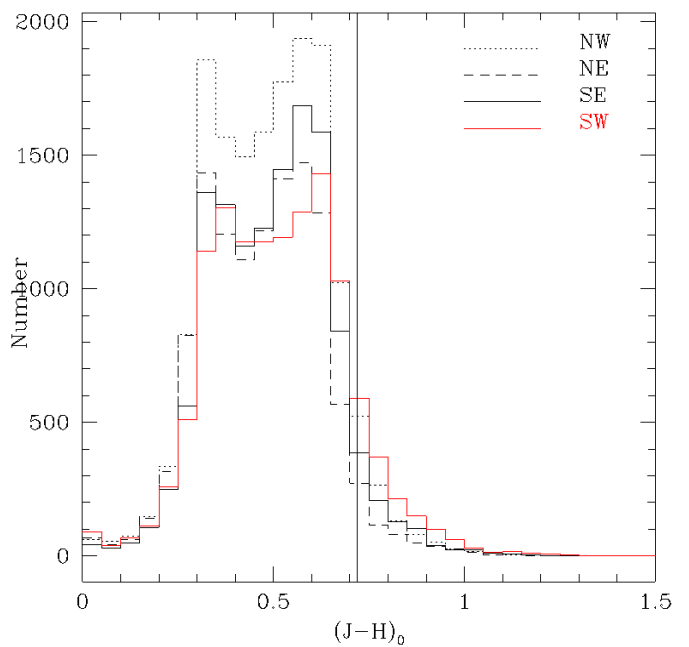}}
\caption{Colour distribution (bin size $0.05$ mag) of sources from the 
outer NE, NW, SE and SW quadrants of the observed area. $8\%$ of sources 
from the SW are redder than $(J-H)_0 = 0.72$ mag - marked by the 
solid vertical line whereas only $4.5\%$ of stars in the other 
quadrants exceed this cut off.}
\label{redcheck}
\end{figure}

The overdensity seen in the SW of the low resolution M-type 
AGB source density plot in Fig. \ref{maps} is not seen C-type 
or RGB populations at any significant level. This would seem to suggest that the overdensity is 
either populated almost exclusively by M-type stars or that it is the result of 
an excess MW foreground with M-star-like NIR colours leaking into our sample. 
The fact that the overdensity is 
not observed in the C-type star density plots is consistent with either proposition, 
as C-type stars do not appear in the MW foreground in any 
significant quantity. Figure \ref{redcheck} 
shows the $J-H$ distribution of sources in a $\sim 30' \times 30'$ field
at the edge of the observed area in the NE, NW, SE and SW (red).  
In the SW the distribution is clearly shifted to the red, and $8\%$ of 
the total number of sources in that region are redder than $(J-H)_0 = 0.72$ mag, compared 
to an average of $4.5\%$ in the NW, NE and SE quadrants. This indicates 
that the leakage of MW stars into the SW quadrant will be almost twice as high as 
in the other quadrants.

\begin{figure} 
\resizebox{\hsize}{!}{\includegraphics[scale = 0.3]{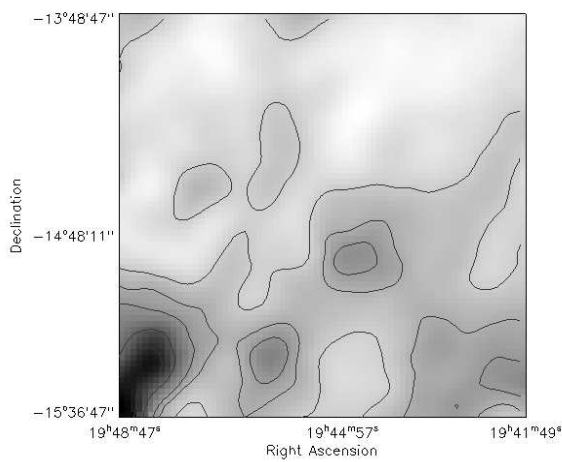}}
\caption{Contour plot of the foreground reddening (E(B-V)) across the observed 
area taken from the extinction maps of \citet{1998ApJ...500..525S}.
Contours are at: $0.20, 0.23, 0.25, 0.30, 0.35$ mag.}
\label{colmap}
\end{figure}

The reason for the difference in colour is harder to discern
but we suggest differential reddening. Our sample has been corrected
for foreground reddening using the \citet{1998ApJ...500..525S} dust
maps (Fig. \ref{colmap}), however,  
the higher resolution of our data means that this correction is imperfect and 
some reddening variation will still be present in the sample. This 
may account for the redder distribution of sources in the SW and 
the consequent leakage of more foreground sources into our sample 
in that region. However, \citet{2012A&A...537A.108K} find only a negligible 
variation in the NIR extinction across the observed area and conclude 
that the overdensity in the SW (based on the same photometric
catalogue) may be a genuine extension of NGC 6822 
but that spectroscopic confirmation is required. \\An examination of the CMD
in the SW overdensity does not show any strong features to
suggest that this structure is made up of genuine NGC 6822 AGB
stars. Although there is a vertical feature extending to brighter
$K$-band magnitudes than in the
SE, it does not represent a strong M-star peak and is probably the
result of the higher density of sources in the SW. Furthermore, the
magnitude distribution of the sources in the SW does not show a strong
TRGB but a rather broad distribution similar to the one seen in the
right-hand panel of Fig. \ref{TRGBdist}. 
Although, due to foreground contamination of the sample outside the densest
region of the galaxy and the difficulty of 
identifying NGC 6822 stars with certainty from photometric data, we 
can not rule out the possibility that the SW overdensity does contain
some outlying AGB stars.

In the context of reddening variations across the galaxy, 
it is also interesting to note the underdensity in the SE 
that was seen in the source density plot of the M-type AGB 
stars (Fig. \ref{maps}), but which is much less apparent in 
the other source density plots. Previously (Sect. \ref{spadis}) 
we suggested that this underdensity may coincide with the 
Super Giant Hole in the HI disk \citep{2000ApJ...537L..95D}. 
This region also coincides roughly with an area of significant 
reddening variation (Fig. \ref{colmap}) and from the colour 
distribution in Fig. \ref{redcheck} it is clear that the decline 
in sources in the SE occurs at bluer colours ($(J-H)_0 \sim 0.65$ mag) 
relative to the SW and NW. We conclude that the underdensity in 
the SE is more likely to be the result of an over subtraction of 
sources due to imperfect reddening-correction in that region
rather than a real feature. 

The removal of foreground contamination via $J-H$ colour alone 
is a useful but imperfect method as the selection of the colour 
criterion used is subjective. Any under- or over-subtraction of 
the foreground will primarily affect the M-type AGB sources rather 
than the redder C-type stars - see Fig. \ref{fore} -  and can affect 
bias the derived C/M ratio; this will be discussed further in Sect. 
\ref{forecol}.

\subsection{C/M ratio and [Fe/H]}
\label{C/Mvar}
Globally we find a [Fe/H] value of
$-1.29 \pm 0.07$ dex (C/M $=0.62 \pm 0.03$) for AGB stars within a $4$
kpc radius of the centre. This value is in good agreement with the
findings of other authors who have derived the iron abundance of NGC
6822 using the C/M ratio. 

Using broad- ($R$ and $I$) and narrow-band ($CN$ and $TiO$) filters to
identify the C-type population in the central $42' \times 28'$ of NGC
6822, \citet{2002AJ....123..832L} derived the C/M ratio to be $1.0 \pm
0.2$. The size of the M-type population in the work of
\citet{2002AJ....123..832L} was estimated by subtracting the estimated
stellar density of the foreground, measured in two strips at the edge
of the observed area that were assumed to contain a negligible number
of genuine NGC 6822 sources, from the total observed sample. No iron
abundance was given by \citet{2002AJ....123..832L} but using the
relation of \citet{2009A&A...506.1137C}, this corresponds to a value
of [Fe/H] $= -1.39$ dex.  \\Using similar techniques to those employed
here ($J$ and $K_{s}$-band photometry) \citet{2005A&A...429..837C}
estimated the C/M ratio in the central $20' \times 20'$ of NGC 6822 to
be $0.32$, with an absolute variation of $6$. This corresponds to
[Fe/H] $=-1.11$ dex and a variation in the iron abundance of
$\Delta$[Fe/H] $=-1.56$ dex using the C/M vs. [Fe/H] relation given in
the same paper. Again using $JHK$ photometric filters,
\citet{2006A&A...454..717K} surveyed the central $3'.6 \times 6'.4$ of
NGC 6822 and reported an overall C/M ratio of $0.27 \pm 0.03$, with
variations between $0.22 \pm 0.03$ and $0.31 \pm 0.04$ in the north
and south respectively. This translates into [Fe/H] $\approx -0.99$
dex globally, using the relations of \citet{2005A&A...434..657B} and
\citet{2005A&A...429..837C} \citep{2006pnbm.conf..108G}, with a
variation of $0.07 \sim 0.09$ dex across the observed area. Using the
more recent relation of \citet{2009A&A...506.1137C}  we derive an
average [Fe/H] of $-1.12$ dex with a spread of $0.07$ dex using the
values of \citet{2006A&A...454..717K}.     

More recently \citet{2012A&A...537A.108K} have presented a C/M ratio
of $\sim 1.05$ with a variation between $0.2$ and $1.8$, based on
their analysis of the original UKIRT catalogue used here but selection
criteria determined from the analysis of a spectroscopic sample. 
This ratio yields a mean [Fe/H] between $-0.90$
and $-1.50$ dex using the relations of \citet{2006pnbm.conf..108G},
\citet{2005A&A...434..657B} and \citep{2009A&A...506.1137C} with
average values of $\sim -1.20$, $\sim -1.30$ and $\sim -1.30$ dex
respectively. These values are in good agreement with our own but the
following points should be noted; firstly, some of the selection
criteria used here, in particular the blue limit, are quite different
from those used by \citet{2012A&A...537A.108K}. Secondly,
\citet{2012A&A...537A.108K} themselves note that their spectroscopic
sample is biased towards C-type stars, which may have affected
their determination of the AGB selection criteria and therefore their
determination of the C/M ratio. Both of these points are discussed in
more detail below (Sect. \ref{blue2}). For our analysis we have
adopted the most recent calibration of \citet{2009A&A...506.1137C}
but for comparison we note that using our C/M ratio ($0.62$) and the
relations of \citet{2003A&A...402..133C}, \citet{2006pnbm.conf..108G}
and \citet{2005A&A...434..657B} we obtain [Fe/H] values of $-0.82$
dex, $-1.14$ dex and $-1.19$ dex, respectively, in the central $4$ kpc
of the galaxy. Therefore any comparison of metallicities derived from
the C/M ratio must take into account the relations that have been used
to derive them.     

Other estimates of the mean metallicity of the old- and intermediate-age 
stars of NGC 6822 that do not rely on the C/M ratio range between $-1.0$ dex
\citep{2001MNRAS.327..918T,2003PASP..115..635D} and $-1.5$ dex
\citep{1996AJ....112.1928G}, with significant
scatter. \citet{2001MNRAS.327..918T} used measurements of the
equivalent width of Ca II triplet lines of individual RGB stars in NGC
6822 to determine a mean metallicity of $-1.0 \pm 0.5$ dex with a
range between $-0.5$ and $-2$ dex. \citet{2001MNRAS.327..918T} comment
that they do not see any evidence for spatial variations but 
they only observed a small area ($5' \times 5'$) near the centre of
the galaxy. \citet{2003PASP..115..635D} used $JHK$ photometry to
investigate the slope of the RGB in three fields ($34' \times 34'$)
across NGC 6822 and determined a mean value of [Fe/H] $= -1.0 \pm 0.3$ dex. 
Therefore the global value of [Fe/H] we present is consistent
with previous findings for AGB and RGB stars from a number of sources
and sits about the mid-point of the range of metallicities proposed
for the galaxy. The consistency between
these populations suggests little chemical evolution during their
genesis but the wider metallicity range seen if genuinely old RR Lyrae
and genuinely young A-type stars are considered (Sect. \ref{intro})
shows that the chemical enrichment of the ISM of NGC 6822 has been a
continual process across multiple stellar generations.

\subsubsection{Sensitivity to the foreground J-H cutoff}
\label{forecol}
The colour ($(J-H)_0 = 0.72$ mag) used to remove foreground contamination was selected as described in 
Sect. \ref{foreground}. \citet{2012A&A...537A.108K} used a similar colour 
($(J-H)_0 = 0.73$ mag) for the removal of the foreground based on their 
spectroscopic analysis of a small subset of the 
photometric catalogue. However, considering the effects of residual MW contamination 
in the photometric sample on our analysis of the C/M ratio and 
the underlying structure of the galaxy after the application of 
our $J-H$ colour criterion, we feel it warrants further discussion. 

Figure \ref{fore} shows the colour-colour diagram of sources from the 
galactic centre and the MW foreground and shows the position of the $J-H$ 
cut off. We see that there is some leakage from the foreground 
above this limit. Based on their spectroscopic sample, \citet{2012A&A...537A.108K} 
examine what percentage of their confirmed MW dwarfs fall in the same 
region as the AGB stars using different $J-H$ cuts. As we would expect 
from Fig. \ref{fore} this percentage is reduced with increasing values 
of $J-H$. Based on Fig. \ref{fore}, a colour selection of $(J-H)_0 = 0.80$ 
mag would reduce foreground leakage into our sample significantly, but it would also 
eliminate many genuine M-type AGB stars. Consequently, 
the C/M ratio is sensitive to any variation in the $J-H$ criterion 
used for the removal of the foreground. A change in the colour criterion 
from $0.72$ to $0.80$ mag, leaving all other selection criteria unchanged and 
with no statistical foreground removal, would reduce the total number
of remaining NGC 6822 stars from $13582$ to $7201$, of which $1880$ are 
AGB stars compared to $3755$ previously. This is very significant for the determination 
of the C/M ratio as it is primarily the number of M-type stars that is 
reduced, i.e. their number decreases from  $2901$ to $1081$ whereas the number of 
C-type stars is only reduced by $55$ from $854$ to $799$. 
The effect on the C/M ratio over the whole observed area is to increase it from 
$0.29 \pm 0.01$ ([Fe/H] $= -1.14 \pm 0.08$ dex) to $0.74 \pm 0.03$ 
([Fe/H] $= -1.33 \pm 0.06$ dex). Within the central $4$ kpc the C/M ratio 
increases from $ 0.48 \pm 0.02$ to $0.87 \pm 0.04$ ([Fe/H]$= -1.36 \pm 0.06$ dex). Some 
statistical foreground removal is still required, although this is reduced to 
$\sim 1.5$ M-type stars and $\sim 0.4$ C-type stars per kpc$^2$. With 
the inclusion of the statistical foreground subtraction, the C/M ratio within $4$
kpc of the galactic centre increases to $0.93 \pm 0.05$ ([Fe/H]$-1.37 = \pm 0.06$ dex).

Thus given the dramatic reduction in the M-type star population and 
the effect on the resultant C/M ratio, and the spectroscopic agreement 
of \citet{2012A&A...537A.108K} with our foreground selection criterion we 
are confident that a $(J-H)_0 > 0.72$ mag colour selection, with the additional 
statistical subtraction detailed in Sect. \ref{addit}, maximises the foreground 
removal whilst minimizing the effects of over subtraction on the C/M ratio 
and the derived metallicity.  
 
\begin{figure}
\resizebox{\hsize}{!}{\includegraphics[scale=0.3]{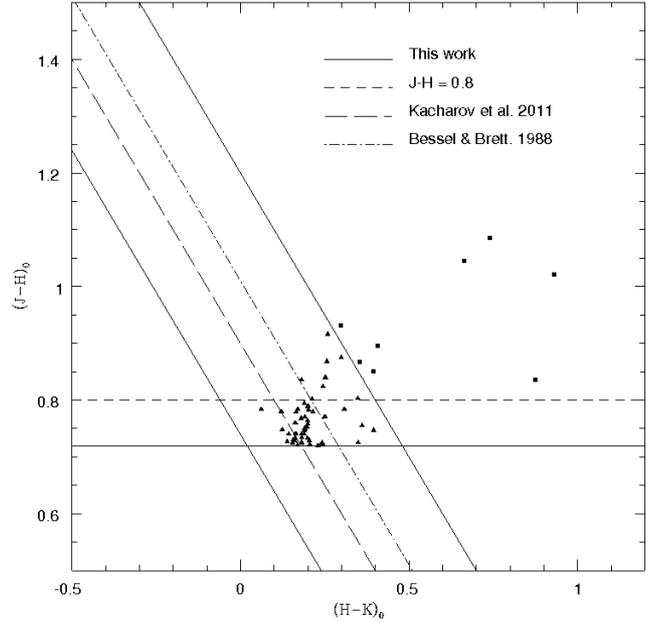}}
\caption{Colour-colour diagram showing the possible contaminant 
sources with $K_0 < 14.75$ mag discussed in the Sect. \ref{foreground}. 
C- and M-type sources are shown as squares and triangles respectively. 
The solid diagonal lines show the colour selection criteria applied 
during this study $(J-K)_0 = 0.74$ and $1.20$ mag. The horizontal 
line at $(J-H)_0 = 0.72$ mag shows the position of the foreground colour boundary. 
The long-dash diagonal line represents the blue $J-K$ boundary 
of \citep{2012A&A...537A.108K} and the dash-dot line the blue-limit of \citep{1988PASP..100.1134B}, 
Sect. \ref{blue2}. The short-dash horizontal line represents the 
alternative $(J-H)_0 = 0.80$ mag colour criterion considered in Sect. \ref{forecol}.}   
\label{sens}
\end{figure}

Fig. \ref{sens} also shows the effectiveness of a more severe 
$J-H$ cutoff in eliminating a number of the very bright sources with $K_0 < 14.75$ mag (Sect. 
\ref{foreground}) which may be MW foreground sources that survived the $(J-H)_0 > 
0.72$ mag colour selection. Interestingly, such a cut does not eliminate all the bright 
sources and as $\sim 20$ of them sit above $(J-H)_0 = 0.80$ mag they may well be genuine 
NGC 6822 sources. This possibility is supported by the sharp decline in the number of bright 
sources seen in the outer CMDs of Fig. \ref{rings}; if these sources belonged to the 
MW foreground we would expect them to be homogeneously distributed across the 
observed area and therefore to see \textit{more} of them in the outer CMDs 
that cover larger areas on the sky. This is not the case; in fact they are 
more prevalent in the central regions where NGC 6822 stars are concentrated, 
which suggests that at least some of these bright sources do belong 
to NGC 6822 - possibly red supergiants younger than the AGB population.

\subsubsection{Sensitivity to the J-K blue limit}
\label{blue2}
The blue limit is also important in the determination of the C/M
ratio. Here a limit of $(J-K)_0 = 0.74$ mag is used to exclude late
K-type stars from the sample whilst preserving as many genuine M-type
AGB's as possible. As with the foreground $J-H$ criterion, the $J-K$
blue limit can impact the number of M-type stars
significantly. \citet{2012A&A...537A.108K} use a blue limit of
$(J-K)_0 = 0.90$ mag, derived from their spectroscopic subset to
further eliminate foreground contamination from MW dwarfs from their
sample. This greatly affects the derived C/M ratio. This limit is
based on the colours of their spectroscopically confirmed
sample. However, whilst \citet{2012A&A...537A.108K} find no stars in
their spectroscopic sample that simultaneously have colours of
$(J-H)_0 > 0.73$ mag and $(J-K)_0 < 0.90$ mag, their original
spectroscopic sample was biased towards C-type stars and $3\%$ of the
photometric catalogue does fall in this region (in Catalogue 1 $10\%$
of our the candidate AGB sources lie in this region). As
  \citet{2012A&A...537A.108K} were unable to classify this $3\%$ they
  have excluded them when deriving the C/M ratio. We prefer to use a
  lower blue limit for the selection of M-type stars as discussed in
  Sect. \ref{blue}, even though a bluer $J-K$ limit may remove
  slightly more `potential' MW contaminants from the sample
  (Fig. \ref{sens}). \\In order to examine the sensitivity of the C/M
ratio to the blue limit used for the selection of M-type stars we have
applied a cutoff of $(J-K)_0 = 0.90$ mag to our AGB sample, leaving
all other selection criteria unchanged and without statistical
foreground subtraction. The resulting C/M ratio for the full observed
area increases to $0.41 \pm 0.02$ ([Fe/H]$= -1.21 \pm 0.07$ dex) and
inside the central $4$ kpc the ratio becomes C/M $= 0.58 \pm 0.03$
([Fe/H]$ = -1.28 \pm 0.07$ dex). With the inclusion of statistical
foreground subtraction (reduced to $\sim 4.8$ M-type stars per
kpc$^2$), the C/M ratio increases to  $0.69 \pm 0.03$ ([Fe/H]$ = -1.31
\pm 0.06$ dex) within $4$ kpc of the centre. \\
\citet{1988PASP..100.1134B} suggest an even redder limit of $(J-K)_0 =
1.01$ mag to completely exclude those stars with a spectral type
earlier than M$0$. Such a severe $J-K$ colour selection would also
expel more of the possible MW remainders (Fig. \ref{sens}) but the
sharp reduction in the number of M-type stars would increase the
global C/M ratio to $1.03 \pm 0.05$ ([Fe/H]$=-1.40 \pm 0.06$ dex)
within a radius of $4$ kpc (with statistical foreground removal). The
results of our sensitivity analysis are summarised in
Table. \ref{tab2}, where all values include statistical foreground
removal and [Fe/H] values have been calculated using the relation
of \citet{2009A&A...506.1137C}.  

\begin{table}   
\centering                  
\begin{tabular}{cccc}
\hline
& \multicolumn{3}{c}{$(J-K)_0$}\\
\hline
$(J-H)_0$& 0.74 &0.90 &1.01\\
\hline
0.72& $\mathrm{0.62 \pm 0.03}$& $\mathrm{0.69 \pm 0.03}$& $\mathrm{1.03 \pm 0.05}$\\
& $\mathrm{-1.29 \pm 0.07}$& $\mathrm{-1.31 \pm 0.06}$& $\mathrm{-1.40 \pm 0.06}$\\ 
0.80& $\mathrm{0.93 \pm 0.05}$& -& -\\
& $\mathrm{-1.37 \pm 0.06}$& -& -\\
\hline\\
\end{tabular}
\caption{C/M (top) and [Fe/H] (bottom) values within 
$4$ kpc of the galactic centre.}  
\label{tab2} 
\end{table}

\subsubsection{K-band and J-band criteria}
\label{jcut}
Among other NIR studies of NGC 6822, there is quite a spread in the
$K$-band TRGB magnitude, the brightest of which is $16.97$ mag \citep{2011AJ....141..194G}. 
As the C-star branch departs from the vertical column of M-type stars in the 
CMD above the TRGB. A brighter TRGB can reduce the number of M-type stars 
without any significant impact on the C-type  star count. \citet{2011AJ....141..194G} 
used a slightly modified Sobel filter with $K$-band photometry, details of 
which are given by \citet{2006ApJ...647.1056G}, to determine a TRGB magnitude 
for NGC 6822, which is much brighter than the TRGB used here. In order to 
examine the sensitivity of the C/M ratio to the TRGB magnitude we
have applied the TRGB of \citet{2011AJ....141..194G} to our sample, 
leaving all other selection criteria unchanged and with no statistical 
foreground removal. This gives a C/M ratio of $0.70 \pm 0.03$ ([Fe/H]$= 
-1.32 \pm 0.06$ dex) in the central $4$ kpc. With the inclusion of statistical 
foreground removal, this value increases to C/M $= 0.83 \pm 0.04$ ([Fe/H] $ = -1.35 \pm 0.06$ dex).

We also examine the sensitivity of the C/M ratio 
to the application of an upper bound on the $K$-band magnitude. As discussed 
in Sect. \ref{foreground}, $61$ sources which survived the $J-H$ foreground 
removal exhibited magnitudes significantly brighter than the C-type 
star branch and were considered as potential foreground interlopers. These 
sources are shown in the colour-colour diagram in Fig. \ref{sens}. It was 
decided not to eliminate these sources from our sample using another magnitude 
cut during our main analysis, although due to the relatively small number 
of sources the effect on our final results would not have been significant if we had. 
Excluding all those sources with $K_0 < 14.75$ mag and leaving all 
other selection criteria unchanged, increases the C/M ratio to $0.64 \pm 0.03$ 
([Fe/H] $= -1.30 \pm 0.06$ dex) within the central $4$ kpc (with statistical 
foreground removal).

Finally we examine the impact of applying $J$-band selection criterion to
eliminate those sources in Fig. \ref{stellCMD} (Sect. \ref{StellDen}) that we 
now believe belong primarily to the foreground but have merged into the 
C-star selection zone from below the TRGB. Based on
Fig. \ref{stellCMD} a $J$-band criteria of $J < 18.0, (J-K)_0 > 1.20$
mag was
considered. Such a cut reduces the C/M ratio to $0.49 \pm 0.02$
([Fe/H]$= -1.24 \pm  0.07$ dex) within $4$ kpc of the centre. However,
the spectroscopic sample of \citet{2012A&A...537A.108K} suggests this
would exclude a number of genuine C-type stars and heavily bias the
ratio. Based on that work, a $J$-band magnitude selection of $J<18.61$ mag  
in the region $(J-K)_0 > 1.20$ mag would seem more appropriate. When applied, this
criteria has little effect on the C/M ratio - inside the $4$ kpc
radial limit the ratio is reduced from $0.62 \pm 0.03$ to $0.58 \pm
0.03$ ([Fe/H]$=-1.28 \pm 0.07$ dex). All values are presented after
statistically foreground removal. Neither of these cuts were 
implemented during our analysis as the work of
\citet{2012A&A...537A.108K} was not available during the determination
of our selection criteria and there was no clear justification based
on our photometric data for the positioning of such a cut. The 
effect of the $J$-band criteria on the C/M ratio does not appear 
to be significant. 

We have carefully analysed the impact of various criteria on
  our selection process and the determination of the C/M ratio. The
  most important factors in selecting C- and M-type AGB candidates are
  the $J-H$ and $J-K$ boundaries and the TRGB magnitude. Using the
  extremes of these three criteria ($(J-H)_0 > 0.80, (J-K)_0 > 1.01$,
  TRGB $K_0 > 17.41$ mag) as well as the $J-$band ($ J < 18.61$ mag)
  and $K$-band ($K_0 > 14.75$ mag) criteria discussed above we would
  estimate the systematic error in our derived values to be
  $^{+0.95}_{-0.04}{(Sys)}\pm 0.03{(Rand)}$ which translates into
  a systematic error on the iron abundance of
  $^{+0.01}_{-0.26}{(Sys)} \pm 0.07{(Rand)}$ dex. However, we feel
  that both the $J-K$ value \citep{1988PASP..100.1134B} and the TRGB
  value \citep{2011AJ....141..194G} are too extreme and seriously bias
  calculated C/M ratio. Therefore, we prefer to calculate the
  systematic error based on the use of the most extreme selection
  criteria that we feel are appropriate ($(J-H)_0 > 0.80, (J-K)_0 >
  0.90$, TRGB $= 17.30, K_0 > 14.75$ and $J < 18.61$ mag) and derive a
  C/M ratio and iron abundance errors of $^{+0.45}_{-0.04}{(Sys)} \pm
  0.03{(Rand)}$ and $^{+0.14}_{-0.01}{(Sys)} \pm 0.07{(Rand)}$ dex,
  respectively.

\subsection{Comparison with other catalogues}
\label{cross}
\begin{figure} 
\resizebox{\hsize}{!}{\includegraphics[scale = 0.5]{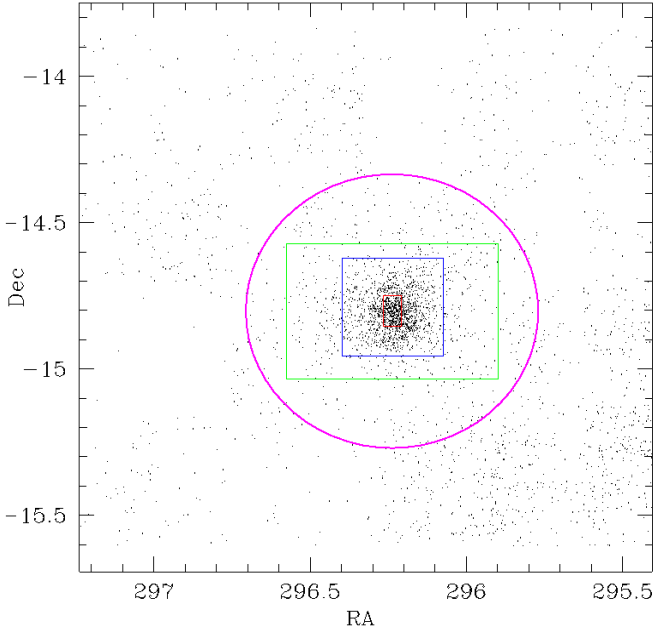}}
\caption{The relative areas observed by ourselves (whole area) and the 
observations of \citet{2006A&A...454..717K} (red rectangle), \citet{2005A&A...429..837C} 
(blue) and \citet{2002AJ....123..832L} (green). The outer circle represents the 
central $4$ kpc, the sources contained in Catalogue 1.}
\label{areas}
\end{figure}

Catalogues of the AGB population of NGC 6822 were also presented 
by \citet{2002AJ....123..832L}, \citet{2005A&A...429..837C} and 
\citet{2006A&A...454..717K} during their studies of the galaxy.
Here we compare their findings with our Catalogue 1. The area observed 
in each study is presented in comparison with the area of our 
observations in Fig. \ref{areas}.  

\subsubsection{\citet{2002AJ....123..832L}}
\label{letarte}
\citet{2002AJ....123..832L} identified $904$ carbon stars mainly
tracing a `halo-like' structure, extending beyond the optical size of
the galaxy, using $(R-I)$ and $(CN-TiO)$ criteria. In the same area 
(Fig. \ref{areas}) we find $2053$ AGB sources, of which we have been 
classified $726$ as C-type stars. The discrepancy of
$\sim 178$ sources has several causes, which we explore below. 

It may partially be the result of our misclassification
of some C-type stars as M-type, due to the inexact $J-K$ colour boundary
between the two spectral types. A comparison with the
catalogue of \citet{2002AJ....123..832L} provides a means by which
to estimate the error in the criteria applied here and
ultimately in the C/M ratio that we derived. After cross-matching 
our AGB catalogue with that of
\citet{2002AJ....123..832L}, we have identified $635$ sources in 
common. Of these $635$ sources, we have classified $80$ as M-type 
stars and $555$ as C-type stars. This may suggest that our C-type count 
should be $ \sim 1.14$ times larger, and our M-type count slightly lower. 
This would increase the final C/M ratio from $0.62 \pm 0.03$ to 
$\sim 0.77 \pm 0.03$ ([Fe/H]$=-1.34 \pm 0.06$ dex) in 
the central $4$ kpc. 

\begin{figure} 
\resizebox{\hsize}{!}{\includegraphics[scale = 0.3]{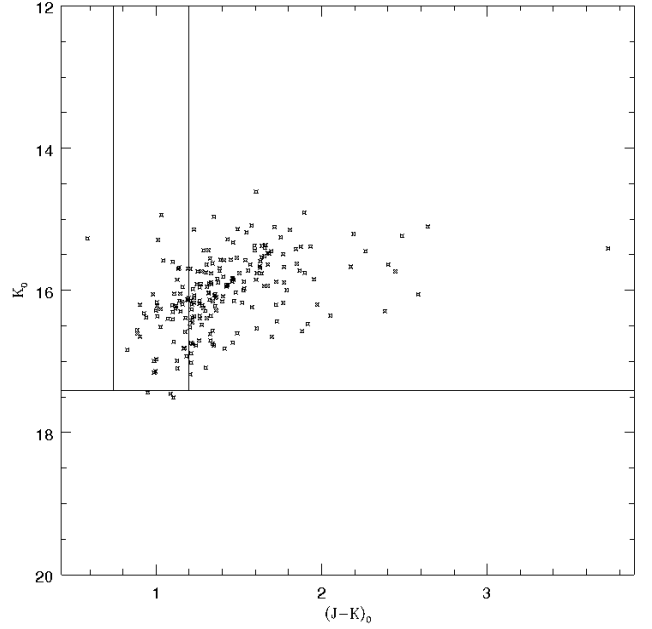}}
\caption{CMD of $217$ of the 
\citet{2002AJ....123..832L} sources we discarded, see text for details. The vertical and
    horizontal lines represent the TRGB ($K=17.41$ mag), blue limit ($(J-K)_0
    = 0.74$ mag) and colour-separation ($(J-K)_0 = 1.20$ mag) respectively.}
\label{LetRemCMD}
\end{figure}

Of the remaining $269$ sources from the catalogue of 
\citet{2002AJ....123..832L} that were not in our AGB catalogue, $235$ 
were identified among the sources that we discarded due to the poorer 
quality of the photometric 
data (i.e. sources that were classified as being something other than 
stellar or probably-stellar in at least one band). Of these $235$ 
sources, $217$ have J and K band magnitudes available that allow
us to classify $54$ as M-type, $162$ as C-type 
stars and one source that falls below the blue limit. From their position on the CMD
(Fig. \ref{LetRemCMD}), these sources are likely to be real C- or
M-type stars, which were discarded from our sample to maintain
photometric reliability. If these stars are genuine C-type stars as 
\citet{2002AJ....123..832L} conclude, then our C-star count should be 
$\sim 1.34$ larger than we claim. Taken together with the multiplier 
$1.14$ above, the $J-K$ criterion we have adopted may
misclassify about $\sim 20\%$ of C-stars as M-type stars. Making a 
correction of this magnitude would increase the C/M ratio from $0.62 \pm 0.03$
to $\sim 0.85 \pm 0.04$ ([Fe/H]$= -1.36 \pm 0.06$ dex) in the 
central $4$ kpc. This is based on the crucial assumption that 
\citet{2002AJ....123..832L} have correctly
classified all of the sources in their catalogue. In order to verify
this assumption spectroscopic data which is not currently
available would be needed. On the other hand, the classification 
by \citet{2002AJ....123..832L} is dependent
solely on colour and it is possible that there have been a number of
misclassifications of objects which are not actually 
C-type stars.

The final $34$ C-type stars found by 
\citet{2002AJ....123..832L}, we are unable to account for. As 
\citet{2002AJ....123..832L} studied the central regions where 
the stellar density is highest, the outstanding $34$ stars may have 
been excluded from our photometric catalogue due to crowding issues.

\subsubsection{\citet{2006A&A...454..717K}}
\label{kang}

\citet{2006A&A...454..717K} used $giJHK$ photometry to identify 
$663$ AGB stars along the bar of NGC 6822. Of these $663$
sources, they classified $522$ as M-type stars and $141$ as C-type stars. 
They used different criteria for the 
selection and identification of their AGB population, using the 
$(g, g-K)$ CMD to separate the AGB stars from the MW
foreground. Examining our AGB catalogue in the same area we have
identified $411$ AGB stars: $160$ C-type and $251$ M-type, i.e. 
a much lower C/M ratio. The difference of $\sim 250$ sources detected by ourselves and
\citet{2006A&A...454..717K} is likely due to the stringent
conditions we placed on the photometry and selection of our sample. 

\begin{figure} 
\resizebox{\hsize}{!}{\includegraphics[scale = 0.3]{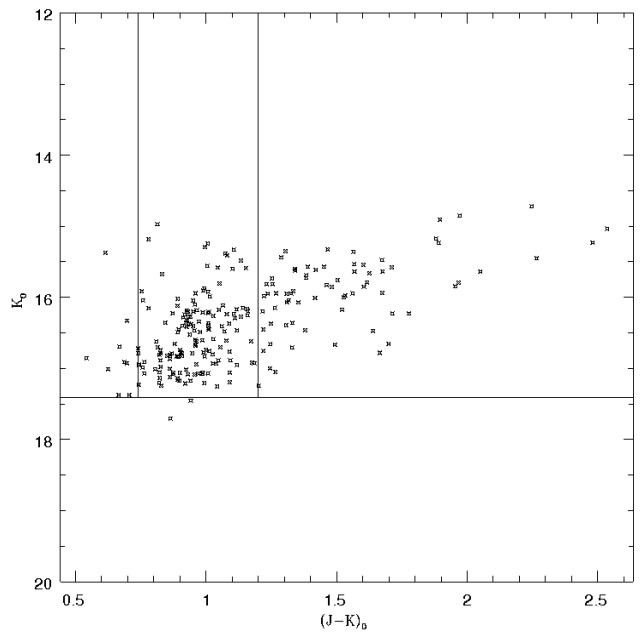}}
\caption{CMD of the
    \citet{2006A&A...454..717K} sources we discarded, see text for
      details. The vertical and horizontal lines are the same as in
      Fig. \ref{LetRemCMD}.}
\label{KangRemCMD}
\end{figure}

Cross-matching our AGB catalogue with the catalogue of
\citet{2006A&A...454..717K} we were able to identify $294$ stars in
common, of which we identified $123$ as C-type stars and $171$ as M-type, 
whilst \citet{2006A&A...454..717K} had identified $81$ C-type and
$213$ M-type stars. Of the remaining $369$ sources in the
\citet{2006A&A...454..717K} catalogue that were not identified in our
AGB catalogue, $63$ were identified among the sources that we
discarded as MW foreground, all of which were M-type AGB's according to
\citet{2006A&A...454..717K}. This is not surprising given our findings 
in Sect. \ref{StellDen} that the majority of the remaining MW interlopers masquerade
as M-type stars not C-type. We suspect therefore that the \citet{2006A&A...454..717K} 
M-type AGB stars include many MW interlopers. A further $266$ of the sources in the
\citet{2006A&A...454..717K} catalogue were identified among the
sources that were discarded from our data set as they were not 
classified as stellar or probably stellar in all bands. Of these 
$266$ sources there were $80$ C-type and $163$ M-type 
stars using our selection criteria ($J$ or $K$ magnitudes were not 
available for $23$ of the sources) and
$53$ C-type and $213$ M-type stars using the selection criteria of
\citet{2006A&A...454..717K}. The discrepancies in the number of stars
of each type found are the result of different colour selection
criteria. The CMD of the \citet{2006A&A...454..717K}
sources identified in our discarded sample (Fig. \ref{KangRemCMD})
shows that the majority have colours consistent with them being 
genuine AGB stars or MW interlopers. Forty sources from \citet{2006A&A...454..717K} 
were not identified 
in our data set; of these \citet{2006A&A...454..717K} identified $7$ 
C-type and $33$ M-type stars.   

In light of the high number of sources that we have matched 
with the catalogues of \citet{2006A&A...454..717K} and \citet{2002AJ....123..832L} 
but that were discarded from our sample due to our strict reliability 
criteria we have presented Catalogue 3 (Sect. \ref{cat}). Catalogue 3 contains the 
sources that were identified as stellar or probably-stellar in only two 
photometric bands, many of these sources are likely to be good AGB 
candidates as shown in Figures \ref{LetRemCMD} \& \ref{KangRemCMD}, but 
were excluded during our initial analysis as we insisted on a stellar 
or probably-stellar classification in all three photometric bands.

\subsubsection{\citet{2005A&A...429..837C}}
\label{cioni}
The catalogue of \citet{2005A&A...429..837C} contains $16,354$ stars
detected in the $J$- and $K_{s}$-bands in the central $20' \times 20'$
of NGC 6822. Without the $H$-band, \citet{2005A&A...429..837C}
could not apply the $J-H$ mag foreground rejection method used here.  
They did attempt to reduce the foreground contamination by selecting 
only those sources with $(J-K)_0 > 0.80$ mag; see Fig. \ref{fore} which 
shows that boundary in the face of our sample, revealing substantial 
leakage of MW stars into a sample selected in that way. 
\citet{2005A&A...429..837C} identified 
$1511$ C-type and $4684$ M-type stars. From our AGB
catalogue we find $1623$ AGB stars in the same area: $600$ C-type and
$1023$ M-type stars. However, we are unable to make
a direct comparison with our AGB catalogue as the poor quality 
of their astrometry and the high source density in this region 
prevents a reliable cross-identification between our 
sources and theirs.

\subsection{The future of NIR selection}
\label{nir}
We have relied on high quality $JHK$ photometry to determine the C/M
ratio of NGC 6822. As AGB stars are amongst the coolest and brightest
sources in the  intermediate age population, the infrared is the most
obvious waveband for the  selection of these sources. However, this
method is at a  disadvantage compared to spectroscopic classification
methods due to the difficulty of distinguishing  between C- and M-type
stars less than $1$ mag brighter than the TRGB  in
$J-K$. \citet{2007A&A...474...35B} comments  and we likewise recognise
that we are still  far from any consistent criteria for the selection
of sources. In particular  C/M ratios determined from NIR colours will
be underestimates due mainly  to the contamination of the M-star count
from hotter C-type stars and also  K-stars if no blue limit is
applied. The  range in colour limits is often attributed to the
metallicity of the different  parent galaxies. As discussed above,
whilst NIR observations are not greatly affected by  reddening they do
suffer from uncertainties due to  metallicity and age in the
population \citep{2005MNRAS.357..669S}.  

Both \citet{2006pnbm.conf..108G} and \citet{2005A&A...434..657B}
consider selective narrow- and broad-band optical photometry
\citep{1986ApJ...305..634C} to be the  best way of selecting C- and
M-type stars in large scale surveys in resolved  galaxies. In a
colour-colour diagram of $CN-TiO$ vs. $V-I$ (or $R-I$), carbon-  and
oxygen-rich stars are clearly separated. However, this technique is
also  flawed as there is no easy way to eliminate foreground M-dwarfs
\citep{2002AJ....123..832L} and it does  not identify bluer AGB stars
well as they merge   with the rest of the stellar population
\citep{2005A&A...434..657B}.  \citet{2006pnbm.conf..108G} also notes
that a lower limit in $V-I$ (or  $R-I$) is usually selected to isolate
oxygen-rich stars of type  M0 or later but that when this same limit
is applied to the C-type population it  neglects the hotter C-type
stars and thereby biases the C/M ratio.  

The use of optical bands is also limited by increasing extinction in a
way that the infrared is not. However, narrow-band survey data is
already available for a number of Local Group galaxies
\citep{2006pnbm.conf..108G}.  Therefore the most logical course of
action would seem to be to use the  method of
\citet{1986ApJ...305..634C} alongside NIR data in the Local Group to
tighten the selection criteria and to gain a better understanding of
the effects of age and metallicity on $JHK$ colours. We refer the
reader to
\citet{2006pnbm.conf..108G,2005A&A...434..657B,2006pnbm.conf..108G}
and references therein for a brief overview of recent surveys of C-
and M-type AGB's in the Local Group and a comparison of selection
techniques.

\section{Conclusions}
\label{concl}
High quality $JHK$ photometry of an area of $\sim 3$ deg$^{2}$ centred on NGC 6822 
has been used to isolate the AGB population and to study the spatial distribution 
of stars and the C/M ratio as a tracer of metallicity. We have 
investigated the spread in the TRGB magnitude, the spatial distribution of 
[Fe/H] as a function of azimuthal angle and radial distance 
and also the sensitivities of the C/M ratio and abundance 
to the applied selection criteria. 

\noindent Our main conclusions are as follows:

\begin{enumerate}
\item The $J-H$ colour used for the star-by-star removal of the foreground MW 
contamination is very valuable but with such heavy foreground contamination as 
we have in the direction of NGC 6822 further statistical foreground subtraction 
is required. We have demonstrated the difficulties involved in isolating the 
M-type AGB population from the MW foreground using only $JHK$ photometric 
colours and have established the sensitivity of the C/M ratio to the $J-H$ cutoff and a number 
of other selection criteria.

\item The TRGB magnitude was found at $K_0 = 17.41 \pm 0.11$ mag. Random statistical scatter 
in the measurements of the TRGB are to be expected and the range of $0.19$ mag that 
we detect is within $\pm 2\sigma$ of the mean value. Our measurements suggest 
that the TRGB magnitude may decline (brighten) as a function of increasing radial distance from 
the galactic centre, possibly due to the outer population being older.  

\item We trace the AGB population out to a radius of $4$ kpc from the 
centre of NGC 6822. Beyond this, genuine NGC 6822 sources cannot be 
cleanly separated from the heavy MW foreground contamination.
\item The colour boundary between C- and M-type stars has a mean value and standard deviation of 
$(J-K)_0 = 1.20 \pm 0.03$ mag ($(J-K)_{2MASS} = 1.28$ mag), with a
spread of $0.1$ mag detected within the AGB population. This is
consistent with previous studies of NGC 6822 but we note, in agreement
with  \citet{2007A&A...474...35B} that this boundary is ill defined 
and that some misclassification of C- and M-type stars occurs. 
Due to the sensitivity of the C/M ratio to this criterion, more 
analysis of spectroscopic data of different metallicity environments is needed to 
constrain this boundary and gain a better understanding of its dependence on metallicity.   

\item The blue limit ($(J-K)_0 = 0.74$ mag) used to isolate M-type AGB stars is also 
important and can severely affect the determination of the C/M ratio. A clearer, 
standardised limit for the exclusion of $K$-type contaminants is needed in order 
to compare C/M ratio determinations from different authors so they can be used to 
better calibrate the C/M vs. [Fe/H] relation.    
 
\item Within a $4$ kpc radius a global [Fe/H] $= -1.29 \pm 0.07$ 
dex was derived from a C/M ratio of $0.62 \pm 0.03$, using the relation 
of \citet{2009A&A...506.1137C}. A spread of $0.18$ dex is found but 
there is no metallicity gradient present either as a function of radial 
distance or as a function of angle. The clumpy distribution of the C/M 
ratio in Fig. \ref{maps} is consistent with the findings of other authors 
\citep{2005A&A...429..837C,2006A&A...454..717K} and is probably real. 
Although whether this relates directly to the metallicity of the
region is not entirely clear given recent findings concerning the
impact of population age of the C/M ratio. A variation in the global
[Fe/H] abundance of $0.11$ dex is seen when individual selection
parameters are varied, or potential much larger if several are altered
at one time (Sect. \ref{jcut}) and for comparison we note a variation of
$0.47$ dex in the global iron abundance when older C/M
vs. [Fe/H] relations are used (Sect. \ref{C/Mvar}).

\item There is a possible error of $\sim 20\%$ in the classification 
of C-type stars, based on a comparison with the work of \citet{2002AJ....123..832L}, 
in the sense that photometrically we misclassify $\sim 25\%$ of C-type 
stars as M-type. Correcting for this would increase our C/M ratio to 
$0.85 \pm 0.04$ ([Fe/H] $= -1.36 \pm 0.06$ dex). A spectroscopic comparison of the 
C- and M-type sources identified using optical and NIR photometry, is needed 
in order to properly constrain the level of error introduced by both methods.  

\item The C/M ratio is a useful tool for gaining a broad overview of the 
metallicity in a distant but resolved galaxy but the correlation between 
C/M and [Fe/H] is not tight, especially at lower metallicities, as demonstrated 
by Fig. B.1 of \citet{2009A&A...506.1137C}, Fig. 3 of \citet{2005A&A...434..657B} 
and Fig. 4 of \citet{2006pnbm.conf..108G}. An improved calibration of the C/M vs. [Fe/H] relation is 
required, as is a better understanding of the other factors that affect the C/M 
ratio such as the age of the population and the effects of foreground contamination 
and differential reddening. For instance \citet{2010MNRAS.tmpL.129F} conclude that a decline in 
the C/M ratio at greater radial distances is more likely to be the 
result of the increasing age of the population, and the resulting 
decline in the number of C-type stars, rather than an increase in 
the population metallicity. As mentioned in Sect. \ref{metgrad}, 
\citet{2010MNRAS.tmp..431H} also note the importance of the 
population age in the interpretation of the C/M ratio. We expect that the 
age (mass) dependence of C-type star formation will become increasingly important in future 
endeavors to better constrain the C/M vs. [Fe/H] relation and must therefore be 
taken into account during the interpretation of any results.\\With a better 
calibration and more uniform 
treatment of the C/M ratio (in the optical and IR) over a range of metallicities and 
the use of spectroscopic indicators, the C/M ratio has the potential to be a more powerful 
tool for the study of metallicity gradients in galaxies that can be
resolved into stars. 
\end{enumerate}

\begin{acknowledgements}
We would like to thank M. Rejkuba of ESO and many colleagues at the
University of Hertfordshire for useful discussions. We would also 
like to extend our thanks to the referee for providing a helpful 
and insightful report which has improved this work.   
\end{acknowledgements}

\bibliographystyle{aa}
\bibliography{liz2}

\end{document}